\documentclass[a4paper,10pt]{article}
\pdfoutput=1 

\usepackage{jheppub} 
\usepackage[export]{adjustbox}
\usepackage[utf8]{inputenc}
\usepackage{multirow}
\usepackage{bbold}
\usepackage[table]{xcolor}
\usepackage{slashed}
\usepackage{bm}
\usepackage{subfig}
\usepackage{float}
\usepackage{fancyvrb}
\usepackage{fvextra}
\usepackage{soul}
\usepackage{comment}
\usepackage{cancel} 
\usepackage{csquotes} 
\usepackage[title]{appendix}

\usepackage{subfig}
\usepackage{color,graphicx,slashed,hyperref}
\usepackage[utf8]{inputenc}
 \usepackage{verbatim}
\usepackage{setspace}

\usepackage{booktabs}
\usepackage{amstext} 
\usepackage{array}   
\newcolumntype{C}{>{$}c<{$}} 
\newcolumntype{L}{>{$}l<{$}} 

\newcommand{\dd}[1][d]{{\rm d}}

\title{The Fate of Black Hole-Induced Moduli Excursions in the Presence of Scalar Potentials}

\author[a]{Karim Benakli,}
\emailAdd{kbenakli@lpthe.jussieu.fr}

\author[a]{Anna Chrysostomou}
\emailAdd{chrysostomou@lpthe.jussieu.fr}

\affiliation[a]{Laboratoire de Physique Th\'eorique et Hautes \'Energies - LPTHE, Sorbonne Universit\'e, CNRS, 4 Place Jussieu, 75005 Paris, France}

\abstract{
Large charged black holes can create macroscopic, locally weakly curved
regions in which moduli take values different from their asymptotic values.
We study how robust this mechanism is once the scalar has a nontrivial
potential.  In four-dimensional Einstein--Maxwell--dilaton theory, the
massless GHS solution provides a finite exterior throat in which the scalar
and the gauge coupling vary logarithmically.  We develop fixed-throat
diagnostics for the competition between the black hole gauge source and a
scalar potential, and compare them with back-reacted exterior evolutions when
needed.  The relevant criterion is not the mere presence of a potential, but
how its force behaves along the scalar trajectory traced by the black hole
throat.  Quadratic stabilizing potentials erase the throat when the Compton
wavelength becomes comparable to the horizon scale.  Runaway, periodic, and
barrier-type potentials instead exhibit distinct failure modes controlled by
their slope, sign, oscillations, or barrier distance along the GHS trajectory.
A quintessence-like scalar remains effectively massless on astrophysical
black hole scales, leaving the throat essentially unobstructed.  If the
charge belongs to a hidden sector, and if the scalar also controls visible
couplings or bulk propagation, such surviving altered-modulus regions could
leave phenomenological imprints in near-horizon accretion or emission.
}


\begin{document}
\maketitle
\flushbottom

\newpage

\section{Introduction}
\label{sec:introduction}

Moduli fields occupy an unusual position in string theory.  They are not
external parameters fixed by hand, but dynamical scalars controlling the
geometry of the compactification and, through it, the masses, couplings and
thresholds of the effective theory.  Their cosmological role, and the
difficulty of stabilizing them in realistic compactifications, are reviewed
for example in Ref.~\cite{Cicoli:2023opf}.  A natural question is therefore
how far such fields can be displaced by physical processes.  Can a
gravitational configuration create a finite region of space in which the
local values of the moduli differ substantially from their asymptotic
values, while remaining within a single underlying theory?

Sen recently proposed a particularly concrete realization of this idea.  In
a theory with massless moduli, a sufficiently large charged black hole can
produce a macroscopic region in which the local curvature, field-strength
invariants, and scalar gradients are small, while the moduli take values
different from those at infinity~\cite{Sen:2025bmj}.  In a more explicit string-theory construction, suitable charged black hole backgrounds can contain regions where the local physics is effectively that of
ten-dimensional type~IIA or type~IIB string theory, or of eleven-dimensional M-theory, even though the asymptotic theory is four-dimensional and asymptotically flat~\cite{Sen:2025ljz}.  The underlying mechanism is simple but striking: the large-size limit makes the background locally weakly-curved, while the moduli can still vary by an order-one amount across the large region.

This perspective is closely related to the broader use of charged black
holes as probes of moduli-space geometry.  In \(4d\), \(\mathcal N=2\)
compactifications, the attractor mechanism fixes the near-horizon values of vector-multiplet moduli in terms of black hole charges.  This allows physical black hole observables to encode data about regions of moduli space far from the asymptotic vacuum~\cite{Ferrara:1995ih,Ferrara:1996dd,Ferrara:1997tw,Moore:1998pn,Denef:2000nb,Delgado:2022dkz}.  In this sense, black holes provide finite-energy probes of field-space regions that would be difficult to access by ordinary local experiments.

This competition has a long history.  Already for charged black holes with a dilaton of mass \(m_\phi >0\), it was shown that adding a potential changes the horizon structure, modifies the extremal limit, and raises the question of whether the infinite throat of the massless solution survives~\cite{Gregory:1992kr,Horne:1992bi}.  In particular, the relevant comparison is between the black hole size and the Compton wavelength of the dilaton: black holes much larger than
\(m_\phi^{-1}\) resemble Reissner--Nordström solutions with the scalar pinned near its minimum, whereas black holes smaller than this scale behave much like the massless dilaton solutions.  Related work has also studied charged black holes in the presence of scalar potentials, moduli stabilisation, and flux compactification effects~\cite{Green:2006nv,Danielsson:2006xw,Delgado:2025crl}.

It is useful to distinguish the different large-region limits involved.  In
Sen's toroidal example, one may write
\[
    a_{\rm Sen}=r_+-r_- \;,
    \qquad
    b_{\rm Sen}=r-r_+ \;,
\]
both dimensionful.  The modulus factor controlling the large internal
radius is
\begin{equation}
    \Xi_{\rm Sen}
    =
    \frac{r_+}{a_{\rm Sen}+b_{\rm Sen}} \;.
    \label{eq:intro-sen-xi}
\end{equation}
During the large-size scaling, this ratio is chosen to be large and held fixed,
while the overall length scales are taken still larger to keep the
higher-dimensional curvature and field-strength invariants small.  Thus, the
large modulus displacement is not produced by large local curvature, but by
a large near-horizon modulus/redshift factor embedded in an even larger,
locally weakly-curved geometry.

Here, we isolate the simplest four-dimensional avatar of this mechanism: a single canonically normalized scalar coupled to a \(U(1)\) gauge kinetic function in Einstein--Maxwell--dilaton theory.  In the absence of a scalar potential, the corresponding static charged solutions are the standard four-dimensional dilatonic black holes~\cite{Gibbons:1987ps,Garfinkle:1990qj}.  We will use the string-theory form of this solution and refer to it as the GHS solution.  It
should be distinguished from the higher-dimensional charged brane solutions used in Sen's explicit string construction~\cite{Horowitz:1991cd}.  In our notation,
\[
    a=\frac{r_+-r_-}{r_+} \;,
    \qquad
    b=\frac{r-r_+}{r_+} \;,
\]
so that \(\Xi_{\rm Sen}=1/(a+b)\) in the corresponding near-horizon regime.
The massless GHS scalar profile is
\begin{equation}
    \phi_{\rm GHS}(b)-\Phi_\infty
    =
    \frac{M_{\rm Pl}}{2\alpha}
    \log\frac{1+b}{a+b} \;,
    \qquad
    \Phi_\infty\equiv \phi(r\to\infty) \;.
    \label{eq:intro-ghs-profile}
\end{equation}
It produces a finite exterior radial region outside the horizon,
\(\Delta r=r-r_+=r_+b>0\), in which the scalar, and therefore the gauge
coupling, differ from their asymptotic values.  If \(r_+\) is large,
this region is macroscopic; if \(a\) is small, the scalar excursion is
logarithmically enhanced. While the analogous modulus ratio is chosen large and held fixed during the large-size scaling in Sen's work, we choose instead to expose the parametric behaviour by taking the near-extremal limit \(a\to0\) in our
four-dimensional analysis.

Our question is then whether this Sen-like exterior scalar region survives
once the scalar is given a potential \(V(\phi)\).  We do not attempt to
reconstruct the full higher-dimensional embedding.  Instead,
we develop local and cumulative diagnostics for a finite GHS throat deformed
by a scalar potential.  The pointwise force ratio tells us where
\(V_{, \phi}(\phi)\) first competes with the gauge source; the local and cumulative
flux measures test whether this competition accumulates across the throat;
the direct deformation calculation gives the induced shift in the
scalar profile and in the local gauge coupling.

We apply these diagnostics to several representative potential shapes:
quadratic stabilizing potentials, shifted exponentials, exponential
runaways, inverse-power runaways, racetrack barriers, axion-like periodic
potentials, and supergravity-inspired corrections.  Whenever possible, we
compare the fixed-throat estimates with back-reacted exterior evolutions of
the coupled Einstein--Maxwell--scalar system.  The goal is not to decide
which complete compactification, if any, realizes the scenario, but to
identify which qualitative features of a potential erase, deform, or allow
a macroscopic charged throat with displaced scalar values.

The paper is organized as follows.
Section~\ref{sec:static-emd-system} sets up the static
Einstein--Maxwell--dilaton (EMD) system and fixes conventions.
Section~\ref{sec:ghs-solution} reviews the massless GHS throat and
introduces the radial coordinate used throughout.
Section~\ref{sec:static-deformations} develops the fixed-throat diagnostics,
including the pointwise ratio \(\eta_{\rm src}\), the local and cumulative
flux measures \(\epsilon_{\rm loc}\) and \(\epsilon_{\rm cum}\), respectively, and the
direct deformation solution.
Section~\ref{sec:static-potential-classes} applies these diagnostics, and
the corresponding back-reacted checks, to each class of potential.
Section~\ref{sec:pheno-outlook} discusses the phenomenological reading of
the results, including hidden-sector charge and quintessence-like scalars.
Section~\ref{sec:conclusions} summarizes the hierarchy of outcomes.

\section{Static Einstein--Maxwell--dilaton system}
\label{sec:static-emd-system}

We begin with the four-dimensional Einstein-frame action,
\begin{equation}
    S =
    \int d^4x \sqrt{-g}
    \left[
        \frac{M_{\rm Pl}^2}{2} R
        - \frac{1}{2}(\partial\phi)^2
        - V(\phi)
        - \frac{1}{4} B(\phi) F_{\mu\nu}F^{\mu\nu}
    \right] \;,
    \label{eq:static-action}
\end{equation}
where \(B(\phi)\) is the gauge kinetic function.  The corresponding local
gauge coupling follows from canonically normalizing the gauge field kinetic
term,
\begin{equation}
    g^2(\phi) = \frac{1}{B(\phi)} \;.
    \label{eq:gauge-coupling}
\end{equation}
For most of the discussion, we have in mind an exponential dilatonic coupling,
\begin{equation}
    B(\phi)=e^{-2\alpha \phi/M_{\rm Pl}} \;,
    \qquad
    g^2(\phi)=e^{2\alpha \phi/M_{\rm Pl}} \;.
    \label{eq:exponential-gauge-kinetic}
\end{equation}

Varying Eq.~\eqref{eq:static-action} with respect to the metric, the gauge
field, and the scalar, respectively, gives
\begin{align}
    M_{\rm Pl}^2 G_{\mu\nu}
    &=
    \partial_\mu\phi\,\partial_\nu\phi
    - \frac{1}{2}g_{\mu\nu}(\partial\phi)^2
    - g_{\mu\nu}V(\phi)
    + B(\phi)
    \left(
        F_{\mu\rho}F_{\nu}{}^{\rho}
        - \frac{1}{4}g_{\mu\nu}F_{\rho\sigma}F^{\rho\sigma}
    \right) \;,
    \label{eq:einstein-eq-general}
    \\
    \nabla_\mu\left(B(\phi)F^{\mu\nu}\right)
    &=0 \;,
    \label{eq:maxwell-eq-general}
    \\
    \Box\phi
    &=
    V_{, \phi}(\phi)
    + \frac{1}{4}B_{, \phi}(\phi)F_{\mu\nu}F^{\mu\nu} \;.
    \label{eq:scalar-eq-general}
\end{align}
The scalar is sourced by the gauge invariant \(F^2\), weighted by the
derivative \(B_{, \phi}(\phi)\).  In the massless case \(V=0\), this source alone
drives a nontrivial radial profile, producing the GHS throat discussed in
Section~\ref{sec:ghs-solution}.  Once a potential is included, it is the
competition between \(B_{, \phi}(\phi)F^2/4\) and \(V_{, \phi}(\phi)\) that determines whether
a throat survives.

We consider static, spherically symmetric configurations, under the
Einstein-frame radial gauge
\begin{equation}
    ds^2
    =
    - e^{2\delta(r)} f(r)\, dt^2
    + \frac{dr^2}{f(r)}
    + r^2 d\Omega_2^2 \;,
    \label{eq:static-spherical-metric}
\end{equation}
which is adapted to non-extremal and extremal static horizons through the
zero structure of \(f(r)\), while \(\delta(r)\) encodes the departure from the
standard Schwarzschild form.  In the explicit GHS solution discussed below, we
will also use a Schwarzschild-like radial coordinate in which the areal radius
is denoted \(R(r)\).  The two descriptions are related by a radial
reparametrization; the source-dominance criterion itself is independent of
this choice of radial gauge.

For a static scalar \(\phi=\phi(r)\), the d'Alembertian reduces to
\begin{equation}
    \Box\phi
    =
    \frac{1}{e^{\delta}r^2}
    \frac{d}{dr}
    \left(
        e^{\delta} r^2 f \phi'
    \right),
    \label{eq:static-box-phi}
\end{equation}
where a prime denotes \(d/dr\), so the radial scalar equation is
\begin{equation}
    \frac{1}{e^{\delta}r^2}
    \frac{d}{dr}
    \left(
        e^{\delta} r^2 f \phi'
    \right)
    =
    V_{, \phi}(\phi)
    + \frac{1}{4}B_{, \phi}(\phi)F_{\mu\nu}F^{\mu\nu}.
    \label{eq:radial-scalar-equation-general}
\end{equation}

\subsection*{Magnetic and electric configurations}

It is useful to treat the magnetic and electric cases separately, primarily
because the sign of \(F^2\) differs between them; this sign controls whether
the gauge source pushes \(\phi\) in the same or opposite direction as the
potential gradient.

For a magnetic monopole of charge \(Q_m\),
\begin{equation}
    F_{\theta\varphi}=Q_m\sin\theta ,
    \qquad
    F_{\mu\nu}F^{\mu\nu}
    =
    \frac{2Q_m^2}{r^4} > 0 \;.
    \label{eq:magnetic-field}
\end{equation}
Substituting into Eq.~\eqref{eq:radial-scalar-equation-general}, the magnetic
scalar equation reads
\begin{equation}
    \frac{1}{e^{\delta}r^2}
    \frac{d}{dr}
    \left(
        e^{\delta} r^2 f \phi'
    \right)
    =
    V_{, \phi}(\phi)
    + \frac{1}{2}B_{, \phi}(\phi)\frac{Q_m^2}{r^4} \;.
    \label{eq:magnetic-radial-scalar-equation}
\end{equation}

For an electric configuration, the \(\nu=t\) component of Maxwell's equation,
Eq.~\eqref{eq:maxwell-eq-general}, requires
\begin{equation}
    \frac{1}{\sqrt{-g}}\,
    \partial_r\!\left(\sqrt{-g}\,B(\phi)\,F^{rt}\right) = 0
    \quad\Longrightarrow\quad
    B(\phi)\,e^{+\delta}\,r^2\,F^{rt} = Q_e \;,
    \label{eq:maxwell-electric}
\end{equation}
where \(Q_e\) is the conserved electric charge.  Writing \(E(r)\equiv F_{tr}\)
for the radial electric field, the metric Eq.~\eqref{eq:static-spherical-metric}
gives \(F^{rt}=e^{-2\delta}E\), so that
\begin{equation}
    E(r)=\frac{Q_e\,e^{\delta}}{B(\phi)\,r^2} \;.
    \label{eq:electric-field}
\end{equation}
Then
\begin{equation}
    F_{\mu\nu}F^{\mu\nu}
    =
    2F_{tr}F^{tr}
    =
    -2e^{-2\delta}E(r)^2
    =
    - \frac{2Q_e^2}{B(\phi)^2 r^4} < 0 \;.
    \label{eq:electric-f2}
\end{equation}
The negative sign, opposite to the magnetic case, is a direct consequence of
the Lorentzian signature: electric and magnetic fields contribute with
opposite signs to \(F^2\).  The electric scalar equation is therefore
\begin{equation}
    \frac{1}{e^{\delta}r^2}
    \frac{d}{dr}
    \left(
        e^{\delta} r^2 f \phi'
    \right)
    =
    V_{, \phi}(\phi)
    -
    \frac{1}{2}
    \frac{B_{, \phi}(\phi)}{B(\phi)^2}
    \frac{Q_e^2}{r^4} \;.
    \label{eq:electric-radial-scalar-equation}
\end{equation}
Defining the inverse gauge kinetic function
\begin{equation}
    \widetilde B(\phi)\equiv\frac{1}{B(\phi)}=g^2(\phi) \;,
    \label{eq:btilde-def}
\end{equation}
the electric source can be written compactly as
\begin{equation}
    \frac{1}{2}\widetilde B_{, \phi}(\phi)\frac{Q_e^2}{r^4} \;,
    \qquad
    \widetilde B_{, \phi}(\phi)=-\frac{B_{, \phi}(\phi)}{B(\phi)^2} \;.
    \label{eq:electric-source-btilde}
\end{equation}
This makes the electric--magnetic duality transparent: magnetic charges
source \(\phi\) through \(B_{, \phi}(\phi)\), while electric charges source it through
\(\widetilde B_{, \phi}(\phi)=(B^{-1})'(\phi)\).  For the exponential coupling
Eq.~\eqref{eq:exponential-gauge-kinetic}, this means that electric and
magnetic sources drive \(\phi\) in opposite directions in field space for the
same sign of \(\alpha\).

In what follows, we work in the magnetic frame; this is a choice of
convenience rather than a restriction.  For magnetic charge, the flux
\(F_{\theta\varphi}=Q_m\sin\theta\) is fixed independently of the scalar
profile, so the gauge source enters the scalar equation directly as
\(B_{, \phi}(\phi)Q_m^2/(2r^4)\).  The electric case instead requires first solving
Gauss' law for \(E(r)\), which trades the field strength for the conserved
charge \(Q_e\), and effectively replaces \(B(\phi)\) by its inverse
\(\widetilde B(\phi)=B(\phi)^{-1}\).  The two cases are therefore related by
electric--magnetic duality combined with \(B\leftrightarrow B^{-1}\), which
for the exponential coupling amounts to \(\alpha\to-\alpha\): the scalar is
pushed in the opposite direction in field space.  Since the source-dominance
criterion below depends only on the local balance between the gauge force and
\(V_{, \phi}(\phi)\), the magnetic frame provides the simplest representative
description.  Electric solutions are obtained by the substitution
\(B\to B^{-1}\), together with the corresponding reversal of the scalar
excursion; for asymmetric potentials this reversal should of course be
applied before comparing the two sides of field space.

\subsection*{Source-dominance criterion}

In the static problem, we impose regularity at the future horizon and a fixed
asymptotic value \(\phi(r\to\infty)=\Phi_{\infty}\).  When \(V(\phi)=0\), the gauge
source alone supports a nontrivial regular scalar profile.  When
\(V(\phi)\neq0\), the profile is determined by the local competition between
the gauge source and the potential gradient.  Because both terms are functions
of \(r\), the competition is resolved pointwise: the gauge source dominates in
a region around the horizon where
\begin{equation}
    \left| 
        \frac{1}{4}B_{, \phi}(\phi)F_{\mu\nu}F^{\mu\nu}
    \right|
    \gg
    \left| V_{, \phi}(\phi) \right| \;,
    \label{eq:source-dominance-general}
\end{equation}
and the potential takes over in the outer region once the gauge-field source,
diluted radially, becomes smaller than the potential force.  In the electric
frame, the same condition holds with \(B\) replaced by
\(\widetilde B = B^{-1}\).

The following sections make this criterion quantitative by using the massless
GHS solution as the benchmark throat geometry, and treating the potential as a
controlled deformation of the massless scalar profile.



\section{The massless GHS throat}
\label{sec:ghs-solution}

Before adding a scalar potential, we recall the massless benchmark.  This
serves two purposes.  First, it shows explicitly how a charged black hole can
support a regular scalar profile even though a minimally coupled static scalar
has no such hair.  Second, it defines the source-dominated throat that we will
later perturb by \(V(\phi)\neq0\).

\subsection{A minimally coupled scalar has no static hair}
\label{sec:static_nohair}

Consider a scalar with no gauge coupling and no potential on a static
black hole exterior.  Setting \(B=\mathrm{const}\) and \(V=0\) in
Eq.~\eqref{eq:radial-scalar-equation-general}, the equation \(\Box\phi=0\) gives
\begin{equation}
    \frac{d}{dr}
    \left(
        e^\delta\, r^2 f\, \phi'
    \right)
    =0 \;.
    \label{eq:nohair-ode}
\end{equation}
Hence,
\begin{equation}
    e^\delta\, r^2 f\, \phi' = C \;,
    \label{eq:nohair-const}
\end{equation}
where \(C\) is an integration constant.  At a regular non-extremal horizon,
\(f(r_h)=0\) while \(e^\delta r^2\) is finite; regularity of \(\phi'\)
therefore forces \(C=0\).  For an extremal horizon, where
\(f\sim(r-r_h)^2\), the same conclusion follows from finiteness of the
stress-energy tensor.\footnote{At an extremal horizon, \(f\sim(r-r_h)^2\)
  implies \(\phi'\sim C/(r-r_h)^2\) from
  Eq.~\eqref{eq:nohair-const}, so \((\partial\phi)^2=f(\phi')^2\sim
  C^2/(r-r_h)^2\) diverges unless \(C=0\).  This is the case of greatest
  interest here, since near-extremal (\(a\ll1\)) black holes are the ones
  with the longest throats.}
In both cases, the scalar is constant throughout the exterior,
\begin{equation}
    \phi(r)=\Phi_\infty \;,
    \qquad
    \Phi_\infty\equiv \phi(r\to\infty) \;.
    \label{eq:nohair-result}
\end{equation}
This is the simplest no-hair baseline.  A nontrivial, regular scalar profile
requires an additional source term on the right-hand side of
Eq.~\eqref{eq:radial-scalar-equation-general}.

\subsection{Gauge sourcing in the massless theory}

The situation changes once \(\phi\) enters the gauge kinetic function.  For
definiteness, we take the magnetic frame and, as introduced in Eq. \eqref{eq:exponential-gauge-kinetic}, the exponential coupling is
\begin{equation}
    B(\phi)=e^{-2\alpha\phi/M_{\rm Pl}} \;,
    \qquad
    g^2(\phi)=e^{2\alpha\phi/M_{\rm Pl}} \;.
    \label{eq:ghs-coupling}
\end{equation}
With \(V=0\), the scalar equation Eq.~\eqref{eq:scalar-eq-general} becomes
\begin{equation}
    \Box\phi
    =
    \frac{1}{4}B_{, \phi}(\phi)F_{\mu\nu}F^{\mu\nu} \;.
    \label{eq:massless-scalar-eq}
\end{equation}
In the magnetic frame, \(F_{\mu\nu}F^{\mu\nu}>0\), while for the exponential
coupling, \(B_{, \phi}(\phi)<0\), so the gauge source has a definite negative sign in
the scalar equation. For a
magnetic monopole of charge \(Q_m\), this source is localized and radially
dependent, and it drives a nontrivial exterior scalar profile.

The resulting static, spherically symmetric solution of the coupled
EMD equations with \(V=0\) is the standard
four-dimensional charged dilatonic black hole~\cite{Gibbons:1987ps,Garfinkle:1991qj}, which we will refer to as the GHS solution.  This solution is not the full higher-dimensional background used in Sen's construction.  Rather, it is the simplest four-dimensional representative of the same basic mechanism: a charged black hole gauge source drives a scalar away from its asymptotic value, over an exterior throat.  In this reduced setting, the physical size of the displaced-scalar region is set by \(r_+\), while the scalar excursion
is controlled by the near-extremality parameter \(a\).
In the Einstein frame, and in the canonical scalar normalization
used in Eq.~\eqref{eq:static-action}, the GHS metric below corresponds to
\(\alpha=1/\sqrt2\).  It takes the form\footnote{The metric,
  Eq.~\eqref{eq:ghs-metric}, uses a non-standard radial coordinate: the
  angular coefficient \(r(r-r_-)\) is not the square of the areal radius
  \(\rho\).  The radial coordinate \(r\) here is therefore not the areal
  radius.  This is why the scalar operator derived below differs from the one
  written in the areal-radius gauge in Eq.~\eqref{eq:static-spherical-metric}, and
  we account for this difference explicitly when deriving the throat
  operator in Section~\ref{sec:throat-operator}.}
\begin{equation}
    ds^2
    =
    -\left(1-\frac{r_+}{r}\right)dt^2
    +\left(1-\frac{r_+}{r}\right)^{-1}dr^2
    + r\,(r-r_-)\,d\Omega_2^2 \;,
    \label{eq:ghs-metric}
\end{equation}
with the dilaton profile
\begin{equation}
    e^{-2\alpha\phi/M_{\rm Pl}}
    =
    e^{-2\alpha\Phi_{\infty}/M_{\rm Pl}}
    \left(1-\frac{r_-}{r}\right) \;.
    \label{eq:ghs-dilaton}
\end{equation}
As before, \(\Phi_{\infty}\) is the asymptotic scalar value, the outer horizon is at
\(r=r_+\), and \(r_-\leq r_+\) is the inner parameter fixed by the magnetic
charge and asymptotic coupling, with a normalization that will not be needed
below.  The extremal limit is \(r_-\to r_+\).

From Eq.~\eqref{eq:ghs-dilaton}, the local gauge coupling satisfies
\begin{equation}
    \frac{g^2(r)}{g_\infty^2}
    =
    \frac{1}{1-r_-/r} \;,
    \label{eq:ghs-gauge-ratio-r}
\end{equation}
where \(g_\infty^2=g^2(\Phi_\infty)\).  In the exterior region
\(r\ge r_+\), this ratio grows monotonically inward and reaches
\begin{equation}
    \frac{g^2(r_+)}{g_\infty^2}
    =
    \frac{1}{1-r_-/r_+}
    =
    \frac{1}{a}
\end{equation}
at the outer horizon.  In the dimensionless throat coordinate \(b=(r-r_+)/r_+\), this becomes
\begin{equation}
    \frac{g^2(b)}{g_\infty^2}
    =
    \frac{1+b}{a+b} \;.
\end{equation}
Thus, a near-extremal black hole, \(r_-\simeq r_+\),
produces a large exterior variation of the local coupling without requiring
large local curvature.  This is the four-dimensional GHS analogue of the
mechanism emphasized by Sen: in a large-mass, large-charge scaling limit,
local invariants can be made small while moduli still vary by an order-one
amount across a macroscopic region.

\subsection{Throat coordinates and the GHS benchmark profile}
\label{sec:throat-coordinates}

\begin{figure}[t]
    \centering
    \includegraphics[width=0.75\linewidth]{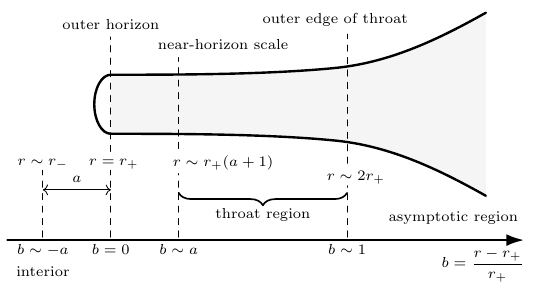}
    \caption{Schematic of the throat geometry in the coordinate
    \(b=(r-r_+)/r_+\), such that the black hole event horizon is mapped to
    \(b=0\).  The coordinate \(a=(r_+-r_-)/r_+\) measures the distance from
    extremality.  The inner horizon is mapped to \(b=-a\); for \(a \ll 1\),
    the exterior region \(b>0\) develops a long throat between \(b\sim a\)
    and \(b \sim 1\).}
    \label{fig:throat-schematic}
\end{figure}

The coordinates and profiles defined in this and the following subsection are used throughout
the rest of the paper, and the reader may wish to keep this section as a
reference point.  For the perturbative analysis of scalar potentials, we reiterate that it is
convenient to use dimensionless coordinates adapted to the GHS throat:
\begin{equation}
    b \equiv \frac{r-r_+}{r_+} \;,
    \qquad
    a \equiv \frac{r_+-r_-}{r_+} \;.
    \label{eq:b-a-def}
\end{equation}
Thus, \(b=0\) at the outer horizon, \(b\to\infty\) asymptotically, and
\(a\in(0,1]\) measures the departure from extremality, with \(a\to0\) serving as the
long-throat limit (see Fig. \ref{fig:throat-schematic}). A direct calculation gives
\begin{equation}
    1-\frac{r_-}{r}
    =
    \frac{a+b}{1+b} \;,
    \label{eq:rmin-r-in-ab}
\end{equation}
and the massless dilaton profile of Eq.~\eqref{eq:ghs-dilaton} becomes
\begin{equation}
    \phi_{\rm GHS}(b)
    =
    \Phi_\infty
    +
    \frac{M_{\rm Pl}}{2\alpha}
    \log\!\left(
        \frac{1+b}{a+b}
    \right) \;,
    \label{eq:phi-ghs-b}
\end{equation}
\noindent for \(\Phi_\infty\equiv \phi(r\to\infty)\). At large \(b\), this returns to \(\Phi_\infty\), while at the horizon, the scalar is
displaced by
\begin{equation}
    \Delta\phi
    \equiv
    \phi_{\rm GHS}(0)-\Phi_\infty
    =
    -\,\frac{M_{\rm Pl}}{2\alpha}\log a \;,
    \label{eq:delta-phi-def}
\end{equation}
which grows logarithmically as \(a\to0\).  Recall that the gauge-coupling ratio is
\begin{equation}
    \frac{g^2(b)}{g_\infty^2}
    =
    \frac{1+b}{a+b} \;,
    \label{eq:g-ratio-ghs-b}
\end{equation}
with horizon value
\begin{equation}
    \frac{g^2(0)}{g_\infty^2}
    =
    \frac{1}{a} \;.
    \label{eq:g-ratio-horizon}
\end{equation}
Thus, \(a\ll1\) defines a strongly displaced near-extremal throat; when \(r_+\) is large, the corresponding exterior region is macroscopic.

In the numerical analyses, we place the inner boundary at a small cutoff of
order the non-extremality scale.  In most of the benchmark scans, we take
\begin{equation}
    b_{\rm min}=a \;.
    \label{eq:bmin-def}
\end{equation}
At this cutoff, the coupling ratio is
\begin{equation}
    \frac{g^2(b_{\rm min})}{g_\infty^2}
    =
    \frac{g^2(a)}{g_\infty^2}
    =
    \frac{1+a}{2a}
    \sim
    \frac{1}{a}
    \qquad (a\ll1) \;,
\end{equation}
which is of the same parametric order as the horizon value \(1/a\), but
does not require evaluating the effective description exactly at \(b=0\).
The cutoff therefore probes the strongly displaced part of the throat while
remaining a finite coordinate distance from the outer horizon,
\[
    r-r_+=r_+a=r_+-r_- \;.
\]
The outer boundary is taken at \(b\gg1\), where
\(\phi_{\rm GHS}\to\Phi_\infty\) and \(g^2/g_\infty^2\to1\).

\subsection{Throat operator and scalar flux}
\label{sec:throat-operator}

The perturbation theory for \(V(\phi)\neq0\) is most transparent when the massless scalar equation is written in a form adapted to the coordinate \(b\).  Since \(b=(r-r_+)/r_+\), the radial operator carries an overall factor \(1/r_+^2\). Thus, the potential enters through the dimensionless
combination \(r_+^2 V_{,\phi}\), or, for a quadratic potential, through
\(\mu^2=(m r_+)^2\). The relevant local expansion parameter is therefore the ratio of this potential force to the gauge source that supports the massless GHS profile.  We now derive the corresponding radial operator directly from the GHS metric,  Eq.~\eqref{eq:ghs-metric}.

In the GHS radial gauge \(\delta=0\), the metric components become
\begin{equation}
    g_{tt}=-f(r) \;,
    \qquad
    g_{rr}=f(r)^{-1},
    \qquad
    g_{\theta\theta}=r(r-r_-) \;,
    \qquad
    f(r)=1-\frac{r_+}{r} \;.
    \label{eq:ghs-components}
\end{equation}
The determinant gives
\begin{equation}
    \sqrt{-g}=r(r-r_-)\sin\theta \;.
\end{equation}
For \(V=0\), the covariant scalar equation reduces to
\begin{equation}
    \frac{1}{r(r-r_-)}
    \frac{d}{dr}
    \left[
        r(r-r_-)\, f(r)\,\frac{d\phi}{dr}
    \right]
    =
    \frac{1}{4}B_{,\phi}(\phi)F_{\mu\nu}F^{\mu\nu}.
    \label{eq:scalar-eq-ghs-coords}
\end{equation}
Changing variables to \(b\), with \(r=r_+(1+b)\), gives
\begin{equation}
    f(r)=\frac{b}{1+b} \;,
    \qquad
    r(r-r_-)=r_+^2(1+b)(a+b) \;.
\end{equation}
Multiplying the scalar equation by \(r_+^2\), and then working in units
\(r_+=1\), one obtains
\begin{equation}
    \frac{1}{(1+b)(a+b)}
    \frac{d}{db}
    \left[
        b(a+b)\frac{d\phi}{db}
    \right]
    =
    \frac{1}{4}B_{,\phi}(\phi)F_{\mu\nu}F^{\mu\nu}\bigg|_{r_+=1} \;.
    \label{eq:scalar-eq-ghs-b}
\end{equation}
Equivalently, before setting \(r_+=1\), the right-hand side is multiplied by
\(r_+^2\).  This is the origin of the dimensionless expansion parameter
\(r_+^2 V_{,\phi}\) when a scalar potential is added.

The radial operator is of Sturm--Liouville form,
\begin{equation}
    \mathcal L_0\phi
    =
    \frac{1}{w(b)}
    \frac{d}{db}
    \left[
        p(b)\frac{d\phi}{db}
    \right] \;,
\end{equation}
with
\begin{equation}
    w(b)=(1+b)(a+b) \;,
    \qquad
    p(b)=b(a+b) \;.
    \label{eq:p-w-ghs}
\end{equation}
The coefficient \(p(b)\) vanishes at the outer horizon, \(p(0)=0\), reflecting
\(f(r_+)=0\).  The endpoint \(b=0\) is therefore a singular endpoint of the
Sturm--Liouville problem.  In practice, we regulate the throat by imposing the
inner boundary condition at \(b_{\rm min}>0\), typically \(b_{\rm min}=a\), as specified in Eq.~\eqref{eq:bmin-def}.

We define the radial scalar flux as the quantity naturally associated with
the homogeneous operator,
\begin{equation}
    \mathcal F_\phi(b)
    \equiv
    p(b)\frac{d\phi}{db}
    =
    b(a+b)\frac{d\phi}{db} \;.
    \label{eq:scalar-flux-def}
\end{equation}
For the GHS profile, Eq.~\eqref{eq:phi-ghs-b},
\begin{equation}
    \frac{d\phi_{\rm GHS}}{db}
    =
    \frac{M_{\rm Pl}}{2\alpha}
    \left(
        \frac{1}{1+b}
        -
        \frac{1}{a+b}
    \right)
    =
    -\,\frac{M_{\rm Pl}}{2\alpha}
    \frac{1-a}{(1+b)(a+b)} \;,
    \label{eq:phi-ghs-prime}
\end{equation}
and therefore
\begin{equation}
    \mathcal F_\phi(b)
    =
    -\,\frac{M_{\rm Pl}}{2\alpha}
    \frac{b(1-a)}{1+b} \;.
    \label{eq:ghs-flux}
\end{equation}
This flux is not constant: it vanishes at the outer horizon and increases in
magnitude monotonically, saturating at large \(b\) to
\begin{equation}
    \mathcal F_\infty
    \equiv
    \lim_{b\to\infty}\mathcal F_\phi(b)
    =
    -\,\frac{M_{\rm Pl}}{2\alpha}(1-a) \;.
    \label{eq:ghs-flux-infty}
\end{equation}
The quantity \(\mathcal F_\infty\) measures the total scalar charge carried
by the massless solution and provides a natural reference scale for the
throat.  Importantly, it remains finite as \(a\to0\), whereas the scalar
excursion in Eq.~\eqref{eq:delta-phi-def} grows as
\(- (M_{\rm Pl}/2\alpha)\log a\).  The large field-space excursion of the
near-extremal throat is therefore not driven by a large scalar charge, but
by the increasing length of the throat.

From Eq.~\eqref{eq:scalar-eq-ghs-b}, we can also read off the signed gauge
source that supports the massless GHS profile:
\begin{equation}
    \mathcal J_{\rm gauge}(b)
    \equiv
    \frac{1}{w(b)}
    \frac{d}{db}
    \left[
        p(b)\frac{d\phi_{\rm GHS}}{db}
    \right] \;.
    \label{eq:signed-gauge-source-def}
\end{equation}
Using Eq.~\eqref{eq:ghs-flux}, this gives
\begin{equation}
    \mathcal J_{\rm gauge}(b)
    =
    -\,\frac{M_{\rm Pl}}{2\alpha}
    \frac{1-a}{(1+b)^3(a+b)} \;.
    \label{eq:signed-gauge-source-explicit}
\end{equation}
The sign is fixed by the magnetic frame.  With
\begin{equation}
    B(\phi)
    =
    B_\infty
    e^{-2\alpha(\phi-\Phi_\infty)/M_{\rm Pl}} \;,
\end{equation}
one has \(B_{,\phi}<0\), while \(F_{\mu\nu}F^{\mu\nu}>0\).  The gauge
contribution to the scalar equation is therefore negative, exactly as in
Eq.~\eqref{eq:signed-gauge-source-explicit}.

For the diagnostic comparisons in the following sections, we mostly need the
magnitude of the GHS gauge source.  We therefore define
\begin{equation}
    \mathcal{S}_{\rm gauge}(b)
    \equiv
    \left|
        \mathcal{J}_{\rm gauge}^{(0)}(b)
    \right|
    =
    \frac{M_{\rm Pl}}{2\alpha}
    \frac{1-a}{(1+b)^3(a+b)} \;.
    \label{eq:gauge-source-magnitude}
\end{equation}
Here, \(\mathcal{J}_{\rm gauge}^{(0)}\) denotes the signed gauge source (where the superscript indicates explicitly that the quantity is evaluated on the massless GHS profile), while its magnitude,
\(\mathcal{S}_{\rm gauge}\), is the corresponding positive reference scale
used in the source-dominance ratios defined in the next subsection.


Once \(V(\phi)\) is turned on, the scalar equation on the fixed GHS
background gains an additional source.  Before setting \(r_+=1\), this source
appears through the dimensionless combination \(r_+^2 V_{,\phi}\).  In units
\(r_+=1\), the fixed-background scalar equation may be written schematically
as
\begin{equation}
    \frac{1}{w(b)}
    \frac{d}{db}
    \left[
        p(b)\frac{d\phi}{db}
    \right]
    =
    \mathcal{J}_{\rm gauge}(\phi;b)
    +
    V_{,\phi}(\phi) \;,
    \label{eq:fixed-background-scalar-schematic}
\end{equation}
where \(\mathcal{J}_{\rm gauge}(\phi;b)\) denotes the gauge contribution on
the fixed magnetic background.  The GHS solution obeys the same equation
with \(V=0\), and
\[
    \mathcal{J}_{\rm gauge}(\phi_{\rm GHS};b)
    =
    \mathcal{J}_{\rm gauge}^{(0)}(b) \;.
\]
Writing
\begin{equation}
    \phi(b)=\phi_{\rm GHS}(b)+\delta\phi(b) \;,
\end{equation}
and subtracting the massless GHS equation gives, to leading order in the
direct potential source,
\begin{equation}
    \frac{d}{db}
    \left[
        p(b) \delta\phi'(b)
    \right]
    =
    w(b)\,
    V_{,\phi}\!\left(\phi_{\rm GHS}(b)\right)
    + \cdots ,
    \label{eq:linearized-potential-source}
\end{equation}
where the prime denotes \(d/db\). The ellipsis denotes the terms obtained by linearizing the gauge source in
\(\delta\phi\), together with higher-order corrections in the deformation.

The local competition is therefore between the signed GHS gauge source
\(\mathcal{J}_{\rm gauge}^{(0)}\), which sustains the massless profile, and
the potential source \(V_{,\phi}(\phi_{\rm GHS})\).  When only the strength
of this competition is needed, we compare the dimensionless potential force
\(r_+^2 |V_{,\phi}|\) to the positive scale
\(\mathcal{S}_{\rm gauge}~=~|\mathcal{J}_{\rm gauge}^{(0)}|\); in the
\(r_+=1\) units used below, this reduces to comparing
\(|V_{,\phi}|\) with \(\mathcal{S}_{\rm gauge}\).  This motivates the local
and cumulative diagnostics introduced in the following section.



\section{Static deformations by a scalar potential}
\label{sec:static-deformations}

We now turn on a scalar potential \(V(\phi)\) and ask when the GHS throat
remains a good approximation.  The goal is not, at this stage, to solve the
fully back-reacted boundary-value problem, but to build a fixed-throat
diagnostic toolkit that measures the strength of the potential force
relative to the gauge-sourced scalar flow of the massless solution.  Four
such diagnostics are developed below, from the most local to the most
directly dynamical:
\begin{itemize}
    \item a pointwise force ratio \(\eta_{\rm src}(b)\)
    (Section~\ref{subsec:local-diagnostic}), comparing the potential slope
    directly to the gauge source at each radius;
    \item a local flux measure \(\epsilon_{\rm loc}(b)\)
    (Section~\ref{subsec:local-diagnostic}), which weights the potential
    source over one logarithmic radial interval;
    \item a cumulative flux measure \(\epsilon_{\rm cum}(b)\)
    (Section~\ref{subsec:cumulative-diagnostic}), integrating the effect
    across the whole throat;
    \item and a direct solution for the linearized deformation \(\delta\phi\)
    (Section~\ref{subsec:linearized-profile}), from which the induced shift
    in the scalar profile and in the local gauge coupling can be read off.
\end{itemize}
Section~\ref{subsec:breakdown-fixed-throat} collects these into a set of
validity criteria for the fixed-throat approximation, used throughout the
potential-by-potential analysis that follows.

The starting point is the scalar equation on the fixed GHS background.  In
the throat coordinate \(b\) derived in Section~\ref{sec:throat-operator},
the radial operator takes the Sturm--Liouville form, using
Eqs.~\eqref{eq:scalar-eq-ghs-b} and \eqref{eq:p-w-ghs},
\begin{equation}
    \frac{1}{w(b)}
    \frac{d}{db}
    \left[
        p(b)\frac{d\phi}{db}
    \right]
    =
    r_+^2
    \left[
        V_{,\phi}(\phi)
        +
        \frac{1}{4}B_{,\phi}(\phi)F_{\mu\nu}F^{\mu\nu}
    \right] \;.
    \label{eq:fixed-throat-scalar-full}
\end{equation}
with
\begin{equation}
    w(b)=(1+b)(a+b) \;,
    \qquad
    p(b)=b(a+b) \;.
    \label{eq:p-w-ghs-repeat}
\end{equation}
The factor \(r_+^2\) appears because \(b=(r-r_+)/r_+\), so the radial
operator written in \(b\)-coordinates is dimensionless.  Thus, the potential
enters through the dimensionless force \(r_+^2V_{,\phi}\).

For \(V=0\), the exact massless profile \(\phi_{\rm GHS}(b)\) satisfies
\begin{equation}
    \frac{1}{w(b)}
    \frac{d}{db}
    \left[
        p(b)\phi_{\rm GHS}'(b)
    \right]
    =
    r_+^2
    \frac{1}{4}B_{,\phi}(\phi_{\rm GHS})F_{\mu\nu}F^{\mu\nu} \;,
    \label{eq:ghs-background-equation}
\end{equation}
where the prime denotes \(d/db\). Writing
\begin{equation}
    \phi(b)=\phi_{\rm GHS}(b)+\delta\phi(b) 
    \label{eq:phi-decomposition}
\end{equation}
and treating the potential as a perturbation around the massless background,
subtracting Eq.~\eqref{eq:ghs-background-equation} from
Eq.~\eqref{eq:fixed-throat-scalar-full} gives
\begin{equation}
    \frac{d}{db}
    \left[
        p(b)\delta\phi'(b)
    \right]
    =
    r_+^2 w(b)\,
    V_{,\phi}\!\left(\phi_{\rm GHS}(b)\right)
    + \cdots .
    \label{eq:linearized-delta-phi-equation}
\end{equation}
The ellipsis denotes terms proportional to \(\delta\phi\), including
\[
    r_+^2 w(b)\,
    V_{,\phi\phi}\!\left(\phi_{\rm GHS}(b)\right)\delta\phi
\]
and the term obtained by linearizing the gauge source \;,
\[
    \frac{r_+^2 w(b)}{4}\,
    \partial_\phi\!\left(B_{,\phi}F^2\right)_{\rm GHS}\,\delta\phi,
\]
together with higher-order corrections in the deformation.  Throughout this
section, we isolate the direct potential source evaluated on the unperturbed
GHS profile, and treat the terms proportional to \(\delta\phi\) as secondary
corrections.  The resulting Eq.~\eqref{eq:linearized-delta-phi-equation} is
the basic fixed-throat deformation equation: it shows that the source driving
the integrated deformation is the radially weighted quantity
\begin{equation}
    r_+^2 w(b)\,
    V_{,\phi}\!\left(\phi_{\rm GHS}(b)\right) \;.
    \label{eq:weighted-potential-source}
\end{equation}
For local validity, however, the relevant dimensionless comparison is the
potential force itself, \(r_+^2 |V_{,\phi}|\), divided by the positive GHS
gauge-source scale, \(\mathcal S_{\rm gauge}\).

\subsection{Pointwise and local flux diagnostics}
\label{subsec:local-diagnostic}

The most direct comparison is pointwise: at each radius, how does the
potential force compare to the gauge source that supports the massless
profile?  Using the positive GHS gauge-source magnitude
\(\mathcal S_{\rm gauge}(b)\) of Eq.~\eqref{eq:gauge-source-magnitude}, we
define
\begin{equation}
    \eta_{\rm src}(b)
    \equiv
    \frac{
        r_+^2\left|V_{,\phi}\!\left(\phi_{\rm GHS}(b)\right)\right|
    }{
        \mathcal S_{\rm gauge}(b)
    }
    =
    \frac{
        r_+^2\,w(b)\left|V_{,\phi}\!\left(\phi_{\rm GHS}(b)\right)\right|
    }{
        \left|\dfrac{d}{db}\left[p(b)\dfrac{d\phi_{\rm GHS}}{db}\right]\right|
    } .
    \label{eq:eta-source-ratio}
\end{equation}
The factor \(r_+^2\) appears because the scalar equation has been written in
the dimensionless coordinate, \(b=(r-r_+)/r_+\).  Thus,
\(\eta_{\rm src}\) compares two dimensionless forces in the fixed-throat
equation.  The condition
\begin{equation}
    \eta_{\rm src}(b)\ll1
\end{equation}
is the pointwise source-dominance criterion: locally, the potential force is
small compared to the gauge source sustaining the GHS profile.

A complementary question is how much scalar flux the potential generates
over a logarithmic radial interval, rather than at a single point.  From
Eq.~\eqref{eq:linearized-delta-phi-equation}, the direct potential source
changes the scalar flux by
\begin{equation}
    d\mathcal F_V
    =
    r_+^2 w(b)\,
    V_{,\phi}\!\left(\phi_{\rm GHS}(b)\right)\,db \;.
\end{equation}
Since \(db=b\,d(\log b)\), the natural measure of the flux injected over one
logarithmic interval is
\begin{equation}
    \epsilon_{\rm loc}(b)
    \equiv
    \frac{
        \left|
            r_+^2\, b\, w(b)\,
            V_{,\phi}\!\left(\phi_{\rm GHS}(b)\right)
        \right|
    }{
        |\mathcal F_\infty|
    } \;,
    \label{eq:epsilon-local-def}
\end{equation}
where \(\mathcal F_\infty\), defined in Eq.~\eqref{eq:ghs-flux-infty}, is
\begin{equation}
    \mathcal F_\infty
    =
    -\,\frac{M_{\rm Pl}}{2\alpha}(1-a) \;.
    \label{eq:flux-infty-repeat}
\end{equation}
The two diagnostics answer different questions.  The pointwise ratio
\(\eta_{\rm src}\) asks whether the potential force competes locally with
the gauge source.  The local flux measure \(\epsilon_{\rm loc}\) asks
whether the same source injects an appreciable fraction of the total GHS
scalar flux over one logarithmic radial interval.  A potential can be small
according to one diagnostic and large according to the other, so both are
useful.

The radial weight entering the flux diagnostic is
\begin{equation}
    b\,w(b)=b(1+b)(a+b).
\end{equation}
It vanishes at the outer horizon, \(b=0\), and grows in the exterior.  This
suppresses \(\epsilon_{\rm loc}\) near \(b=0\) unless the potential slope
itself becomes parametrically large there.  Since \(d(\log b)\) is
ill-defined exactly at the horizon, we regulate the throat by imposing the
inner boundary at \(b_{\rm min}>0\), typically \(b_{\rm min}=a\), as expressed in Eq.~\eqref{eq:bmin-def}.  This
cutoff probes the strongly displaced part of the near-extremal throat while
remaining a finite coordinate distance outside the outer horizon.

The interpretation of \(\epsilon_{\rm loc}\) is then straightforward.  If
\(\epsilon_{\rm loc}(b)\ll1\) throughout the throat, the potential changes
the scalar flux only adiabatically over each logarithmic radial interval,
and the GHS profile is locally stable against this direct source.  If
instead \(\epsilon_{\rm loc}(b)\gtrsim1\) somewhere in the exterior, the
potential injects an \(O(1)\) fraction of the GHS scalar flux over one
logarithmic interval, signalling the onset of a significant local
deformation.  The radius at which \(\epsilon_{\rm loc}\) peaks is itself
informative: in some examples, the maximum lies near the inner throat, where
\(\phi_{\rm GHS}\) is most displaced; in others, it lies at intermediate
radius, where the growth of \(b\,w(b)\) outpaces the decay of the potential
slope.  The location of this peak therefore diagnoses \emph{how} the
potential affects the throat, not just whether it does.



\subsection{Cumulative deformation diagnostic}
\label{subsec:cumulative-diagnostic}

A potential can produce a small correction at every radius, and nevertheless
accumulate into a sizeable deformation over many logarithmic intervals, even
when \(\epsilon_{\rm loc}\) never becomes large.  Integrating the direct
potential source in Eq.~\eqref{eq:linearized-delta-phi-equation} from the
inner cutoff \(b_{\rm min}\) to a point \(b\) gives the cumulative flux
deformation,
\begin{equation}
    \Delta\mathcal F_V(b)
    =
    r_+^2
    \int_{b_{\rm min}}^{b}
    dx\,
    w(x)\,
    V_{,\phi}\!\left(\phi_{\rm GHS}(x)\right) \;,
    \label{eq:cumulative-flux-deformation}
\end{equation}
and the corresponding dimensionless diagnostic defined with respect to Eq.~\eqref{eq:ghs-flux-infty},
\begin{equation}
    \epsilon_{\rm cum}(b)
    \equiv
    \frac{
        \left|
            \Delta\mathcal F_V(b)
        \right|
    }{
        |\mathcal F_\infty|
    } \;,
    \qquad
    \epsilon_{\rm cum}^{\rm max}
    \equiv
    \max_{b_{\rm min}\leq b\leq b_{\rm max}}
    \epsilon_{\rm cum}(b) \;.
    \label{eq:epsilon-cumulative-def}
\end{equation}
For potentials whose slope vanishes at the asymptotic value,
\(V_{,\phi}(\Phi_\infty)=0\), the integrand decays at large \(b\) and the
integral can saturate, so that one may take \(b_{\rm max}\to\infty\) when
the falloff is sufficiently fast.  For potentials with a non-negligible
asymptotic slope, or with slow falloff along the GHS tail, the outer boundary
\(b_{\rm max}\gg1\) must instead be specified explicitly; we state this
choice in each example below.

This diagnostic matters most for shallow or slowly varying potentials.  An
inverse-power or exponential tail can be weak at every individual radius yet
act coherently over a broad interval, in which case \(\epsilon_{\rm loc}\)
underestimates the total deformation while \(\epsilon_{\rm cum}\) captures
it.  The converse can also happen: a steep potential localized near the
inner throat may give a sharp peak in \(\epsilon_{\rm loc}\) but only a
modest cumulative change if the affected interval is narrow.  Comparing
\(\epsilon_{\rm loc}\) and \(\epsilon_{\rm cum}\) side by side is therefore
often more informative than either diagnostic alone.



\subsection{Linearized profile and boundary conditions}
\label{subsec:linearized-profile}

The flux-based diagnostics above are all built from the potential source
evaluated on the unperturbed background; none of them directly gives the
size of the resulting profile deformation \(\delta\phi\).  To determine
that deformation, we solve directly for \(\delta\phi\).  In the fixed-throat approximation, the leading direct potential source gives
\begin{equation}
    \frac{d}{db}
    \left[
        p(b)\delta\phi'(b)
    \right]
    =
    r_+^2 w(b)\,
    V_{,\phi}\!\left(\phi_{\rm GHS}(b)\right) \;,
    \label{eq:delta-phi-linearized-final}
\end{equation}
where a prime here denotes \(d/db\). This equation is first order in the deformation flux
\(p(b)\delta\phi'(b)\), but second order in the scalar deformation
\(\delta\phi(b)\) itself:
\[
    \frac{d}{db}\left[p(b)\delta\phi'(b)\right]
    =
    p(b)\delta\phi''(b)+p'(b)\delta\phi'(b) \;.
\]
It is therefore closed by two boundary conditions.  At large radius, the
scalar approaches the prescribed asymptotic value \(\Phi_\infty\).  Since
\(\phi_{\rm GHS}\to\Phi_\infty\) in this large-radius regime, this fixes
\begin{equation}
    \delta\phi(b_{\rm max})=0 \;.
    \label{eq:delta-phi-outer-bc}
\end{equation}
At the inner end, we impose a zero deformation-flux condition at the cutoff,
\begin{equation}
    p(b_{\rm min})\,\delta\phi'(b_{\rm min})=0 \;.
    \label{eq:regular-flux-bc}
\end{equation}
Eqs~\eqref{eq:delta-phi-outer-bc} and
\eqref{eq:regular-flux-bc} define the fixed-throat boundary-value scheme
used below.  The inner condition is imposed at the regulated endpoint
\(b_{\rm min}>0\), rather than at the horizon itself, and should be viewed
as a choice of matching prescription for the deformation.  Changing this
inner flux condition adds a homogeneous solution with constant scalar flux,
\begin{equation}
    p(b)\,\delta\phi_h'(b)=\text{constant},
    \qquad
    \delta\phi_h'(b)\propto \frac{1}{p(b)} \;,
\end{equation}
which can shift the detailed profile by a cutoff-dependent amount
throughout the exterior.  The flux-based diagnostics of
Sections~\ref{subsec:local-diagnostic} and
\ref{subsec:cumulative-diagnostic}, however, are fixed directly by the
potential source evaluated on the unperturbed GHS background and are
therefore insensitive to this matching choice at leading order.

From the solution, we extract two direct measures of the deformation: the
maximum shift in the scalar profile,
\begin{equation}
    \Delta\phi_{\rm max}
    \equiv
    \max_{b_{\rm min}\leq b\leq b_{\rm max}}
    |\delta\phi(b)|\;,
    \label{eq:max-delta-phi}
\end{equation}
and the maximum fractional change in the local gauge coupling,
\begin{equation}
    \Delta_g^{\rm max}
    \equiv
    \max_{b_{\rm min}\leq b\leq b_{\rm max}}
    \left|
        \frac{g^2(\phi_{\rm GHS}+\delta\phi)}
             {g^2(\phi_{\rm GHS})}
        -1
    \right| \;.
    \label{eq:max-g-deformation}
\end{equation}
For the exponential gauge kinetic function,
\begin{equation}
    B(\phi)
    =
    B_\infty
    e^{-2\alpha(\phi-\Phi_\infty)/M_{\rm Pl}},
    \qquad
    g^2(\phi)=B(\phi)^{-1},
\end{equation}
one has
\begin{equation}
    \frac{g^2(\phi_{\rm GHS}+\delta\phi)}
         {g^2(\phi_{\rm GHS})}
    =
    \exp\!\left(\frac{2\alpha\,\delta\phi}{M_{\rm Pl}}\right) \;.
\end{equation}
Thus, for small deformations,
\begin{equation}
    \Delta_g^{\rm max}
    \simeq
    \frac{2\alpha}{M_{\rm Pl}}\,
    \Delta\phi_{\rm max} \;.
    \label{eq:delta-g-linear}
\end{equation}



\subsection{Validity and breakdown of the fixed-throat approximation}
\label{subsec:breakdown-fixed-throat}

The fixed-throat approximation is reliable only while the
potential-induced deformation stays small. The diagnostics above express
this requirement in complementary ways.  Pointwise, one requires
\begin{equation}
    \eta_{\rm src}^{\rm max}
    \equiv
    \max_{b_{\rm min}\leq b\leq b_{\rm max}}
    \eta_{\rm src}(b)
    \ll 1 \;,
    \label{eq:validity-source-pointwise}
\end{equation}
which demands that the dimensionless potential force
\(r_+^2 |V_{,\phi}|\) remain small compared with the gauge source that
supports the massless throat.  Over logarithmic radial intervals, one
requires
\begin{equation}
    \epsilon_{\rm loc}^{\rm max}
    \equiv
    \max_{b_{\rm min}\leq b\leq b_{\rm max}}
    \epsilon_{\rm loc}(b)
    \ll 1 \;,
    \label{eq:validity-local}
\end{equation}
and for the accumulated flux deformation,
\begin{equation}
    \epsilon_{\rm cum}^{\rm max}\ll1 \;.
    \label{eq:validity-cumulative}
\end{equation}
At the level of the field profile itself, one requires
\begin{equation}
    \Delta\phi_{\rm max}
    \ll
    \Delta\phi_{\rm throat} \;,
    \qquad
    \Delta\phi_{\rm throat}
    \equiv
    \phi_{\rm GHS}(b_{\rm min})-\Phi_\infty \;,
    \label{eq:fixed-throat-profile-condition}
\end{equation}
or, more conservatively, that the induced fractional shift in the local
gauge coupling remain small,
\begin{equation}
    \Delta_g^{\rm max}\ll1 \;.
    \label{eq:validity-gauge}
\end{equation}

These criteria play complementary roles rather than duplicating one another.
The pointwise ratio \(\eta_{\rm src}\) locates where the potential force
first competes with the gauge source.  The flux diagnostics
\(\epsilon_{\rm loc}\) and \(\epsilon_{\rm cum}\) measure how much scalar
flux this force injects, respectively, per logarithmic interval and in total.
Finally, \(\Delta\phi_{\rm max}\) and \(\Delta_g^{\rm max}\) check directly
whether these effects add up to an order-one deformation of the scalar
profile or of the observable gauge coupling.

These order-one conditions define diagnostic-dependent thresholds.  As we shall demonstrate in the next section, for a
potential with an overall strength parameter, such as \(m\), \(\lambda\), or
\(A\), we quote the value at which a given diagnostic becomes order one,
\begin{equation}
    \eta_{\rm src}^{\rm max}\sim1 \;,
    \qquad
    \epsilon_{\rm loc}^{\rm max}\sim1 \;,
    \qquad
    \epsilon_{\rm cum}^{\rm max}\sim1 \;,
    \qquad
    \text{or}
    \qquad
    \Delta_g^{\rm max}\sim1 \;.
    \label{eq:critical-parameter-def}
\end{equation}
Which threshold is most physically relevant depends on the shape of the
potential.  For monotonic potentials, the local, cumulative, and profile
diagnostics often track the same physical breakdown.  For oscillatory
potentials, however, a large local force can average to a much smaller
integrated deformation.  We therefore treat
Eq.~\eqref{eq:critical-parameter-def} as a set of diagnostic criteria, not
as a single universal definition of breakdown.

Within the fixed-throat approximation, all these criteria probe the same
underlying expectation: the GHS throat ceases to be a good approximation
once the potential-induced force, flux, or profile deformation becomes
comparable to the gauge-sourced scalar flow generated by the black hole
gauge field.

Finally, a caveat that recurs throughout the examples below.  If the
potential has a non-negligible slope at the asymptotic value,
\begin{equation}
    V_{,\phi}(\Phi_\infty)\neq0 \;,
\end{equation}
then \(\Phi_\infty\) is not itself a static extremum of the potential, and
the outer boundary condition of Eq.~\eqref{eq:delta-phi-outer-bc} at
\(b_{\rm max}\to\infty\) is not strictly consistent with a genuinely static
asymptotically flat solution.  In that case, the fixed-throat problem should
be understood as posed on a finite outer boundary \(b_{\rm max}\),
representing the region over which the black hole throat is embedded in a
slowly varying, or externally fixed, modulus background.



\section{Static potential classes}
\label{sec:static-potential-classes}

We now apply the fixed-throat diagnostics of
Section~\ref{sec:static-deformations} to several representative scalar
potentials.  In each case, the goal is the same: does the potential-induced
force stay weak enough for the gauge-sourced GHS throat to survive, or does
it deform or erase it?  We work through the following potentials, in order
of increasing structure:
\begin{itemize}
    \item a \emph{quadratic} stabilizing potential
    (Section~\ref{subsec:quadratic-potential}), the cleanest benchmark for
    the breakdown of the gauge-sourced throat;
    \item \emph{shifted-exponential} and \emph{pure exponential runaway}
    potentials (Sections~\ref{subsec:shifted-exponential-static} and
    \ref{subsec:exponential-runaway}), which test a restoring force that
    saturates and a force that grows or decays monotonically along the
    throat;
    \item an \emph{inverse-power} runaway potential
    (Section~\ref{subsec:inverse-power-runaway}), where the force weakens
    down the throat rather than strengthening;
    \item a \emph{racetrack} potential
    (Section~\ref{subsec:racetrack-stabilization}), where the relevant
    question is not the size of a perturbative deformation but whether the
    modulus stays trapped behind a finite barrier;
    \item an \emph{axion-like periodic} potential
    (Section~\ref{subsec:axion-like-periodic-potentials}), where the force
    changes sign along the throat and can be washed out by oscillatory
    cancellation;
    \item a pair of \emph{supergravity-inspired} potentials
    (Section~\ref{subsec:sugra-inspired-inverse-power}), used as a toy
    diagnostic of how sensitive a regulated throat profile is to the sign of
    an exponential correction.
\end{itemize}

For each class, we proceed in two stages.  We first analyze the deformation
in the fixed-throat approximation, treating the GHS geometry and scalar
profile as a fixed background.  This yields the local and cumulative
diagnostics of Section~\ref{sec:static-deformations} and, when useful, a
direct solution for the linearized deformation \(\delta\phi\).  We then
compare, where appropriate, with a corresponding back-reacted check,
analytically where possible and numerically otherwise, to see which
conclusions survive once the geometry is allowed to respond, and which are
artifacts of holding the throat fixed. 

We emphasize that throughout this section, the unperturbed scalar profile is
\begin{equation}
    \phi_{\rm GHS}(b)
    =
    \Phi_\infty
    +
    \frac{M_{\rm Pl}}{2\alpha}
    \log\!\left(
        \frac{1+b}{a+b}
    \right),
    \label{eq:phi-ghs-examples}
\end{equation}
with gauge-coupling ratio
\begin{equation}
    \frac{g^2(b)}{g_\infty^2}
    =
    \frac{1+b}{a+b} .
    \label{eq:g-ratio-ghs-examples}
\end{equation}
In the fixed-throat approximation the direct deformation is sourced by
\begin{equation}
    r_+^2\,w(b)\,
    V_{,\phi}\!\left(\phi_{\rm GHS}(b)\right),
    \qquad
    w(b)=(1+b)(a+b).
    \label{eq:weighted-source-examples}
\end{equation}
Unless stated otherwise we work in units \(r_+=1\).  Equivalently, masses
enter through combinations such as \(m r_+\), while potential amplitudes
enter through the appropriate dimensionless combinations multiplying
\(r_+^2 V_{,\phi}\).  In the benchmark fixed-throat scans the inner boundary
is placed at \(b_{\rm min}=a\).  The outer boundary \(b_{\rm max}\) is
specified explicitly whenever a diagnostic depends on it.

In the back-reacted comparisons we use the same potential normalization and
the same asymptotic scalar value \(\Phi_\infty\).  Since the fixed-throat
scalar diagnostics depend only on \(V_{,\phi}\), they are insensitive to an
overall additive shift in \(V\).  The gravitational equations, however,
respond to \(V(\phi)\) itself: a nonzero value \(V(\Phi_\infty)\) at the
prescribed asymptotic modulus acts as an effective local cosmological
constant and would prevent the exterior region from being asymptotically
flat.  Whenever \(V(\Phi_\infty)\neq0\) for a given potential, we therefore
subtract this constant,
\[
    V(\phi)\longrightarrow V(\phi)-V(\Phi_\infty),
\]
before solving the back-reacted equations.  The fixed-throat diagnostics are
unaffected by this choice.

\subsection{Quadratic stabilizing potential}
\label{subsec:quadratic-potential}

The simplest way to give the scalar a restoring force is to expand the
potential around the prescribed asymptotic value \(\Phi_\infty\),
\begin{equation}
    V_{\rm quad}(\phi)
    =
    \frac{1}{2}m^2(\phi-\Phi_\infty)^2 \;.
    \label{eq:quadratic-potential}
\end{equation}
Here, \(\phi\) is the same canonically normalized scalar field used
throughout Sections~\ref{sec:static-emd-system}--\ref{sec:static-potential-classes},
with GHS profile \(\phi_{\rm GHS}(b)\).  We denote by
\[
    \delta\phi\equiv \phi-\phi_{\rm GHS}
\]
the deformation away from the massless profile, and by
\[
    \Delta\phi_{\rm throat}
    \equiv
    \phi_{\rm GHS}(b_{\rm min})-\Phi_\infty
\]
the scalar excursion sampled by the regulated throat. 

This example is deliberately elementary.  Its role is to calibrate the
diagnostics against the standard massive-dilaton expectation before applying
them to potentials with more structure.

Eq.~\eqref{eq:quadratic-potential} is the universal local form of any
smooth stabilizing potential near a minimum, and the most transparent first
test of the competition between the gauge-field source, which drives the
GHS scalar profile, and a scalar potential, which tries to pin the field to
its asymptotic value.  The potential introduces one dimensionless mass
parameter,
\begin{equation}
    \mu \equiv m r_+ ,
    \label{eq:quad-mu-def}
\end{equation}
equivalently the inverse Compton wavelength in horizon units,
\(\lambda_c\equiv1/\mu\).

This quadratic toy model is close in spirit to previous analyses of charged
black holes coupled to a massive dilaton, and to later studies of stabilized
moduli in black hole flows~\cite{Gregory:1992kr,Horne:1992bi,Delgado:2025crl}.
Recall that the basic physical dichotomy is the same: if \(m r_+\ll1\), the scalar is
effectively light on the black hole scale and the massless dilatonic
behaviour can persist; if \(m r_+\gg1\), the scalar is pinned close to its
minimum and the solution approaches the constant-coupling
Reissner--Nordstr\"om regime.  Our fixed-throat analysis resolves this
competition across the near-extremal GHS exterior, and distinguishes the
local Compton-scale criterion from the more stringent fixed-background
criterion obtained by demanding that the deformation remain small over a
long regulated radial interval.

One qualification matters from the start.  The near-extremal GHS throat can
scan a large field range,
\begin{equation}
    \Delta\phi_{\rm throat}
    =
    \phi_{\rm GHS}(b_{\rm min})-\Phi_\infty
    =
    \frac{M_{\rm Pl}}{2\alpha}
    \log\!\left(
        \frac{1+b_{\rm min}}{a+b_{\rm min}}
    \right) \;.
    \label{eq:quadratic-throat-excursion-exact}
\end{equation}
For the benchmark choice \(b_{\rm min}=a\), this becomes
\begin{equation}
    \Delta\phi_{\rm throat}
    =
    \frac{M_{\rm Pl}}{2\alpha}
    \log\!\left(
        \frac{1+a}{2a}
    \right)
    \simeq
    \frac{M_{\rm Pl}}{2\alpha}
    \log\frac{1}{2a}
    \qquad (a\ll1) \;.
    \label{eq:quadratic-throat-excursion}
\end{equation}
As long as this displacement stays within the radius of validity of the
Taylor expansion of some microscopic potential,
Eq.~\eqref{eq:quadratic-potential} is the leading local approximation to a
generic stabilizing force.  In a long throat, however,
\(\Delta\phi_{\rm throat}\) can be order one or larger in Planck units, and
the purely quadratic model is then better regarded as a controlled toy model
for a globally harmonic restoring force, not as a universal model of
stabilized moduli at large field distance.

The force generated by Eq.~\eqref{eq:quadratic-potential} is
\begin{equation}
    V_{{\rm quad},\phi}(\phi)
    =
    m^2(\phi-\Phi_\infty) \;.
\end{equation}
Evaluated on the massless GHS profile,
\begin{equation}
    \phi_{\rm GHS}(b)
    =
    \Phi_\infty
    +
    \frac{M_{\rm Pl}}{2\alpha}
    \log\!\left(\frac{1+b}{a+b}\right) \;,
    \label{eq:quadratic-ghs-profile}
\end{equation}
this gives
\begin{equation}
    V_{{\rm quad},\phi}\!\left(\phi_{\rm GHS}(b)\right)
    =
    \frac{m^2M_{\rm Pl}}{2\alpha}
    \log\!\left(
        \frac{1+b}{a+b}
    \right).
    \label{eq:quadratic-force-ghs}
\end{equation}
Since \(\phi_{\rm GHS}>\Phi_\infty\) throughout the exterior in the magnetic
frame, \(V_{{\rm quad},\phi}>0\): the quadratic potential acts as a restoring
force opposing the gauge-sourced displacement.

We first analyze the quadratic potential in the fixed-throat approximation,
neglecting its back-reaction on both the geometry and the leading GHS scalar
profile.  In this approximation, the direct source entering the linearized
scalar equation is
\begin{equation}
    r_+^2 w(b)\,
    V_{{\rm quad},\phi}\!\left(\phi_{\rm GHS}(b)\right)
    =
    \frac{\mu^2M_{\rm Pl}}{2\alpha}
    (1+b)(a+b)
    \log\!\left(
        \frac{1+b}{a+b}
    \right) \;.
    \label{eq:quadratic-weighted-source}
\end{equation}
Because the deformation equation is linear in \(V_{,\phi}\), the induced
correction \(\delta\phi\), the pointwise force ratio, the local flux
diagnostic, and the cumulative diagnostic all scale linearly with
\(\mu^2=m^2r_+^2\).  This makes the quadratic potential a useful calibration
case: the response only needs to be computed once and can then be rescaled
to any mass.

Using the positive gauge-source magnitude
\(\mathcal S_{\rm gauge}\) of Eq.~\eqref{eq:gauge-source-magnitude}, the
pointwise force ratio is
\begin{equation}
    \eta_{\rm src}^{\rm quad}(b)
    =
    \frac{
        \mu^2(1+b)^3(a+b)
        \log\!\left(\dfrac{1+b}{a+b}\right)
    }{
        1-a
    } \;,
    \label{eq:quadratic-eta-source}
\end{equation}
the direct comparison between the quadratic restoring force and the gauge
source that supports the massless GHS throat.  The complementary local flux
diagnostic, normalized to the total scalar flux scale
\(|\mathcal F_\infty|=M_{\rm Pl}(1-a)/(2\alpha)\), is
\begin{equation}
    \epsilon_{\rm loc}^{\rm quad}(b)
    =
    \frac{
        \mu^2\, b(1+b)(a+b)
        \log\!\left(\dfrac{1+b}{a+b}\right)
    }{
        1-a
    } \;.
    \label{eq:quadratic-local-diagnostic}
\end{equation}
At the outer edge of the source-dominated throat, \(b\sim1\), these give
\begin{align}
    \epsilon_{\rm loc}^{\rm quad}(1)
    &=
    \mu^2
    \frac{2(1+a)\log\!\left(\dfrac{2}{1+a}\right)}{1-a}
    \simeq
    2\log2\, \mu^2 \;, \nonumber \\\eta_{\rm src}^{\rm quad}(1)
    &=
    \mu^2
    \frac{8(1+a)\log\!\left(\dfrac{2}{1+a}\right)}{1-a}
    \simeq
    8\log2\,\mu^2 \;,
    \label{eq:quadratic-local-b1}
\end{align}
for \(a\ll1\).  Both become order one only for \(m r_+\sim O(1)\): the
local flux diagnostic at \(\mu=O(1)\), and the pointwise ratio somewhat
earlier, at \(\mu=O(0.4)\).  This is close to the scale found below in the
back-reacted exterior test, and matches the expected local criterion: a
scalar with Compton wavelength of order the horizon radius directly
competes with the gauge source at the outer edge of the throat.

Neither diagnostic behaves well if the fixed GHS background is extended to
arbitrarily large radius.  For \(b\gg1\),
\begin{equation}
    \log\frac{1+b}{a+b}
    =
    \frac{1-a}{b}
    +
    O(b^{-2}) \;,
\end{equation}
so \(\epsilon_{\rm loc}^{\rm quad}(b)\sim \mu^2 b^2\) grows without bound.
This does not mean the throat is infinitely sensitive to an arbitrarily
small mass; it means that a quadratic mass term controls the asymptotic
massive tail, whereas the fixed-throat diagnostic is meant to test a chosen
finite exterior region.

The cumulative diagnostic,
\begin{equation}
    \epsilon_{\rm cum}^{\rm quad}(b)
    =
    \frac{
        \mu^2
        \displaystyle
        \int_{b_{\rm min}}^{b}
        d\bar b\,
        (1+\bar b)(a+\bar b)
        \log\!\left(\frac{1+\bar b}{a+\bar b}\right)
    }{
        1-a
    } \;,
    \label{eq:quadratic-cumulative-diagnostic}
\end{equation}
has a positive integrand for all \(\bar b>0\), so it grows monotonically and
peaks at the chosen outer boundary.  For \(\bar b\gg1\), the integrand behaves
as \((1-a)\bar b+O(1)\), so
\(\epsilon_{\rm cum}^{\rm quad}\) grows as \(b^2\) in the asymptotic region;
its value is meaningful only once the radial interval under examination is
specified.

We can also solve directly for the fixed-throat profile deformation.  The
boundary-value problem is
\begin{equation}
    \frac{d}{db}
    \left[
        b(a+b)\,\delta\phi'(b)
    \right]
    =
    r_+^2(1+b)(a+b)\,
    m^2
    \left[
        \phi_{\rm GHS}(b)-\Phi_\infty
    \right] \;,
    \label{eq:quadratic-delta-phi-equation}
\end{equation}
with boundary conditions
\begin{equation}
    \delta\phi(b_{\rm max})=0 \;,
    \qquad
    b_{\rm min}(a+b_{\rm min})\,\delta\phi'(b_{\rm min})=0 \;.
    \label{eq:quadratic-boundary-conditions}
\end{equation}
The first boundary condition fixes the asymptotic scalar value at the chosen outer boundary,
while the second imposes a vanishing deformation flux through the regulated inner
boundary.  From the solution, we monitor
\begin{equation}
    \Delta\phi_{\rm max}
    =
    \max_{b_{\rm min}\leq b\leq b_{\rm max}}
    |\delta\phi(b)|,
    \qquad
    \Delta_g^{\rm max}
    =
    \max_{b_{\rm min}\leq b\leq b_{\rm max}}
    \left|
        \exp\!\left(\frac{2\alpha\, \delta\phi}{M_{\rm Pl}}\right)-1
    \right|.
    \label{eq:quadratic-profile-diagnostics}
\end{equation}

Restricted to the source-dominated throat, \(b_{\rm max}\sim1\), the profile
deformation becomes order one only for \(m^2r_+^2\sim O(1)\), consistent with
\(\epsilon_{\rm loc}(b\sim1)\sim m^2r_+^2\).  A more stringent perturbative
test extends the fixed GHS background over the long exterior interval
\(b_{\rm min}=a\), \(b_{\rm max}=1/a\), where the deformation accumulates
over a large radial range.  Denoting by \(m_{\rm crit}\) the scalar mass at
which this long-tail profile deformation becomes order one, the
fixed-throat scan gives
\begin{equation}
    m_{\rm crit} r_+
    \simeq
    c_{\rm quad}\,
    \sqrt{a\log(1/a)} \;,
    \qquad
    c_{\rm quad}\simeq 1.35\text{--}1.40 \;,
    \label{eq:quadratic-critical-sqrt-alog}
\end{equation}
for the benchmark choices
\begin{equation}
    \Phi_\infty=2M_{\rm Pl} \;,
    \qquad
    2\alpha^2=1 \;,
    \qquad
    b_{\rm min}=a \;,
    \qquad
    b_{\rm max}=1/a \;.
\end{equation}
Equivalently,
\begin{equation}
    m_{\rm crit}^2 r_+^2
    \simeq
    c_{\rm quad}^2\,a\log(1/a)
    \sim
    2\,a\log(1/a) \;.
\end{equation}
The coefficient \(c_{\rm quad}\) is not universal: it depends on the
regulated domain and on the chosen definition of order-one breakdown.  The
\(a\log(1/a)\) scaling of this long-tail fixed-background criterion is the
robust feature.  In terms of the critical Compton wavelength,
\begin{equation}
    \lambda_{c,{\rm crit}}
    =
    \frac{1}{m_{\rm crit}r_+}
    \sim
    \frac{1}{\sqrt{a\log(1/a)}} \;,
    \label{eq:fixed-throat-compton-scale}
\end{equation}
which is parametrically larger than the horizon radius.  The logarithm
appears because the deformation is compared to the full GHS scalar
excursion, \(\Delta\phi_{\rm throat}\propto\log(1/a)\), over the regulated
exterior domain.

We now compare the fixed-throat estimate with a back-reacted exterior
evolution of the Einstein--Maxwell--scalar system.  This is not a full
asymptotic shooting problem: we initialize the system at a regulated inner
cutoff using the corresponding near-extremal GHS data, and integrate the
coupled equations outward to a fixed UV point, with the scalar potential
included directly in the metric and scalar equations rather than treated as
a perturbing source on a fixed GHS background.

For the back-reacted exterior integration, it is convenient to use the
dimensionless radial coordinate,
\begin{equation}
    x \equiv \frac{r}{r_+}=1+b \;.
    \label{eq:x-coordinate-backreacted}
\end{equation}
Thus, the regulated inner point is
\[
    x_0=1+b_{\rm cut} \;,
\]
and the exterior solution is evolved to a fixed UV point \(x_{\rm UV}\).

The inner boundary condition is set at \(x_0=1+b_{\rm cut}\), and the system
is evolved to \(x_{\rm UV}=2\), a fixed endpoint corresponding to the outer edge of the regulated throat.  The diagnostic is the maximum fractional
deviation of the local gauge coupling from the corresponding \(V=0\) GHS
solution,
\begin{equation}
    \Delta_g^{\rm max}
    =
    \max_{x_0\leq x\leq x_{\rm UV}}
    \left|
        \frac{g^2_{\;V \neq 0}(x)}{g^2_{\;V=0}(x)}-1
    \right| \;,
    \label{eq:backreacted-g-diagnostic}
\end{equation}
and we define the critical mass by the deliberately moderate threshold,
\(\Delta_g^{\rm max}=0.1\). As such, the analysis is focused on when the coupling profile is changed at the ten-percent level, not when it is destroyed entirely.

For the benchmark
\begin{equation}
    \Phi_\infty=2M_{\rm Pl} \;,
    \qquad
    2\alpha^2=1 \;,
    \qquad
    x_{\rm UV}=2 \;,
    \qquad
\end{equation}
we ran two complementary scans.  First, keeping \(a=10^{-4}\) fixed and
varying \(b_{\rm cut}\), we obtain the values in
Table~\ref{tab:quadratic-backreacted-bcut-scan}. 
The values are all order one in horizon units, 
\begin{equation}
    mr_+\sim0.25-0.35 \;, \qquad
\lambda_c\sim3 - 4 \;.
\end{equation}
The residual \(b_{\rm cut}\)-dependence reflects
the fact that moving the inner boundary deeper into the throat increases the
local field excursion sampled by the solution; the critical Compton
wavelength nonetheless remains only a few times the horizon radius, remaining comparable to the black hole scale.

\begin{table}[t]
\centering
\begin{tabular}{ccc}
\toprule
    \(b_{\rm cut}\)
    &
    \(\mu_{\rm crit}=m_{\rm crit}r_+\)
    &
    \(\lambda_{c,{\rm crit}}=1/\mu_{\rm crit}\)
    \\
    \midrule
    \(10^{-2}\)        & \(0.2723\) & \(3.67\) \\
    \(3\times10^{-3}\) & \(0.2610\) & \(3.83\) \\
    \(10^{-3}\)        & \(0.2638\) & \(3.79\) \\
    \(3\times10^{-4}\) & \(0.2839\) & \(3.52\) \\
    \(10^{-4}\)        & \(0.3178\) & \(3.15\)\\
    \bottomrule
\end{tabular}
\caption{Back-reacted exterior scan at fixed near-extremality
\(a=10^{-4}\), varying the regulated inner cutoff \(b_{\rm cut}\). The
critical mass is defined by \(\Delta_g^{\rm max}=0.1\) on
\(1+b_{\rm cut}\leq x\leq2\).}
\label{tab:quadratic-backreacted-bcut-scan}
\end{table}

Second, we ran a fixed-ratio near-extremal scan with \(b_{\rm cut}=10a\),
shown in Table~\ref{tab:quadratic-backreacted-a-scan}. The scan is not
monotonic in \(a\). This non-monotonicity is not a numerical artifact: repeating the calculation
with the logarithmic-coordinate step density increased by factors of two and
four left the extracted critical values unchanged at the displayed precision
(see Table~\ref{tab:quadratic-backreacted-convergence}). The \(V=0\) baseline
error decreased by orders of magnitude with resolution, while the critical
values stayed fixed. The non-monotonicity therefore reflects the cutoff
prescription, not a loss of numerical convergence: in the fixed-ratio
prescription, taking \(a\to0\) also moves the inner boundary deeper into the
throat, mixing the longer-throat limit with a changing radial domain.

\begin{table}[h]
\centering
\begin{tabular}{cccc}
\toprule
    \(a\) & \(b_{\rm cut}=10a\)
      & \(\mu_{\rm crit}=m_{\rm crit}r_+\)
      & \(\lambda_{c,{\rm crit}}=1/\mu_{\rm crit}\)
    \\
    \midrule
    \(10^{-2}\)        & \(10^{-1}\)        & \(0.3508\) & \(2.85\) \\
    \(3\times10^{-3}\) & \(3\times10^{-2}\) & \(0.2964\) & \(3.37\) \\
    \(10^{-3}\)        & \(10^{-2}\)        & \(0.2715\) & \(3.68\) \\
    \(3\times10^{-4}\) & \(3\times10^{-3}\) & \(0.2600\) & \(3.85\) \\
    \(10^{-4}\)        & \(10^{-3}\)        & \(0.2638\) & \(3.79\) \\
    \(3\times10^{-5}\) & \(3\times10^{-4}\) & \(0.2947\) & \(3.39\) \\
    \(10^{-5}\)        & \(10^{-4}\)        & \(0.4158\) & \(2.41\)\\
    \bottomrule
\end{tabular}
\caption{Back-reacted exterior scan with fixed ratio \(b_{\rm cut}=10a\),
again defining the critical mass by \(\Delta_g^{\rm max}=0.1\) on
\(1+b_{\rm cut}\leq x\leq2\). The non-monotonicity reflects the cutoff
prescription rather than a loss of numerical convergence.}
\label{tab:quadratic-backreacted-a-scan}
\end{table}

\begin{table}[h]
\centering
\begin{tabular}{cccc}
\toprule
    \(a\) & \(b_{\rm cut}\)
      & \(\mu_{\rm crit}\)
      & \(\lambda_{c,{\rm crit}}\)
    \\
    \midrule
    \(10^{-4}\)        & \(10^{-3}\)        & \(0.263809\) & \(3.790620\) \\
    \(3\times10^{-5}\) & \(3\times10^{-4}\) & \(0.294689\) & \(3.393403\) \\
    \(10^{-5}\)        & \(10^{-4}\)        & \(0.415763\) & \(2.405214\) \\
    \bottomrule
\end{tabular}
\caption{Resolution check for the non-monotonic part of the fixed-ratio
scan. Increasing the logarithmic-coordinate step density leaves the
extracted critical values unchanged at the displayed precision.}
\label{tab:quadratic-backreacted-convergence}
\end{table}

Over the back-reacted exterior interval \(1+b_{\rm cut}\leq x\leq2\), the
critical scale therefore remains
\[
    m_{\rm crit}r_+=O(0.1\text{--}1) \;,
    \qquad
    \lambda_{c,{\rm crit}}=O(1\text{--}10) \;,
\]
rather than following the fixed-throat long-tail scaling
\(\sqrt{a\log(1/a)}\).

Our back-reacted exterior scan is a local, quantitative version of this
criterion for the near-extremal exterior segment.  Defining the onset of
breakdown by a ten-percent change in the local gauge-coupling profile, we
find
\begin{equation}
    m_{\rm crit} r_+
    \simeq
    0.25\text{--}0.4 \;,
    \qquad
    \lambda_{c,{\rm crit}}
    \simeq
    2.5\text{--}4 \;.
    \label{eq:hh-comparison-critical}
\end{equation}
The transition therefore occurs when the Compton wavelength is a few horizon
radii, precisely the order-one crossover anticipated by the massive-dilaton
analyses of Refs \cite{Gregory:1992kr,Horne:1992bi}.  Note that the numerical coefficient in
Eq.~\eqref{eq:hh-comparison-critical} is not universal: it depends on the
near-extremality, the cutoff prescription, the UV comparison point, and the
chosen ten-percent threshold.  What is robust is the scaling
\(m_{\rm crit}r_+\sim O(1)\) for the back-reacted exterior segment.

This comparison also clarifies why the fixed-throat long-tail criterion
\[
    m_{\rm crit}r_+\sim\sqrt{a\log(1/a)}
\]
should not be identified with the massive-dilaton crossover.  The two
criteria answer different questions.

The massive-dilaton crossover is a physical, back-reacted statement.  It
compares the scalar Compton wavelength \(m^{-1}\) with the black hole size
\(r_+\).  If
\[
    m r_+\ll1 \;,
\]
the scalar is light on the scale of the exterior geometry, and the solution
can behave approximately like a massless dilatonic black hole.  If instead
\[
    m r_+\gg1 \;,
\]
the scalar is heavy on the same scale, its variation is suppressed, and the
solution is driven toward a pinned, approximately Reissner--Nordstr\"om
regime.  Thus, the physical crossover is expected at
\[
    m r_+ \sim O(1).
\]

The fixed-throat long-tail criterion is different.  There, the GHS geometry
and the GHS scalar profile are kept fixed by hand, and one asks how small
the mass term must be for the perturbation \(\delta\phi\) not to accumulate
over a prescribed radial interval.  If this interval is taken to be the long
regulated tail
\[
    b_{\rm min}=a,
    \qquad
    b_{\rm max}=1/a \;,
\]
then the perturbation is integrated over a region whose size grows as
\(a\to0\).  The resulting criterion is therefore more stringent:
\[
    m_{\rm crit}r_+
    \sim
    \sqrt{a\log(1/a)} \;.
\]
This scaling does not mean that the physical black hole crossover occurs at
a parametrically smaller mass.  It means that preserving the \emph{fixed}
massless GHS profile over an increasingly long asymptotic tail requires an
increasingly light scalar.

The back-reacted exterior calculation tests a different and more local
question: whether a self-consistent exterior solution, evolved from a
regulated near-horizon cutoff to a fixed physical UV point \(x_{\rm UV}=2\),
still exhibits a region in which the local \(U(1)\) gauge coupling
\(g^2(\phi)=B(\phi)^{-1}\) differs appreciably from its asymptotic value
\(g_\infty^2\).  In that problem, the radial
domain is not enlarged as \(1/a\), and the geometry is allowed to respond to
the scalar potential.  The result therefore follows the expected
Compton-wavelength criterion,
\[
    m_{\rm crit}r_+=O(0.1\text{--}1) \;,
\]
with the ten-percent deformation threshold giving
\(m_{\rm crit}r_+\simeq0.25\text{--}0.4\) in the scans above.  Thus, the
back-reacted calculation restores the physical expectation that the throat
segment is significantly modified only when the scalar mass is of order inverse black hole size.

The fixed-throat and back-reacted exterior analyses therefore answer
different questions.  The fixed-throat calculation asks how large a mass
term can be before a potential-induced perturbation accumulates over a
prescribed GHS exterior domain; extended to \(b_{\rm max}=1/a\), it gives
\(m_{\rm crit}r_+\sim\sqrt{a\log(1/a)}\), the correct criterion for
preserving the fixed GHS profile over that long regulated interval.  The
back-reacted exterior calculation instead asks whether a self-consistent
black hole exterior, evolved to a fixed physical UV point \(x_{\rm UV}=2\),
still exhibits the GHS-like coupling excursion.  The relevant criterion is
local to the exterior segment tested, and the numerical result,
\(m_{\rm crit}r_+\sim0.25\)--\(0.4\),
\(\lambda_{c,{\rm crit}}\sim2.5\)--\(4\), sits much closer to the local
source-dominance estimates \(\eta_{\rm src}(b\sim1)\sim1\) and
\(\epsilon_{\rm loc}(b\sim1)\sim m^2r_+^2\) than to the long-tail
fixed-background scaling.

There is no contradiction here.  The fixed-throat long-tail criterion is
stringent because it demands that the massive scalar perturbation stay small
over an increasingly large asymptotic interval.  The back-reacted exterior
calculation instead tests the existence of a self-consistent
altered-coupling region outside the horizon, for which the quadratic
potential need only have a Compton wavelength a few times larger than the
horizon radius to modify significantly the throat segment.

The quadratic model therefore gives the following hierarchy of criteria:
\begin{align}
    m r_+ \ll \sqrt{a\log(1/a)}
    &\quad\Longrightarrow\quad
    \text{fixed GHS profile preserved over the long regulated tail},
    \label{eq:quad-fixed-throat-preserved}
    \\
    m r_+ \sim O(0.1\text{--}1)
    &\quad\Longrightarrow\quad
    \text{\(\sim\)10\% deformation of the back-reacted exterior segment},
    \label{eq:quad-backreacted-breakdown}
    \\
    m r_+ \gtrsim O(1)
    &\quad\Longrightarrow\quad
    \text{scalar pinned on the scale of the horizon}.
    \label{eq:quad-pinned}
\end{align}
The first is a perturbative fixed-background result; the second is the
relevant criterion for the back-reacted exterior region studied here; the
third is the expected local Compton-wavelength criterion.

This comparison clarifies what is, and is not, universal about the
quadratic potential.  The scaling \(m_{\rm crit}^2r_+^2\sim a\log(1/a)\) is
a property of the fixed-throat, long-tail perturbation problem, and should
not be used as the criterion for the existence of a back-reacted
altered-coupling region near the black hole.  The back-reacted exterior
solutions show that such a region survives until the scalar mass is large
enough that its Compton wavelength is only a few horizon radii.

The quadratic potential is therefore best viewed as a calibration case rather
than as a new mechanism.  It reproduces the expected massive-scalar
criterion: a black hole exterior is significantly modified when the scalar
Compton wavelength becomes comparable to the horizon scale.  The additional
lesson of the fixed-throat analysis is to distinguish this physical
back-reacted crossover from the more stringent long-tail perturbative
criterion obtained by requiring the massless GHS profile to remain accurate
over an artificially extended regulated exterior interval.  This distinction
will be useful below, where less elementary potentials can produce local,
cumulative, or oscillatory effects that do not reduce to a simple Compton
wavelength estimate.


\subsection{Shifted exponential stabilizing potential}
\label{subsec:shifted-exponential-static}

We next consider a stabilizing exponential potential whose minimum lies at
the prescribed asymptotic scalar value.  It is useful to write the
displacement from this value as
\begin{equation}
    \varphi(b)\equiv \phi(b)-\Phi_\infty \;,
    \label{eq:shifted-exp-varphi-def}
\end{equation}
so that the minimum sits at \(\varphi=0\).  This is only a field parametrization for the
same canonically normalized EMD scalar field used in the previous subsections; since
\(d\varphi=d\phi\), derivatives with respect to \(\varphi\) and \(\phi\)
are identical.  On the GHS background,
\begin{equation}
    \varphi_{\rm GHS}(b)
    =
    \frac{M_{\rm Pl}}{2\alpha}
    \log\!\left(
        \frac{1+b}{a+b}
    \right)
    >0 \;.
    \label{eq:shifted-exp-varphi-ghs}
\end{equation}

The shifted exponential potential is
\begin{equation}
    V_{\rm sh.exp.}(\varphi)
    =
    \frac{\mu^2 M_{\rm Pl}^2}{\lambda^2 r_+^2}
    \left[
        e^{-\lambda\varphi/M_{\rm Pl}}
        -1
        +
        \lambda\frac{\varphi}{M_{\rm Pl}}
    \right] \;,
    \qquad
    \lambda>0 \;,
    \label{eq:shifted-exp-potential}
\end{equation}
where
\[
    \mu \equiv m r_+
\]
is the dimensionless small-field mass parameter.  The potential satisfies
\begin{equation}
    V_{\rm sh.exp.}(0)=0 \;,
    \qquad
    V_{{\rm sh.exp.},\varphi}(0)=0 \;,
    \qquad
    V_{{\rm sh.exp.},\varphi\varphi}(0)=\frac{\mu^2}{r_+^2}=m^2 \;.
\end{equation}
Near the minimum,
\begin{equation}
    V_{\rm sh.exp.}(\varphi)
    =
    \frac12\,\frac{\mu^2}{r_+^2}\,\varphi^2
    -
    \frac16\,\frac{\lambda\mu^2}{M_{\rm Pl}r_+^2}\,
    \varphi^3
    +
    O(\lambda^2\varphi^4) \;.
    \label{eq:shifted-exp-small-field}
\end{equation}
Thus, the quadratic potential of
Section~\ref{subsec:quadratic-potential} is recovered in the limit
\(\lambda\to0\) at fixed \(\mu\), while \(\lambda\) controls the leading
departure from the harmonic approximation.

The force entering the scalar equation is
\begin{equation}
    V_{{\rm sh.exp.},\phi}
    =
    V_{{\rm sh.exp.},\varphi}
    =
    \frac{\mu^2 M_{\rm Pl}}{\lambda r_+^2}
    \left[
        1-e^{-\lambda\varphi/M_{\rm Pl}}
    \right] \;.
    \label{eq:shifted-exp-force}
\end{equation}
Along the magnetic GHS throat, \(\varphi_{\rm GHS}>0\), so
\(V_{{\rm sh.exp.},\phi}>0\).  The potential therefore acts as a restoring
force opposing the gauge-sourced displacement.  Unlike the quadratic force,
however, it saturates at large positive displacement,
\begin{equation}
    V_{{\rm sh.exp.},\phi}
    \longrightarrow
    \frac{\mu^2 M_{\rm Pl}}{\lambda r_+^2}
    \qquad
    \left(\lambda\varphi/M_{\rm Pl}\gg1\right) \;,
\end{equation}
rather than growing without bound.

Using the general definition of Eq.~\eqref{eq:eta-source-ratio}, the
pointwise force ratio evaluated on the fixed, unperturbed GHS profile is
\begin{equation}
    \eta_{\rm src}^{\rm sh.exp.}(b)
    =
    \frac{
        r_+^2
        \left|
            V_{{\rm sh.exp.},\phi}
            \!\left(\varphi_{\rm GHS}(b)\right)
        \right|
    }{
        \mathcal S_{\rm gauge}(b)
    } \;,
    \label{eq:shifted-exp-eta-def}
\end{equation}
where recall that \(V_{,\phi}=V_{,\varphi}\) because
\(\varphi=\phi-\Phi_\infty\).  Substituting
Eq.~\eqref{eq:shifted-exp-force} gives
\begin{equation}
    \eta_{\rm src}^{\rm sh.exp.}(b)
    =
    \frac{2\alpha\mu^2}{\lambda}
    \frac{(1+b)^3(a+b)}{1-a}
    \left[
        1-
        \left(
            \frac{a+b}{1+b}
        \right)^{\lambda/(2\alpha)}
    \right] \;.
    \label{eq:shifted-exp-eta-ghs}
\end{equation}
This reduces to the quadratic result in the small-field regime
\(\lambda\varphi_{\rm GHS}/M_{\rm Pl}\ll1\), since
\begin{equation}
    1-e^{-\lambda\varphi_{\rm GHS}/M_{\rm Pl}}
    =
    \lambda\frac{\varphi_{\rm GHS}}{M_{\rm Pl}}
    +
    O(\lambda^2\varphi_{\rm GHS}^2) \;.
\end{equation}

On the throat interval \(a\leq b\leq1\), for the benchmark choices
\(\lambda=0.1\) and \(2\alpha^2=1\), the maximum of
\(\eta_{\rm src}^{\rm sh.exp.}\) lies at the outer edge of the throat,
\(b=1\), throughout the near-extremal range probed below.  We define
\(\mu_{\rm cross}^{\rm throat}\) to be the value of
\[
    \mu\equiv m r_+
\]
for which this maximum first becomes order one:
\begin{equation}
    \max_{a\leq b\leq1}
    \eta_{\rm src}^{\rm sh.exp.}(b)
    =
    1 \;.
    \label{eq:shifted-exp-cross-definition}
\end{equation}
Equivalently, \(\mu_{\rm cross}^{\rm throat}\) is the local fixed-throat
mass scale at which the shifted-exponential force becomes comparable to the
GHS gauge source somewhere inside the throat.  It is not obtained from a
global boundary-value problem, and it does not mean that the entire scalar
profile has already been erased.  It is a local source-comparison criterion.

For example, at a reference value \(\mu_{\rm ref}=0.05\), we find
\[
    \max_{a\leq b\leq1}
    \eta_{\rm src}^{\rm sh.exp.}(b)
    =
    1.353\times10^{-2}
    \qquad
    (a=10^{-4}) \;.
\]
Since the shifted-exponential force is proportional to \(\mu^2\), the
crossover value follows simply by rescaling:
\begin{equation}
    \mu_{\rm cross}^{\rm throat}
    =
    \mu_{\rm ref}
    \left[
        \max_{a\leq b\leq1}
        \eta_{\rm src}^{\rm sh.exp.}(b;\mu_{\rm ref})
    \right]^{-1/2}
    =
    0.429862
\end{equation}
for
\[
    \lambda=0.1 \;,\qquad a=10^{-4},\qquad 2\alpha^2=1 \;.
\]
Changing \(a\) between \(10^{-4}\) and \(10^{-2}\) changes this number by
less than one percent.  Thus, the local throat criterion gives an order-one
condition on \(m r_+\), rather than a parametric power of the
near-extremality parameter.

This local throat criterion should not be confused with diagnostics that
extend the fixed GHS background far outside the throat.  The throat region
used above is \(a\leq b\leq1\).  If instead one evaluates the same pointwise
ratio out to a much larger radius, for example to an outer cutoff that grows
as \(b_{\max}\sim1/a\), the maximum no longer measures the local throat
region.  It is dominated by the asymptotic tail.

The reason is simple.  Deep in the tail, \(b\gg1\), the GHS displacement is
small,
\begin{equation}
    \varphi_{\rm GHS}(b)
    =
    \frac{M_{\rm Pl}}{2\alpha}
    \frac{1-a}{b}
    +
    O(b^{-2}) \;,
\end{equation}
while the gauge source falls faster,
\begin{equation}
    \mathcal S_{\rm gauge}(b)
    \sim
    \frac{M_{\rm Pl}}{2\alpha}\,
    \frac{1-a}{b^4} .
\end{equation}
In this small-field regime, the shifted-exponential potential has already
reduced to a quadratic mass term:
\begin{equation}
    V_{{\rm sh.exp.},\phi}(\varphi_{\rm GHS})
    \simeq
    \frac{\mu^2}{r_+^2}\,\varphi_{\rm GHS} \;.
\end{equation}
Therefore, the pointwise ratio behaves as
\begin{equation}
    \eta_{\rm src}^{\rm sh.exp.}(b)
    \sim
    \mu^2 b^3
    \qquad
    (b\gg1) \;.
\end{equation}
The ratio grows in the tail not because the potential force is becoming
large, but because the GHS gauge source used in the denominator is becoming
very small.  Consequently, if the outer endpoint is allowed to scale with
\(a\), the resulting ``critical'' value of \(\mu\) is controlled by the
chosen asymptotic cutoff.  This produces cutoff-dependent powers of \(a\),
including \(\sqrt a\)-type scalings.

These long-tail scalings answer a different question: how far out can one
force the massless GHS profile to remain a good approximation in a region
where a massive scalar should already be relaxing back to its asymptotic
value?  They do not diagnose the local breakdown of the throat itself.  The
local throat breakdown is instead captured by
Eq.~\eqref{eq:shifted-exp-cross-definition}, and gives
\(\mu_{\rm cross}^{\rm throat}=O(1)\).

We also solved the fixed-throat boundary-value problem directly, keeping the
nonlinear dependence of the shifted exponential force on the deformation.
It is convenient to introduce
\begin{equation}
    y\equiv -\,\frac{\varphi}{M_{\rm Pl}} \;,
    \qquad
    y_{\rm GHS}(b)
    =
    -\,\frac{1}{2\alpha}
    \log\!\left(\frac{1+b}{a+b}\right) \;,
    \label{eq:shifted-exp-y-def}
\end{equation}
so that \(y_{\rm GHS}<0\) in the throat.  In units \(M_{\rm Pl}=r_+=1\),
the shifted exponential force gives
\begin{equation}
    -\,V_{{\rm sh.exp.},\phi}
    =
    \frac{\mu^2}{\lambda}
    \left(
        e^{\lambda y}-1
    \right) \;.
\end{equation}
Note that the minus sign appears because \(\varphi=-M_{\rm Pl}y\), so the equation for
\(y\) is the negative of the equation for \(\varphi\).

Then, by writing
\begin{equation}
    y(b)=y_{\rm GHS}(b)+\delta y(b) \;,
\end{equation}
and keeping the direct shifted-exponential source nonlinear, the
fixed-throat deformation equation becomes
\begin{equation}
    \frac{d}{db}
    \left[
        p(b)\,\delta y'(b)
    \right]
    =
    w(b)\,
    \frac{\mu^2}{\lambda}
    \left[
        e^{\lambda(y_{\rm GHS}(b)+\delta y(b))}-1
    \right] \;.
    \label{eq:shifted-exp-nonlinear-bvp}
\end{equation}

The boundary conditions are then
\begin{equation}
    \delta y(b_{\rm max})=0 \;,
    \qquad
    p(b_{\rm min})\,\delta y'(b_{\rm min})=0 \;.
    \label{eq:shifted-exp-bvp-bc}
\end{equation}
The first boundary condition fixes the prescribed asymptotic value at the chosen outer
boundary, while the second imposes a zero deformation flux through the
regulated inner endpoint.

To quantify the nonlinear fixed-throat deformation, we monitor the shift
\(\delta y(b_{\rm min})\) at the inner cutoff and the corresponding change
in the local gauge coupling relative to the undeformed GHS profile.  Since
\[
    g^2(\phi)
    =
    g_\infty^2
    \exp\!\left[
        \frac{2\alpha(\phi-\Phi_\infty)}{M_{\rm Pl}}
    \right]
    =
    g_\infty^2 e^{-2\alpha y} \;,
\]
the gauge-coupling ratio at the cutoff is
\begin{equation}
    S_g
    \equiv
    \frac{
        g^2(y(b_{\rm min}))
    }{
        g^2(y_{\rm GHS}(b_{\rm min}))
    }
    =
    \exp\!\left[
        -2\alpha\,\delta y(b_{\rm min})
    \right] \;.
    \label{eq:shifted-exp-gauge-survival-factor}
\end{equation}
A stabilizing force reduces the displacement
\(\varphi=-M_{\rm Pl}y\) relative to its massless GHS value.  Equivalently,
\(y(b_{\rm min})>y_{\rm GHS}(b_{\rm min})\), so
\(\delta y(b_{\rm min})>0\) and \(S_g<1\).  Thus, \(S_g\) measures the
suppression of the local gauge coupling relative to its undeformed GHS
value at the cutoff, with \(S_g=1\) recovered when the potential is absent. Table~\ref{tab:shifted-exp-nonlinear-scan} shows the resulting scan at the benchmark
\begin{equation}
2\alpha^2=1 \;, \qquad a=10^{-4} \;, \qquad b_{\rm min}=a\;, \qquad b_{\rm max}=1\;, \qquad \lambda=0.1 \;.
\end{equation}

\begin{table}[t]
\centering
\begin{tabular}{cccc}
\toprule
    \(\mu\) & \(\eta_{\rm src}^{\rm max}\) & \(S_g\) & \(\delta y(b_{\rm min})\) \\
    \midrule
    \(0.10\) & \(0.054\) & \(0.990\) & \(0.00715\) \\
    \(0.20\) & \(0.216\) & \(0.961\) & \(0.0283\)  \\
    \(0.30\) & \(0.487\) & \(0.915\) & \(0.0627\)  \\
    \(0.40\) & \(0.866\) & \(0.857\) & \(0.109\)   \\
    \(0.43\) & \(1.001\) & \(0.838\) & \(0.125\)   \\
    \(0.50\) & \(1.353\) & \(0.791\) & \(0.166\) \\
    \bottomrule
\end{tabular}
\caption{Nonlinear fixed-throat scan for the shifted exponential potential,
at \(a=10^{-4}\), \(b_{\rm min}=a\), \(b_{\rm max}=1\), \(\lambda=0.1\),
and \(2\alpha^2=1\).  The maximum of \(\eta_{\rm src}\) occurs at \(b=1\),
where the boundary condition \(\delta y(1)~=~0\) makes it coincide with the
fixed-profile value; hence the displayed \(\eta_{\rm src}^{\rm max}\) scales
as \(\mu^2\).  The quantity
\(S_g~=~\exp[-2\alpha\,\delta y(b_{\rm min})]\) is the local gauge coupling
relative to the undeformed GHS value at the cutoff.}
\label{tab:shifted-exp-nonlinear-scan}
\end{table}

Interpolating to \(\eta_{\rm src}^{\rm max}=1\) gives
\begin{equation}
    \mu_{\rm cross}^{\rm throat}
    =
    0.429862 \;,
    \label{eq:shifted-exp-mu-cross-bvp}
\end{equation}
in agreement with the fixed-profile estimate above.  At this crossover, the
inner scalar displacement is still only mildly changed.  For the benchmark
parameters,
\begin{equation}
    \frac{\delta y(b_{\rm min})}{|y_{\rm GHS}(b_{\rm min})|}
    \simeq
    2\times10^{-2} \;,
\end{equation}
while the local gauge coupling at the cutoff remains
\[
    S_g\simeq0.84
\]
of its undeformed GHS value.  Thus,
\(\eta_{\rm src}^{\rm max}\sim1\) marks the onset of local competition
between the potential force and the gauge source near the outer edge of the
throat.  It does not correspond to the erasure of the scalar excursion, nor
to the complete loss of the local gauge-coupling enhancement.

Finally, we compare the fixed-throat estimate with a back-reacted exterior
evolution of the Einstein--Maxwell--scalar equations.  As in the quadratic
case, this is not a full asymptotic shooting problem: we initialize the
solution at a regulated inner cutoff using the near-extremal GHS data, and
integrate the coupled equations outward to a fixed UV point,
\[
    x_{\rm UV}=2 \;,
    \qquad
    x\equiv \frac{r}{r_+}=1+b \;.
\]
We make a choice of new variables \(Y\) and \(W\) as a
convenient rewriting of the same back-reacted radial equations.  The point is
to isolate the metric combination that appears naturally in the static
spherically symmetric Einstein equations, and the scalar flux that replaces
the fixed-throat Sturm--Liouville flux.  We define
\[
    Y\equiv fR^2 \;,
    \qquad
    W\equiv Y\phi'=Y\varphi' \;.
\]
Thus, \(Y\) packages the redshift function and the areal radius into the
combination controlled directly by the radial Einstein equation, while
\(W\) is the back-reacted scalar flux.  In the fixed-throat limit, \(W\)
plays the same role as the flux \(p(b)\phi'(b)\) used in the
Sturm--Liouville analysis above.

In these variables the effect of the shifted exponential potential is
particularly transparent.  Schematically, the metric equation and the scalar
flux equation take the form
\begin{equation}
    Y''=2-2R^2V(\phi),
    \qquad
    W'=W'_{\rm gauge}+R^2V_{,\phi}(\phi) \;.
    \label{eq:shifted-exp-backreacted-schematic}
\end{equation}
The potential itself therefore enters the metric equation as a local
energy-density source, while its derivative enters the scalar equation as an
additional force competing with the gauge source.  Here
\(V(\Phi_\infty)=0\), so unlike some of the potentials discussed below, no
vacuum-energy subtraction is needed before solving the back-reacted
equations.

We use the benchmark,\begin{equation}
2\alpha^2=1 \;, \qquad a=10^{-4}\;, \qquad b_{\rm cut}=10^{-3} \; \qquad x_{\rm UV}=2\;, \qquad \lambda=0.1 \;, 
\end{equation}
and the same diagnostic
as in the quadratic case,
\begin{equation}
    \Delta_g^{\rm max}
    =
    \max_{x_0\leq x\leq x_{\rm UV}}
    \left|
        \frac{g_{\rm full}^2(x)}{g_{V=0}^2(x)}-1
    \right| \;,
    \qquad
    x_0=1+b_{\rm cut} \;.
    \label{eq:shifted-exp-backreacted-deltag}
\end{equation}
The scan is shown in Table~\ref{tab:shifted-exp-backreacted-scan}.

\begin{table}[t]
\centering
\begin{tabular}{cc}
\toprule
    \(\mu\) & \(\Delta_g^{\rm max}\) \\
    \midrule
    \(0.10\) & \(0.0112\) \\
    \(0.20\) & \(0.0461\) \\
    \(0.30\) & \(0.109\)  \\
    \(0.35\) & \(0.154\)  \\
    \(0.40\) & \(0.210\)  \\
    \(0.43\) & \(0.249\)  \\
    \(0.45\) & \(0.279\)  \\
    \(0.50\) & \(0.365\)  \\
    \(0.60\) & \(0.611\) \\
    \bottomrule
\end{tabular}
\caption{Back-reacted exterior scan for the shifted exponential potential,
at \(a=10^{-4}\), \(b_{\rm cut}=10^{-3}\), \(x_{\rm UV}=2\), \(\lambda=0.1\),
\(2\alpha^2=1\).  The critical mass is defined by
\(\Delta_g^{\rm max}=0.1\).}
\label{tab:shifted-exp-backreacted-scan}
\end{table}

Interpolating to \(\Delta_g^{\rm max}=0.1\) gives
\begin{equation}
    \mu_{\rm crit}^{\rm back}
    =
    0.288292 \;,
    \label{eq:shifted-exp-mucrit-back}
\end{equation}
so the back-reacted exterior begins to deviate at the ten-percent level for
\(\mu=O(0.3)\); this is the same order-one range found in the quadratic benchmark,
\(\mu_{\rm crit}\simeq0.25\)--\(0.4\).

The fixed-throat and back-reacted thresholds need not coincide exactly,
since they measure different quantities.  The source ratio
\(\eta_{\rm src}\) marks where the potential force first becomes locally
comparable to the gauge source on the unperturbed profile, whereas
\(\Delta_g^{\rm max}\) measures the accumulated fractional change in the
local gauge coupling along the back-reacted exterior.  Even so, both
diagnostics point to the same conclusion for the shifted exponential
potential: the throat/exterior crossover is set by an order-one value of
\(m r_+\), not by a power of the near-extremality parameter \(a\).



\subsection{Exponential runaway potentials}
\label{subsec:exponential-runaway}

We now consider pure exponential runaway potentials.  This case differs
qualitatively from the quadratic and shifted-exponential stabilizing
potentials: there is no preferred minimum at the asymptotic scalar value,
and therefore no fixed Compton wavelength.  Instead, the local slope and
curvature of the potential vary along the throat.

We use the displacement from the asymptotic scalar value,
\begin{equation}
    \varphi(b) \equiv \phi(b)-\Phi_\infty \;,
    \label{eq:exp-runaway-varphi-def}
\end{equation}
chosen so that the magnetic GHS throat corresponds to a positive inward
displacement.  Following the conventions of Section~\ref{sec:throat-coordinates}, the
gauge kinetic function is
\begin{equation}
    B(\varphi)
    =
    B_\infty
    \exp\!\left(
        -2\alpha\frac{\varphi}{M_{\rm Pl}}
    \right),
    \qquad
    \frac{g^2(\varphi)}{g_\infty^2}
    =
    \exp\!\left(
        2\alpha\frac{\varphi}{M_{\rm Pl}}
    \right) \;.
    \label{eq:exp-runaway-gauge-coupling}
\end{equation}
Thus, increasing \(\varphi\) corresponds to increasing the local gauge
coupling.  On the GHS background,
\begin{equation}
    \frac{g^2(b)}{g_\infty^2}
    =
    \frac{1+b}{a+b},
    \qquad
    \varphi_{\rm GHS}(b)
    =
    \frac{M_{\rm Pl}}{2\alpha}
    \log\!\left(
        \frac{1+b}{a+b}
    \right) \;.
    \label{eq:exp-runaway-ghs-profile}
\end{equation}
The scalar therefore moves toward larger \(\varphi\), and hence toward
larger \(g^2\), as one goes inward along the magnetic throat.

We take the exponential potential to be
\begin{equation}
    V(\varphi)
    =
    V_\infty
    \exp\!\left[
        -\sigma\lambda
        \frac{\varphi}{M_{\rm Pl}}
    \right] \;,
    \qquad
    \lambda>0 \;,
    \qquad
    \sigma=\pm1 \;.
    \label{eq:exp-runaway-potential}
\end{equation}
The sign \(\sigma\) is meaningful only relative to the direction of the
GHS scalar flow.  Since the magnetic throat drives \(\varphi\) upward,
\(\sigma=+1\) means that the scalar moves down the exponential as it enters
the throat, whereas \(\sigma=-1\) means that it climbs an exponential wall.
We will refer to these as the favourable and dangerous orientations,
respectively.

For \(\sigma=+1\),
\begin{equation}
    V(\varphi)=V_\infty e^{-\lambda\varphi/M_{\rm Pl}} \;.
\end{equation}
The potential decreases in the same direction in which the GHS throat
increases the gauge coupling.  Its force,
\begin{equation}
    V_{,\phi}
    =
    V_{,\varphi}
    =
    -\,\frac{\lambda}{M_{\rm Pl}}V \;,
\end{equation}
therefore points toward smaller \(\varphi\), opposing the increase of \(g^2\) driven by the black hole.  However, the magnitude of this force is exponentially
suppressed along the GHS trajectory.  This is why this orientation is
favourable: although the force has the restoring sign, it becomes weaker in
the deep throat and does not create an exponentially growing obstruction.

For \(\sigma=-1\), by contrast,
\begin{equation}
    V(\varphi)=V_\infty e^{+\lambda\varphi/M_{\rm Pl}} \;,
\end{equation}
so the GHS trajectory climbs an exponential wall.  The potential force has
the opposite sign and its magnitude grows inward.  This is the dangerous
orientation: even a small asymptotic normalization \(V_\infty\) can become
important in the displaced part of a near-extremal throat.

The local effective mass is not constant:
\begin{equation}
    m_{\rm eff}^2(\varphi)
    \equiv
    V_{,\varphi\varphi}
    =
    \frac{\lambda^2}{M_{\rm Pl}^2}V(\varphi) \;.
    \label{eq:exp-runaway-meff}
\end{equation}
Thus, the exponential potential cannot be characterized by a single analogue
of \(m r_+\).  The relevant obstruction is instead the running potential
force evaluated along the would-be GHS profile.

Across the finite throat interval \(a\leq b\leq1\), the GHS displacement is
\begin{equation}
    \Delta\varphi_{\rm throat}
    =
    \varphi_{\rm GHS}(a)-\varphi_{\rm GHS}(1)
    =
    \frac{M_{\rm Pl}}{2\alpha}
    \log\!\left(
        \frac{(1+a)^2}{4a}
    \right)
    \simeq
    \frac{M_{\rm Pl}}{2\alpha}
    \log\frac{1}{4a}
    \qquad
    (a\ll1) \;.
    \label{eq:exp-runaway-throat-excursion}
\end{equation}
Note that the factor \(4a\) is an endpoint effect: here, the comparison is between the
regulated bottom of the throat, \(b=a\), and the throat exit, \(b=1\), rather
than between the horizon and infinity.

Evaluated on the GHS profile, the exponential potential becomes
\begin{equation}
    V(\varphi_{\rm GHS}(b))
    =
    V_\infty
    \left(
        \frac{1+b}{a+b}
    \right)^{-\sigma q} \;,
    \qquad
    q\equiv\frac{\lambda}{2\alpha} \;.
    \label{eq:exp-runaway-V-on-ghs}
\end{equation}
Its force is
\begin{equation}
    V_{,\varphi}
    =
    -\,\sigma\frac{\lambda}{M_{\rm Pl}}V \;,
\end{equation}
and hence
\begin{equation}
    \left|
        V_{,\varphi}(\varphi_{\rm GHS})
    \right|
    =
    \frac{\lambda V_\infty}{M_{\rm Pl}}
    \left(
        \frac{1+b}{a+b}
    \right)^{-\sigma q} \;.
    \label{eq:exp-runaway-force-on-ghs}
\end{equation}
Near the bottom of the regulated throat, comparing the potential value and
force at \(b=a\) to their values at the outer edge \(b=1\) gives
\begin{equation}
    \frac{V(\varphi_{\rm GHS}(a))}
         {V(\varphi_{\rm GHS}(1))}
    \sim
    (4a)^{\sigma q},
    \qquad
    \frac{
        \left|V_{,\varphi}(\varphi_{\rm GHS}(a))\right|
    }{
        \left|V_{,\varphi}(\varphi_{\rm GHS}(1))\right|
    }
    \sim
    (4a)^{\sigma q}
    \label{eq:exp-runaway-bottom-to-edge-ratio}
\end{equation}
for \(a\ll1\).  For \(\sigma=+1\), the potential energy and force are
suppressed at the bottom of the throat relative to the outer edge.  For
\(\sigma=-1\), they are amplified.

The first diagnostic is the pointwise source ratio.  Here, the fixed-throat scalar equation of Eq.~\eqref{eq:fixed-throat-scalar-full} has the weighted form
\begin{equation}
    \frac{d}{db}
    \left[
        p(b)\frac{d\varphi}{db}
    \right]
    =
    r_+^2 w(b)
    \left[
        V_{,\varphi}(\varphi)
        +
        \frac14 B_{,\varphi}F^2
    \right] \;.
    \label{eq:exp-runaway-fixed-throat-eq}
\end{equation}
Using the positive GHS gauge-source magnitude
\(\mathcal S_{\rm gauge}\) of Eq.~\eqref{eq:gauge-source-magnitude}, the
pointwise ratio is
\begin{equation}
    \eta_{\rm src}^{\rm exp}(b)
    =
    \frac{
        r_+^2
        \left|
            V_{,\varphi}(\varphi_{\rm GHS}(b))
        \right|
    }{
        \mathcal S_{\rm gauge}(b)
    } \;.
    \label{eq:exp-runaway-eta-def}
\end{equation}
Substituting Eq.~\eqref{eq:exp-runaway-force-on-ghs} gives
\begin{equation}
    \eta_{\rm src}^{\rm exp}(b)
    =
    \nu
    (1+b)^3(a+b)
    \left(
        \frac{1+b}{a+b}
    \right)^{-\sigma q} \;,
    \qquad
    q\equiv\frac{\lambda}{2\alpha} \;,
    \label{eq:exp-runaway-eta}
\end{equation}
where
\begin{equation}
    \nu
    \equiv
    \frac{2\alpha\lambda r_+^2 V_\infty}
         {M_{\rm Pl}^2(1-a)} \;.
    \label{eq:exp-runaway-nu-def}
\end{equation}
The parameter \(\nu\) is the dimensionless normalization of the exponential
force relative to the GHS gauge-source scale.  The remaining \(b\)-dependence
in Eq.~\eqref{eq:exp-runaway-eta} determines whether this force is enhanced
or suppressed along the throat.

In the deep throat, \(a\lesssim b\ll1\), Eq.~\eqref{eq:exp-runaway-eta}
reduces to
\begin{align}
    \sigma=+1 \,:
    \qquad
    \eta_{\rm src}^{(+)}(b)
    &\sim
    \nu\,(a+b)^{1+q} \;,
    \label{eq:exp-runaway-eta-plus-deep}
    \\
    \sigma=-1 \,:
    \qquad
    \eta_{\rm src}^{(-)}(b)
    &\sim
    \nu\,(a+b)^{1-q} \;.
    \label{eq:exp-runaway-eta-minus-deep}
\end{align}
Thus, the favourable orientation never produces a parametrically enhanced
local bottom source as \(a\to0\).  The dangerous orientation does so only
when
\begin{equation}
    q>1
    \qquad\Longleftrightarrow\qquad
    \lambda>2\alpha \;.
    \label{eq:exp-runaway-pointwise-threshold}
\end{equation}
For \(0<q<1\), the potential grows along the GHS flow, but not steeply
enough to generate a parametrically large pointwise source at the regulated
bottom of a long throat.

The local crossover is defined by
\begin{equation}
    \eta_{\rm src}^{\rm exp}(b_{\rm pt})\sim1 \;.
    \label{eq:exp-runaway-bpt-def}
\end{equation}
For the dangerous orientation with \(q>1\), the deep-throat estimate gives
\begin{equation}
    a+b_{\rm pt}
    \sim
    \nu^{1/(q-1)} \;,
    \qquad
    \sigma=-1,\quad q>1 \;,
    \label{eq:exp-runaway-bpt-dangerous}
\end{equation}
provided the solution lies in the regime \(a\lesssim b_{\rm pt}\ll1\).  The
region below \(b_{\rm pt}\) is locally potential-dominated, while the region
above \(b_{\rm pt}\) remains locally GHS-like.

We use the benchmark
\begin{equation}
    2\alpha^2=1 \;
    \qquad
    a=10^{-4} \;,
    \qquad
    b_{\rm min}=a \;,
    \qquad
    b_{\rm max}=10^3 \;.
    \label{eq:exp-runaway-uvmatched-benchmark}
\end{equation}
The main observable is
\begin{equation}
    \Delta_g(b)
    =
    \frac{g_{\rm exp}^2(b)}
         {g_{\rm GHS}^2(b)}
    -1
    =
    \exp\!\left[
        \frac{2\alpha}{M_{\rm Pl}}
        \left(
            \varphi_{\rm exp}(b)-\varphi_{\rm GHS}(b)
        \right)
    \right]
    -1 \;.
    \label{eq:exp-runaway-Deltag-fixed-throat}
\end{equation}
This directly measures the physical change in the local gauge coupling
relative to the undeformed GHS profile.

The results are summarized in Table~\ref{tab:exp-runaway-uvmatched}.  We
write
\[
    b_{\rm IR}\equiv b_{\rm min}
\]
for the regulated inner endpoint.  For \(q=3\), the nonlinear
boundary-value problem can have more than one UV-matched solution for the
same value of \(\nu\).  We distinguish the weak branch, continuously
connected to the GHS solution as \(V_\infty\to0\), from strongly deformed
branches with a much larger IR displacement.

\begin{table}[t]
\centering
\begin{tabular}{c c c c c c c}
\toprule
\(\sigma\) & \(q\) & \(\nu\) & Branch &
\(\Delta\chi_{\rm IR}^{\rm num}\) &
\(\Delta_g(b_{\rm IR})\) &
\(\Delta_g(1)\)
\\
\midrule
\(+1\) & \(1\) & \(10^{-4}\) & weak
& \(-4.75\times10^{-2}\)
& \(6.94\times10^{-2}\)
& \(6.94\times10^{-2}\)
\\
\(-1\) & \(1\) & \(10^{-4}\) & weak
& \(-5.84\times10^{-2}\)
& \(8.62\times10^{-2}\)
& \(8.07\times10^{-2}\)
\\
\(-1\) & \(2\) & \(10^{-4}\) & no weak branch found
& -- & -- & --
\\
\(-1\) & \(2\) & \(10^{-6}\) & weak
& \(-6.72\times10^{-2}\)
& \(9.97\times10^{-2}\)
& \(1.07\times10^{-2}\)
\\
\(-1\) & \(3\) & \(10^{-6}\) & no branch found
& -- & -- & --
\\
\(-1\) & \(3\) & \(10^{-10}\) & weak
& \(-1.78\times10^{-2}\)
& \(2.54\times10^{-2}\)
& \(2.05\times10^{-3}\)
\\
\(-1\) & \(3\) & \(4\times10^{-10}\) & weak
& \(-9.92\times10^{-2}\)
& \(1.51\times10^{-1}\)
& \(1.22\times10^{-2}\)
\\
\(-1\) & \(3\) & \(4\times10^{-10}\) & strong
& \(-4.86\times10^{-1}\)
& \(9.90\times10^{-1}\)
& \(5.81\times10^{-2}\)
\\
\(-1\) & \(3\) & \(5\times10^{-10}\) & weak
& \(-1.58\times10^{-1}\)
& \(2.50\times10^{-1}\)
& \(1.97\times10^{-2}\)
\\
\(-1\) & \(3\) & \(6\times10^{-10}\) & no weak branch found
& -- & -- & --
\\
\(-1\) & \(3\) & \(10^{-9}\) & no branch found in scan
& -- & -- & --
\\
\bottomrule
\end{tabular}
\caption{UV-matched fixed-throat exponential deformations in the canonical
normalization \(2\alpha^2=1\), for \(a=10^{-4}\), \(b_{\rm min}=a\), and
\(b_{\rm max}=10^3\).  The auxiliary numerical field is
\(\chi_{\rm num}\equiv\Phi_\infty-\phi=-\varphi\).  The quantity
\(\Delta\chi_{\rm IR}^{\rm num}\) denotes its shift relative to the GHS
value at the IR endpoint.  Negative values correspond to a larger physical
displacement \(\varphi\), and hence to a larger local gauge coupling than in
the undeformed GHS profile.  The weak branch is the branch continuously
connected to the GHS solution as \(V_\infty\to0\).}
\label{tab:exp-runaway-uvmatched}
\end{table}

Three cases illustrate the hierarchy.  For \(q=1\), both orientations remain
controlled at \(\nu=10^{-4}\).  The favourable orientation gives
\[
    \Delta_g(b_{\rm IR})\simeq 0.069 \;,
\]
while the dangerous orientation gives
\[
    \Delta_g(b_{\rm IR})\simeq 0.086 \;.
\]
The dangerous wall is already more deforming, but only mildly so for
\(q=1\).

For \(q=2\), the dangerous orientation is substantially more restrictive.
No weak UV-matched branch is found at \(\nu=10^{-4}\) within the shooting
window, while a regular weak branch is found at \(\nu=10^{-6}\).  In the
latter case,
\[
    \Delta_g(b_{\rm IR})\simeq0.100 \;,
    \qquad
    \Delta_g(1)\simeq1.07\times10^{-2} \;.
\]
The deformation is therefore concentrated near the bottom of the throat.

For \(q=3\), the UV-matched problem becomes highly restrictive and
nonlinear.  A weak branch is found at \(\nu=10^{-10}\), with only a
few-percent IR gauge-coupling deformation.  By
\(\nu=4\times10^{-10}\), the weak branch has grown to
\[
    \Delta_g(b_{\rm IR})\simeq0.151 \;.
\]
At the same value of \(\nu\), a second, strongly deformed branch is also
found, with
\[
    \Delta_g(b_{\rm IR})\simeq0.99 \;.
\]
The weak branch persists at \(\nu=5\times10^{-10}\), where
\[
    \Delta_g(b_{\rm IR})\simeq0.25 \;,
\]
but no weak branch is found at \(6\times10^{-10}\) in the same shooting
window.  A broader scan at \(\nu=10^{-9}\) does not reveal a branch in the
interval
\[
    \Delta\chi_{\rm IR}^{\rm num}\in[-5,0] \;.
\]
Thus, for this setup, the weakly deformed branch connected to GHS appears
to terminate, or at least to leave the shooting window, around
\begin{equation}
    \nu_{\rm end}^{\rm weak}(q=3)
    \sim
    (5\text{--}6)\times10^{-10} \;.
    \label{eq:exp-runaway-q3-weak-endpoint}
\end{equation}

The appearance of multiple branches is a useful warning.  The fixed-throat
boundary-value problem is nonlinear: for the same UV modulus and the same
potential parameters, there can be more than one scalar profile satisfying
the boundary conditions.  The branch relevant for perturbing the GHS throat
is the weak branch, because it is continuously connected to the
\(V_\infty=0\) GHS solution.  Strong branches should instead be interpreted
as large deformations of the fixed-throat scalar profile, and not as small
corrections to the GHS throat.

Most importantly, the disappearance or strong deformation of the weak branch
is not necessarily triggered by a pointwise source ratio of order one at
the bottom of the throat.  For example, in the \(q=3\),
\(\nu=4\times10^{-10}\) case,
\[
    \eta_{\rm src}(b_{\rm IR})
    \simeq
    1.0\times10^{-2} \;,
\]
while the weak branch already gives a fifteen-percent gauge-coupling
deviation at the bottom, and a strongly deformed branch also exists.  The
limitation is therefore global: the exponential force must be accommodated
while preserving the same finite-radius UV modulus.  A small local source is
necessary for local consistency, but it is not sufficient to guarantee the
existence of a weakly deformed UV-matched fixed-throat branch.

The UV-matched scan nevertheless agrees with the analytic hierarchy:
\[
    q=1 \quad \text{mild} \;,
    \qquad
    q=2 \quad \text{more restrictive} \;,
    \qquad
    q=3 \quad \text{highly restrictive} \;.
\]
It also shows that the fixed-throat problem can fail globally before the
local pointwise diagnostic itself becomes order one.

A full UV-matched back-reacted boundary-value problem is beyond the scope of
the present analysis, but it is useful to state what such a problem would
involve.  Since a pure exponential potential has no static minimum, a
strictly static, asymptotically flat solution with
\(\varphi\to0\), or equivalently \(\phi\to\Phi_\infty\) at infinity, should
not be expected unless the exponential is only a local approximation to a
larger potential, or unless \(V_\infty\) is negligible over the region of
interest.  The appropriate local problem is instead a finite-radius matching
problem,
\begin{equation}
    r_+
    \ll
    r_{\rm UV}
    \ll
    H_\infty^{-1} \;,
    \qquad
    H_\infty^2
    \simeq
    \frac{V_\infty}{3M_{\rm Pl}^2} \;,
    \qquad
    \varphi(r_{\rm UV})=0 \;.
    \label{eq:exp-runaway-uv-matching}
\end{equation}
Here, \(H_\infty^{-1}\) denotes the curvature scale associated with the local
potential energy at the matching point.  The condition
\(r_{\rm UV}\ll H_\infty^{-1}\) ensures that the geometry over the matching
region can be treated as approximately static and asymptotically flat for
the purposes of the local black hole calculation.

The corresponding back-reacted theory is
\begin{equation}
    S
    =
    \int d^4x\sqrt{-g}
    \left[
        \frac{M_{\rm Pl}^2}{2}R
        -\frac12(\partial\varphi)^2
        -V_\infty e^{-\sigma\lambda\varphi/M_{\rm Pl}}
        -\frac14
        B_\infty e^{-2\alpha\varphi/M_{\rm Pl}}
        F_{\mu\nu}F^{\mu\nu}
    \right] \;.
    \label{eq:exp-runaway-backreacted-action}
\end{equation}
For a static, spherically symmetric magnetic solution one may use the same
radial ansatz as in Section~\ref{sec:static-emd-system},
\begin{equation}
    ds^2
    =
    -e^{-2\delta(r)}N(r)dt^2
    +
    \frac{dr^2}{N(r)}
    +
    R(r)^2d\Omega_2^2 \;,
    \qquad
    F_{\theta\phi}=Q\sin\theta \;.
    \label{eq:exp-runaway-backreacted-ansatz}
\end{equation}
Equivalently, one may work in the areal gauge \(R=r\).  The scalar equation
has the schematic form
\begin{equation}
    \nabla^2\varphi
    =
    V_{,\varphi}
    +
    \frac14 B_{,\varphi}F^2 \;,
    \label{eq:exp-runaway-backreacted-scalar-schematic}
\end{equation}
with
\begin{equation}
    V_{,\varphi}
    =
    -\sigma\frac{\lambda}{M_{\rm Pl}}V \;,
    \qquad
    B_{,\varphi}
    =
    -2\frac{\alpha}{M_{\rm Pl}}B \;.
    \label{eq:exp-runaway-backreacted-derivatives}
\end{equation}
The metric equations are sourced by the scalar gradient, the potential
energy, and the magnetic energy density \(B(\varphi)Q^2/R^4\).  At a regular
non-extremal horizon one fixes
\[
    N(r_+)=0 \;,
    \qquad
    \varphi(r_+)=\varphi_+ \;,
\]
with regularity determining \(\varphi'(r_+)\) in terms of
\(\varphi_+\), \(Q\), and \(V_\infty\).  At the outer matching radius, one
fixes
\begin{equation}
    \varphi(r_{\rm UV})=0 \;,
    \qquad
    \delta(r_{\rm UV})=0 \;,
    \label{eq:exp-runaway-backreacted-uv-bc}
\end{equation}
where the second condition sets the normalization of the time coordinate.
The natural observable is again the physical gauge-coupling deviation,
\begin{equation}
    \Delta_g(r)
    =
    \frac{
        g_{\rm br}^2(r)
    }{
        g_{\rm GHS}^2(r)
    }
    -1 \;,
    \label{eq:exp-runaway-backreacted-Deltag}
\end{equation}
or the same quantity expressed in the throat coordinate \(b\).  Here,
\(g_{\rm GHS}^2\) denotes the undeformed GHS gauge-coupling profile with the
same reference asymptotic value and near-horizon parameters.

Here, we do not attempt this full finite-radius boundary-value problem.
Instead, as a first back-reacted check, we perform the same local initial
value test used for the quadratic and shifted-exponential potentials.  The
initial data are taken to be the near-extremal GHS data at a regulated
cutoff,
\[
    b_{\rm cut}=10a \;,
\]
and the coupled Einstein--Maxwell--scalar equations are integrated outward
to a fixed radius
\[
    x_{\rm UV}=2 \;,
    \qquad
    x\equiv\frac{r}{r_+}=1+b \;.
\]
This is not a full UV-matched boundary-value problem.  Rather, it measures
how quickly a near-GHS throat begins to depart from the massless solution
once the exponential potential is included in both the scalar and metric
equations.  We use the benchmark
\begin{equation}
2\alpha^2=1 \;, \qquad a=10^{-4}\;, \qquad b_{\rm cut}=10^{-3} \; \qquad x_{\rm UV}=2\;, \qquad \sigma=-1 \;. 
\end{equation}
The diagnostic is the physical gauge-coupling deviation,
\begin{equation}
    \Delta_g^{\rm max}
    =
    \max_{x_0\le x\le x_{\rm UV}}
    \left|
        \frac{g_{\rm br}^2(x)}{g_{V=0}^2(x)}-1
    \right| \;
    \qquad
    x_0=1+b_{\rm cut} \;.
    \label{eq:exp-runaway-backreacted-ivp-deltag}
\end{equation}

For \(q=1\), the scan gives Table~\ref{tab:exp-runaway-backreacted-q1}. For \(q=2\), the scan gives Table~\ref{tab:exp-runaway-backreacted-q2}.
\begin{table}[t]
\centering
\begin{tabular}{cc}
\toprule
\(\nu\) & \(\Delta_g^{\rm max}\) \\
\midrule
\(10^{-6}\) & \(8.29\times10^{-5}\) \\
\(3\times10^{-6}\) & \(2.49\times10^{-4}\) \\
\(10^{-5}\) & \(8.29\times10^{-4}\) \\
\(3\times10^{-5}\) & \(2.49\times10^{-3}\) \\
\(10^{-4}\) & \(8.30\times10^{-3}\) \\
\(3\times10^{-4}\) & \(2.49\times10^{-2}\) \\
\(10^{-3}\) & \(8.36\times10^{-2}\) \\
\bottomrule
\end{tabular}
\caption{Back-reacted local initial-value scan for the dangerous exponential with
\(\sigma=-1\), \(q=1\), \(a=10^{-4}\), \(b_{\rm cut}=10^{-3}\),
\(x_{\rm UV}=2\), and \(2\alpha^2=1\).}
\label{tab:exp-runaway-backreacted-q1}
\end{table}

\begin{table}[t]
\centering
\begin{tabular}{cc}
\toprule
\(\nu\) & \(\Delta_g^{\rm max}\) \\
\midrule
\(10^{-8}\) & \(4.62\times10^{-4}\) \\
\(3\times10^{-8}\) & \(1.39\times10^{-3}\) \\
\(10^{-7}\) & \(4.62\times10^{-3}\) \\
\(3\times10^{-7}\) & \(1.39\times10^{-2}\) \\
\(10^{-6}\) & \(4.62\times10^{-2}\) \\
\(3\times10^{-6}\) & \(1.39\times10^{-1}\) \\
\(10^{-5}\) & \(4.65\times10^{-1}\) \\
\bottomrule
\end{tabular}
\caption{Back-reacted local initial-value scan for the dangerous exponential with
\(\sigma=-1\), \(q=2\), \(a=10^{-4}\), \(b_{\rm cut}=10^{-3}\),
\(x_{\rm UV}=2\), and \(2\alpha^2=1\).}
\label{tab:exp-runaway-backreacted-q2}
\end{table}

Interpolating to \(\Delta_g^{\rm max}=0.1\), we find
\begin{equation}
    \nu_{\rm crit}^{(q=1)}
    \simeq
    1.19\times10^{-3} \;,
    \qquad
    \nu_{\rm crit}^{(q=2)}
    \simeq
    2.16\times10^{-6} \;.
    \label{eq:exp-runaway-backreacted-nucrit}
\end{equation}
Thus, the \(q=2\) dangerous exponential is more than two orders of magnitude
more restrictive than the \(q=1\) case in this back-reacted initial-value problem test:
\[
    \frac{\nu_{\rm crit}^{(q=1)}}{\nu_{\rm crit}^{(q=2)}}
    \simeq
    5.5\times10^2 \;.
\]

The initial-value problem and UV-matched problems are not identical.  In the fixed-throat
UV-matched scan, the UV modulus is held fixed and the IR scalar value is
tuned, so the deformation is constrained globally by the outer matching
condition.  In the initial-value problem test, by contrast, the near-GHS data are held fixed
at the cutoff and the solution is allowed to drift outward.  Correspondingly,
the maximum of \(\Delta_g\) occurs at the outer endpoint \(x_{\rm UV}=2\).
Nevertheless, both tests support the same hierarchy: \(q=1\) is mild, while
\(q=2\) is already much more restrictive.



\subsection{Inverse-power runaway potentials}
\label{subsec:inverse-power-runaway}

We next consider inverse-power runaway potentials.  Since fractional powers
require a positive argument, it is useful to separate the canonically
normalized GHS displacement from the positive dimensionless variable entering
the potential.  We define
\begin{equation}
    u(b)
    \equiv
    u_\infty+\frac{\varphi(b)}{M_{\rm Pl}} \;,
    \qquad
    u_\infty>0 \;,
    \label{eq:inverse-power-u-def}
\end{equation}
where, as before,
\[
    \varphi(b) \equiv \phi(b)-\Phi_\infty \;.
\]
We consider
\begin{equation}
    V(u)
    =
    \Lambda M_{\rm Pl}^4 u^{-q} \;,
    \qquad
    q>0 \; \;.
    \label{eq:inverse-power-potential}
\end{equation}
Here, \(\Lambda\) is dimensionless\footnote{The exponent \(q\) used in this inverse-power potential is an
independent parameter.  It should not be confused with the dimensionless
exponential slope \(q=\lambda/(2\alpha)\) in
Subsection~\ref{subsec:exponential-runaway}, nor with the racetrack
exponents used in Subsection~\ref{subsec:racetrack-stabilization}.}.  Potentials of this type are standard
examples of runaway scalar potentials in quintessence and tracker models;
we do not assume any particular microscopic origin, and use them only as
simple local models for a shallow modulus potential near a charged
dilatonic black hole.

The magnetic GHS throat drives the scalar toward larger \(\varphi\), and
hence toward larger \(u\).  Along this direction, the inverse-power force
decreases.  Since \(du/d\varphi=1/M_{\rm Pl}\), 
\begin{equation}
    V_{,\varphi}
    =
    -\,q\Lambda M_{\rm Pl}^3 u^{-q-1} \;,
    \qquad
    \left|V_{,\varphi}\right|
    =
    q\Lambda M_{\rm Pl}^3 u^{-q-1} \;.
    \label{eq:inverse-power-force}
\end{equation}
This is the main qualitative difference from the dangerous exponential wall
of the previous subsection: inverse powers become weaker down the throat
rather than stronger.  As such, they  are a useful test of whether shallow runaway
potentials can coexist with a GHS-like gauge-dominated throat at all.

On the GHS background,
\begin{equation}
    u_{\rm GHS}(b)
    =
    u_\infty
    +
    \frac{1}{2\alpha}
    \log\!\left(
        \frac{1+b}{a+b}
    \right) \;,
    \label{eq:inverse-power-u-ghs}
\end{equation}
where the canonical displacement has been divided by \(M_{\rm Pl}\).  The
numerical examples below use the canonical GHS normalization,
\(2\alpha^2=1\).

The fixed-throat scalar equation has the same weighted form as in the
previous examples,
\begin{equation}
    \frac{d}{db}
    \left[
        p(b)\frac{d\varphi}{db}
    \right]
    =
    r_+^2 w(b)
    \left[
        V_{,\varphi}(\varphi)
        +
        \frac14 B_{,\varphi}F^2
    \right] \;.
    \label{eq:inverse-power-fixed-throat-eq}
\end{equation}
Using the positive GHS gauge-source magnitude
\(\mathcal S_{\rm gauge}(b)\) of Eq.~\eqref{eq:gauge-source-magnitude}, the
pointwise obstruction parameter is
\begin{equation}
    \eta_{\rm src}^{\rm inv}(b)
    =
    \frac{
        r_+^2
        \left|V_{,\varphi}(u_{\rm GHS}(b)) \right|
    }{
        \mathcal S_{\rm gauge}(b)
    }
    =
    \nu_q
    (1+b)^3(a+b)
    u_{\rm GHS}(b)^{-q-1} \;,
    \label{eq:inverse-power-eta}
\end{equation}
where
\begin{equation}
    \nu_q
    \equiv
    \frac{
        2\alpha q \Lambda M_{\rm Pl}^2 r_+^2
    }{
        1-a
    } \;.
    \label{eq:inverse-power-nuq-def}
\end{equation}
This is the inverse-power analogue of the dimensionless force amplitudes
used in the previous subsections.

For \(a\ll b\leq1\), Eq.~\eqref{eq:inverse-power-eta} simplifies to
\begin{equation}
    \eta_{\rm src}^{\rm inv}(b)
    \simeq
    \nu_q
    (1+b)^3 b
    \left[
        u_\infty
        +
        \frac{1}{2\alpha}
        \log\!\left(
            \frac{1+b}{b}
        \right)
    \right]^{-q-1},
    \label{eq:inverse-power-eta-interior}
\end{equation}
which is independent of \(a\).  Here, \((1+b)^3b\) increases with \(b\), while
the bracketed logarithmic factor decreases, so its inverse power also
increases with \(b\).  The local obstruction is therefore largest at the
outer edge of any finite throat interval, not at the bottom.

On the full interval \(a\leq b\leq1\), the local crossover is therefore set
at \(b=1\).  Solving \(\eta_{\rm src}^{\rm inv}(1)=1\) for the amplitude
gives
\begin{equation}
    \Lambda_{\rm cross}(a,q)
    =
    \frac{1-a}
         {16\alpha q(1+a)M_{\rm Pl}^2 r_+^2}
    \left[
        u_\infty
        +
        \frac{1}{2\alpha}
        \log\!\left(
            \frac{2}{1+a}
        \right)
    \right]^{q+1} \;,
\end{equation}
and hence
\begin{equation}
    \Lambda_{\rm cross}^{(0)}(q)
    =
    \frac{1}
         {16\alpha q M_{\rm Pl}^2 r_+^2}
    \left(
        u_\infty+\frac{\log2}{2\alpha}
    \right)^{q+1}
    \qquad
    (a\to0) \;.
    \label{eq:inverse-power-Lambda-cross-exact}
\end{equation}
This is an \(O(1)\) constant independent of the throat length, in sharp
contrast with the dangerous exponential wall, where the obstruction can
become power-law enhanced at the bottom of a long throat.  For
\(u_\infty=2\), \(2\alpha^2=1\), and \(M_{\rm Pl}r_+=1\),
Table~\ref{tab:inverse-power-cross-full} gives the limiting values.

\begin{table}[t]
\centering
\begin{tabular}{ccc}
\toprule
\(q\) &
\(\Lambda_{\rm cross}(a=10^{-5})\) &
\(\Lambda_{\rm cross}^{(0)}\)
\\
\midrule
\(1/8\) & \(1.97344\) & \(1.97349\) \\
\(1/4\) & \(1.10592\) & \(1.10594\) \\
\(1/3\) & \(0.894956\) & \(0.894977\) \\
\(1/2\) & \(0.694620\) & \(0.694637\) \\
\(2/3\) & \(0.606520\) & \(0.606535\) \\
\(1\)   & \(0.548059\) & \(0.548073\) \\
\(2\)   & \(0.682367\) & \(0.682387\) \\
\bottomrule
\end{tabular}
\caption{Local crossing amplitudes for inverse-power potentials, defined by
\(\eta_{\rm src}^{\rm inv}(1)=1\), in the canonical normalization
\(2\alpha^2=1\), with \(u_\infty=2\) and \(M_{\rm Pl}r_+=1\).}
\label{tab:inverse-power-cross-full}
\end{table}

The same reasoning extends to smaller subregions
\(a\leq b\leq b_{\rm cut}\).  For \(a\ll b_{\rm cut}\), the maximum of
\(\eta_{\rm src}^{\rm inv}\) on this restricted interval occurs at
\(b=b_{\rm cut}\), and the local crossing amplitude is
\begin{equation}
    \Lambda_{\rm cross}(q;b_{\rm cut})
    \simeq
    \frac{
        1
    }{
        2\alpha q\,M_{\rm Pl}^2 r_+^2
        (1+b_{\rm cut})^3 b_{\rm cut}
    }
    \left[
        u_\infty
        +
        \frac{1}{2\alpha}
        \log\!\left(
            \frac{1+b_{\rm cut}}{b_{\rm cut}}
        \right)
    \right]^{q+1} \;.
    \label{eq:inverse-power-Lambda-cross-bcut}
\end{equation}
In the limit \(b_{\rm cut}\to0\), this scales as
\begin{equation}
    \Lambda_{\rm cross}(q;b_{\rm cut})
    \propto
    b_{\rm cut}^{-1}
    \left(
        \log\frac1{b_{\rm cut}}
    \right)^{q+1}.
    \label{eq:inverse-power-Lambda-cross-bcut-asymptotic}
\end{equation}
Thus, pushing the cutoff inward requires a rapidly growing amplitude to
compete locally with the gauge source.  The deep throat is therefore
parametrically safer than the outer matching region.  This behaviour is
shown explicitly in Table~\ref{tab:inverse-power-cross-bcut} for \(q=1/3\).

\begin{table}[t]
\centering
\begin{tabular}{ccc}
\toprule
\(b_{\rm cut}\) &
\(\Lambda_{\rm cross}(q=1/3;b_{\rm cut})\) &
\(\Lambda_{\rm cross}(b_{\rm cut})/\Lambda_{\rm cross}(1)\)
\\
\midrule
\(1\)    & \(0.894977\) & \(1\) \\
\(0.3\)  & \(14.1543\) & \(15.8152\) \\
\(0.1\)  & \(91.0618\) & \(101.748\) \\
\(0.03\) & \(480.813\) & \(537.235\) \\
\(0.01\) & \(1885.08\) & \(2106.28\) \\
\bottomrule
\end{tabular}
\caption{Local crossing amplitude on subregions
\(a\le b\le b_{\rm cut}\) for \(q=1/3\), in the canonical normalization
\(2\alpha^2=1\), with \(u_\infty=2\) and \(M_{\rm Pl}r_+=1\).  The rapid
growth toward small \(b_{\rm cut}\) shows that the deep throat is
increasingly gauge dominated.}
\label{tab:inverse-power-cross-bcut}
\end{table}

This local diagnostic only compares the potential force to the GHS gauge
source at each radius; it does not by itself determine the size of the
resulting profile deformation.  To check this, we solve the fixed-throat
scalar equation directly.  It is useful to return to Eq.~\eqref{eq:inverse-power-fixed-throat-eq} and write the equation explicitly in terms of
the dimensionless variable \(u\) using the definition provided in Eq.~\eqref{eq:inverse-power-u-def}, \(u=u_\infty+\varphi/M_{\rm Pl}\). The scalar equation divided by \(M_{\rm Pl}\) becomes
\begin{equation}
    {\cal L}_a[u]
    =
    {\cal L}_a[u_{\rm GHS}]
    -
    q\Lambda M_{\rm Pl}^2 r_+^2\,u^{-q-1},
    \qquad
    {\cal L}_a[u]
    \equiv
    \frac1{w(b)}
    \frac{d}{db}
    \left[
        p(b)\frac{du}{db}
    \right].
    \label{eq:inverse-power-deformed-equation}
\end{equation}
The minus sign follows from
\(V_{,\varphi}=-q\Lambda M_{\rm Pl}^3u^{-q-1}\).  Thus, for
\(\Lambda>0\), the inverse-power force points in the same direction as the
magnetic GHS gauge source: it further drives the scalar toward larger
\(u\) as one moves inward, but with a force that weakens down the throat.

In the numerical examples below, we set \(M_{\rm Pl}r_+=1\), so the last term
in Eq.~\eqref{eq:inverse-power-deformed-equation} reduces to
\(-q\Lambda u^{-q-1}\).  We integrate outward from the regulated inner edge
using the GHS initial data,
\begin{equation}
    u(b_{\rm min})=u_{\rm GHS}(b_{\rm min}) \;,
    \qquad
    u'(b_{\rm min})=u'_{\rm GHS}(b_{\rm min}) \;,
    \qquad
    b_{\rm min}=a \;,
    \label{eq:inverse-power-deformed-ic}
\end{equation}
with prime denoting \(d/db\). At \(\Lambda=\Lambda_{\rm cross}\), where the pointwise source ratio reaches
one at the throat exit \(b=1\), we monitor the accumulated deformation,
\begin{equation}
    \Delta u_{\rm exit}
    \equiv
    u_\Lambda(1)-u_{\rm GHS}(1) \;.
    \label{eq:inverse-power-delta-u-exit-def}
\end{equation}
Table~\ref{tab:inverse-power-delta-u-exit} gives this quantity for the benchmark
\begin{equation}
2\alpha^2=1\;, \qquad M_{\rm Pl}r_+=1 \qquad a=10^{-5}\;, \qquad u_\infty=2 \;. 
\end{equation}
Even at the crossing amplitude, where the local obstruction reaches unity
at the outer edge, the integrated deformation remains only at the few-percent
level.  It is not parametrically suppressed by \(a\), but it is also far
from an order-one distortion of the profile.

\begin{table}[t]
\centering
\begin{tabular}{ccc}
\toprule
\(q\) && \(\Delta u_{\rm exit}\) at
\(\Lambda=\Lambda_{\rm cross}\)
\\
\midrule
\(1/8\) && \(4.65\times10^{-2}\) \\
\(1/4\) && \(4.54\times10^{-2}\) \\
\(1/3\) && \(4.47\times10^{-2}\) \\
\(1/2\) && \(4.33\times10^{-2}\) \\
\(2/3\) && \(4.20\times10^{-2}\) \\
\(1\)   && \(3.96\times10^{-2}\) \\
\(2\)   && \(3.36\times10^{-2}\) \\
\bottomrule
\end{tabular}
\caption{Fixed-throat deformation at the throat exit for inverse-power
potentials in the canonical normalization \(2\alpha^2=1\), with
\(u_\infty=2\), \(M_{\rm Pl}r_+=1\), and \(a=10^{-5}\).  The deformation is
\(\Delta u_{\rm exit}=u_\Lambda(1)-u_{\rm GHS}(1)\).  The amplitude is set
to the local crossing value \(\Lambda=\Lambda_{\rm cross}(a,q)\).  The
numerical checks \(\eta_{\rm src}(b=1)=1\) and
\(\max_{a\le b\le1}\eta_{\rm src}=1\) are satisfied in every row.}
\label{tab:inverse-power-delta-u-exit}
\end{table}

This confirms the local diagnostic with an explicit fixed-throat solution:
inverse-power potentials do not generate an IR-enhanced deformation at the
bottom of the throat.  

We now check whether the same qualitative behaviour
appears in a local back-reacted evolution.  As for the previous potentials,
this is not a full UV-matched boundary-value problem.  We fix the initial
data to their GHS values at
\begin{equation}
    a=10^{-4} \;,
    \qquad
    b_{\rm cut}=10^{-3} \;,
    \qquad
    2\alpha^2=1 \;,
\end{equation}
and integrate the coupled Einstein--Maxwell--scalar equations outward to
\begin{equation}
    x_{\rm UV}=2 \;,
    \qquad
    x\equiv \frac{r}{r_+}=1+b \;, 
    \qquad
    x_0=1+b_{\rm cut} \;.
\end{equation}

The diagnostic is the
maximum fractional change in the local gauge coupling relative to the
corresponding \(V=0\) GHS evolution,
\begin{equation}
    \Delta_g^{\rm max}
    =
    \max_{x_0\le x\le x_{\rm UV}}
    \left|
        \frac{g_{\rm br}^2(x)}{g_{V=0}^2(x)}-1
    \right| \;,
    \label{eq:inverse-power-backreacted-Deltagmax}
\end{equation}
where \(g_{\rm br}\) denotes the local gauge coupling on the back-reacted solution with the non-zero potential. For each value of \(q\), the potential amplitude is normalized to the
fixed-throat local crossing value \(\Lambda_{\rm cross}(a,q)\), defined by
\(\eta_{\rm src}^{\rm inv}(b=1)=1\).  Table~\ref{tab:inverse-power-backreacted-ivp}
shows the scan.  In every case the maximum of \(\Delta_g\) occurs at the
outer endpoint \(x_{\rm UV}=2\), not at the bottom of the throat.
This is the back-reacted analogue of the fixed-throat result demonstrating that the
inverse-power force weakens as the GHS trajectory moves inward.

We also monitor the metric deformation through
\begin{equation}
    \Delta_R^{\rm max}
    =
    \max_{x_0\le x\le x_{\rm UV}}
    \left|
        \frac{R_{\rm br}(x)}{R_{V=0}(x)}-1
    \right| \;.
    \label{eq:inverse-power-backreacted-DeltaRmax}
\end{equation}

\begin{table}[t]
\centering
\begin{tabular}{cccc}
\toprule
\(q\) &
\(\Lambda/\Lambda_{\rm cross}\) &
\(\Delta_g^{\rm max}\) &
\(\Delta_R^{\rm max}\)
\\
\midrule
\(1/3\) & \(0.05\) & \(2.21\times10^{-2}\) & \(7.47\times10^{-3}\) \\
\(1/3\) & \(0.10\) & \(4.41\times10^{-2}\) & \(1.50\times10^{-2}\) \\
\(1/3\) & \(0.20\) & \(8.78\times10^{-2}\) & \(3.04\times10^{-2}\) \\
\(1/3\) & \(0.30\) & \(1.31\times10^{-1}\) & \(4.62\times10^{-2}\) \\
\(1/3\) & \(0.50\) & \(2.16\times10^{-1}\) & \(7.92\times10^{-2}\) \\
\(1/3\) & \(1.00\) & \(4.18\times10^{-1}\) & \(1.71\times10^{-1}\) \\
\hline
\(1\) & \(0.05\) & \(8.08\times10^{-3}\) & \(2.30\times10^{-3}\) \\
\(1\) & \(0.10\) & \(1.61\times10^{-2}\) & \(4.61\times10^{-3}\) \\
\(1\) & \(0.20\) & \(3.22\times10^{-2}\) & \(9.25\times10^{-3}\) \\
\(1\) & \(0.30\) & \(4.81\times10^{-2}\) & \(1.39\times10^{-2}\) \\
\(1\) & \(0.50\) & \(7.98\times10^{-2}\) & \(2.35\times10^{-2}\) \\
\(1\) & \(1.00\) & \(1.57\times10^{-1}\) & \(4.80\times10^{-2}\) \\
\hline
\(2\) & \(0.05\) & \(4.39\times10^{-3}\) & \(1.01\times10^{-3}\) \\
\(2\) & \(0.10\) & \(8.78\times10^{-3}\) & \(2.01\times10^{-3}\) \\
\(2\) & \(0.20\) & \(1.75\times10^{-2}\) & \(4.04\times10^{-3}\) \\
\(2\) & \(0.30\) & \(2.63\times10^{-2}\) & \(6.08\times10^{-3}\) \\
\(2\) & \(0.50\) & \(4.37\times10^{-2}\) & \(1.02\times10^{-2}\) \\
\(2\) & \(1.00\) & \(8.67\times10^{-2}\) & \(2.07\times10^{-2}\) \\
\bottomrule
\end{tabular}
\caption{Back-reacted local initial-value scan for inverse-power potentials,
with \(a=10^{-4}\), \(b_{\rm cut}=10^{-3}\), \(x_{\rm UV}=2\),
\(2\alpha^2=1\), \(u_\infty=2\), and \(M_{\rm Pl}r_+=1\).  The amplitude is
measured in units of the fixed-throat local crossing value
\(\Lambda_{\rm cross}(a,q)\), defined by
\(\eta_{\rm src}^{\rm inv}(b=1)=1\).  The quantity
\(\Delta_g^{\rm max}\) is the maximum fractional deviation of the local gauge
coupling from the corresponding \(V=0\) GHS evolution, while
\(\Delta_R^{\rm max}\) denotes the maximum fractional deviation of the
areal-radius function from the same reference solution.  In all rows the
maximum gauge-coupling deviation occurs at the outer endpoint of the
integration interval, not in the deep throat.}
\label{tab:inverse-power-backreacted-ivp}
\end{table}

Even when \(\Lambda\) is raised to the fixed-throat local crossing value,
the back-reacted deformation remains modest, especially for larger \(q\).
Moreover, the maximum of \(\Delta_g\) is always attained at the outer end of
the interval.  This confirms the qualitative fixed-throat picture: the
inverse-power force does not create an IR obstruction to the GHS throat;
its largest effect comes from the outer matching region, where the scalar
has not yet run far enough for the inverse-power slope to be strongly
suppressed. Interpolating to \(\Delta_g^{\rm max}=0.1\) gives the critical amplitudes in
Table~\ref{tab:inverse-power-backreacted-crit}.

\begin{table}[t]
\centering
\begin{tabular}{cccc}
\toprule
\(q\) &
\(\Lambda_{\rm crit}/\Lambda_{\rm cross}\) &
\(\Lambda_{\rm cross}\) &
\(\Lambda_{\rm crit}\)
\\
\midrule
\(1/3\) & \(0.228\) & \(0.8948\) & \(0.204\) \\
\(1\)   & \(0.629\) & \(0.5479\) & \(0.344\) \\
\(2\)   & \(1.156\) & \(0.6822\) & \(0.788\) \\
\bottomrule
\end{tabular}
\caption{Back-reacted critical amplitudes for inverse-power potentials,
defined by \(\Delta_g^{\rm max}=0.1\), for the same local initial-value
setup as in Table~\ref{tab:inverse-power-backreacted-ivp}.  The amplitudes
are compared with the fixed-throat local crossing value
\(\Lambda_{\rm cross}\), defined by \(\eta_{\rm src}^{\rm inv}(b=1)=1\).}
\label{tab:inverse-power-backreacted-crit}
\end{table}

The back-reacted local initial-value test becomes sensitive to inverse-power
potentials at amplitudes below the fixed-throat crossing value for shallow
powers, most notably \(q=1/3\).  This is not a deep-throat instability.  It
reflects the accumulation of a small but persistent inverse-power force in
the outer part of the integration region, where the scalar has not yet run
far enough for the inverse-power slope to be strongly suppressed.  The
fixed-throat conclusion is therefore refined, not reversed: inverse-power
potentials do not generate an IR-enhanced obstruction at the bottom of the
throat, but their back-reaction can still be visible in the outer matching
region.

The physical picture behind these results is simple.  Because
\[
    |V_{,\varphi}|
    =
    q\Lambda M_{\rm Pl}^3u^{-q-1}
\]
decreases along the GHS flow, the local obstruction is smallest in the deep
throat and largest near its outer edge.  Lengthening the throat therefore
makes the inner region more, not less, gauge dominated.  The condition for
local gauge dominance throughout the full interval \(a\leq b\leq1\) is
\begin{equation}
    \Lambda
    \ll
    \Lambda_{\rm cross}^{(0)}(q)
    =
    \frac{1}
         {16\alpha q M_{\rm Pl}^2 r_+^2}
    \left(
        u_\infty+\frac{\log2}{2\alpha}
    \right)^{q+1}.
    \label{eq:inverse-power-gauge-dominated-condition}
\end{equation}
This is an \(O(1)\) bound independent of \(a\): making the throat longer does
not lower the crossover amplitude.  Inverse powers are mild in the specific
sense that they do not attack the bottom of the throat.  The back-reacted
local initial-value test shows, however, that they are not completely
negligible: a finite fraction of the fixed-throat crossing amplitude can
still leave a visible imprint in the outer matching region.


\subsection{Racetrack stabilization and barrier crossing}
\label{subsec:racetrack-stabilization}

So far, turning on a potential meant asking how much it deforms the GHS
scalar profile.  A racetrack potential changes the question.  It has a
local minimum, separated from a runaway direction by a finite barrier, so
the relevant issue is not only the size of the deformation, but whether the
scalar excursion driven by the black hole carries the modulus over the
barrier.  In this subsection, we make this criterion precise, first in the
fixed-throat approximation and then with a back-reacted check.

The notation involves two coordinates on the same one-dimensional scalar
trajectory.  The canonically normalized EMD field is \(\phi\), with
asymptotic value \(\Phi_\infty\), and we write
\[
    \varphi(b)\equiv \phi(b)-\Phi_\infty \;.
\]
This same displacement controls the gauge kinetic function,
\begin{equation}
    B(\phi)
    =
    B_\infty
    \exp\!\left[
        -2\alpha
        \frac{\phi-\Phi_\infty}{M_{\rm Pl}}
    \right]
    =
    B_\infty
    \exp\!\left[
        -2\alpha
        \frac{\varphi}{M_{\rm Pl}}
    \right] \;.
    \label{eq:racetrack-emd-gauge-coupling}
\end{equation}
The racetrack potential is written in terms of a dimensionless modulus
coordinate \(\chi\), related to the same scalar trajectory by
\begin{equation}
    \chi
    =
    \chi_\infty
    +
    \beta_{\rm rt}
    \frac{\phi-\Phi_\infty}{M_{\rm Pl}}
    =
    \chi_\infty
    +
    \beta_{\rm rt}\frac{\varphi}{M_{\rm Pl}} \;.
    \label{eq:racetrack-beta-embedding}
\end{equation}
Thus, \(\chi\) is not an independent scalar field in the present effective
model; it is the coordinate in which the racetrack potential is expressed.
The same scalar displacement both changes the local gauge coupling and moves
along the racetrack potential.  Here \(\chi_\infty\) is chosen to be the
metastable minimum of the racetrack potential, while \(\Phi_\infty\) is the
asymptotic value of the canonically normalized EMD field.  The constant
\(\beta_{\rm rt}\) fixes the relative normalization between the canonical
EMD displacement and the racetrack coordinate.

Throughout this subsection, \(\varphi\) denotes the canonical EMD
displacement, while \(\chi\) denotes the racetrack coordinate on the same
scalar trajectory.  We use the canonical GHS normalization
\begin{equation}
    2\alpha^2=1 \;,
    \qquad
    \alpha=\frac{1}{\sqrt2} \;,
  \qquad
  \frac{M_{\rm Pl}}{2\alpha}=\frac{1}{\sqrt2}M_{\rm Pl} \;,
    \label{eq:racetrack-alpha-convention}
\end{equation}
and set \(M_{\rm Pl}=r_+=1\).

Following Eq.~\eqref{eq:fixed-throat-scalar-full}, the fixed-throat scalar equation can be written as an equation for
\(\chi\),
\begin{equation}
    \frac{1}{w(b)}
    \frac{d}{db}
    \left[
        p(b)\frac{d\chi}{db}
    \right]
    =
    \mathcal J_{\rm gauge}^{(\beta)}(b)
    +
    \beta_{\rm rt}^2 V_{{\rm rt},\chi}(\chi) \;,
    \label{eq:racetrack-fixed-throat-equation}
\end{equation}
with \(p(b)\) and \(w(b)\) as in Section~\ref{sec:throat-operator}.  The
factor \(\beta_{\rm rt}^2\) follows directly from
Eq.~\eqref{eq:racetrack-beta-embedding}: one power of \(\beta_{\rm rt}\)
converts the equation of motion from \(\varphi\) to \(\chi\), and the other
converts \(\partial V/\partial\varphi\) to
\(\partial V_{\rm rt}/\partial\chi\).

The first term on the right-hand side is the signed gauge source inherited
from the GHS solution,
\begin{equation}
    \mathcal J_{\rm gauge}^{(\beta)}(b)
    =
    -c_\beta
    \frac{1-a}{(1+b)^3(a+b)},
    \qquad
    c_\beta\equiv\frac{\beta_{\rm rt}}{2\alpha} \;.
    \label{eq:racetrack-gauge-source-beta}
\end{equation}
In the magnetic convention, this source is negative throughout the exterior
and drives the scalar away from its UV value as one moves down the throat.
The second term is the racetrack force.  When
\(V_{{\rm rt},\chi}(\chi)>0\) for \(\chi>\chi_\infty\), it opposes the
inward displacement driven by the black hole gauge source.  This competition
between the gauge source and the restoring racetrack force is the mechanism
studied below.

For the shooting problem, we fix the outer boundary at a large but finite
radius,
\begin{equation}
    a=10^{-4},
    \qquad
    b_{\rm out}=10^3,
    \qquad
    \chi(b_{\rm out})=\chi_\infty \;.
    \label{eq:racetrack-basic-parameters}
\end{equation}
Thus, the racetrack coordinate is placed at its metastable minimum far from
the black hole, and we ask whether the throat pushes it past the barrier.
With no racetrack force, \(V_{\rm rt}=0\), the profile is just the GHS
solution re-expressed in \(\chi\) units and normalized at \(b_{\rm out}\),
\begin{equation}
    \chi_{\rm GHS}^{(b_{\rm out})}(b)
    =
    \chi_\infty
    +
    c_\beta
    \left[
        \log\frac{1+b}{a+b}
        -
        \log\frac{1+b_{\rm out}}{a+b_{\rm out}} 
    \right] \;.
    \label{eq:racetrack-finite-outer-ghs}
\end{equation}
The subtracted term is numerically small, but it matters conceptually: the
shooting problem is normalized at a finite radius, not at infinity.

It is convenient to use the logarithmic coordinate
\begin{equation}
    b(\tau)=b_{\rm out}e^{-\tau} \;,
    \label{eq:racetrack-tau-coordinate}
\end{equation}
which runs from \(\tau=0\) at the outer boundary to
\[
    \tau_{\rm max}=\log\frac{b_{\rm out}}{a}
\]
at the inner cutoff.  In this coordinate,
Eq.~\eqref{eq:racetrack-fixed-throat-equation} becomes
\begin{equation}
    \chi_{\tau\tau}
    -
    \frac{b}{a+b}\,\chi_\tau
    =
    b(1+b)
    \left[
        \mathcal J_{\rm gauge}^{(\beta)}(b)
        +
        \beta_{\rm rt}^2 V_{{\rm rt},\chi}(\chi)
    \right] \;,
    \label{eq:racetrack-tau-equation}
\end{equation}
where subscripts \(\tau\) denote derivatives with respect to \(\tau \). Eq.~\eqref{eq:racetrack-tau-equation} is the form integrated numerically.

\begin{figure}[t]
    \centering
    \includegraphics[width=0.6\linewidth]{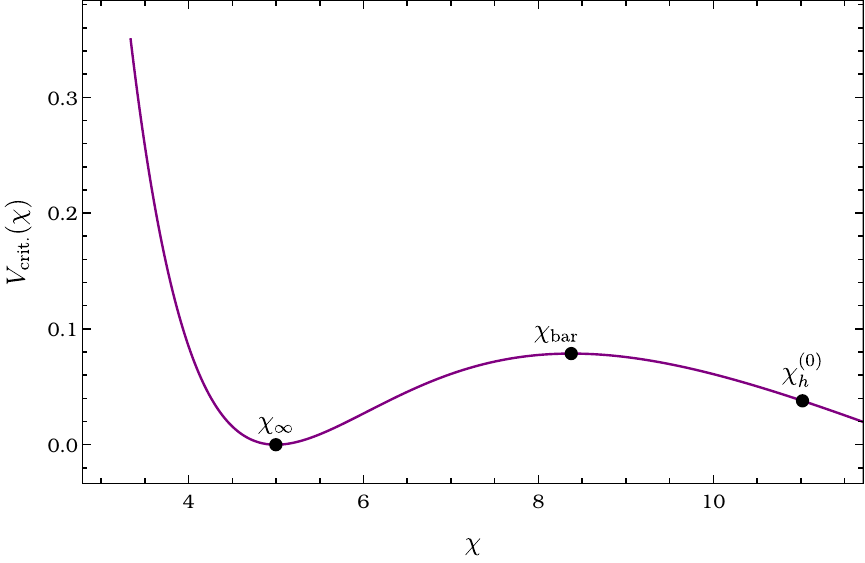}
    \caption{The critical racetrack potential
    \(V_{\rm crit}(\chi)=T_{\rm crit}[\widehat V_{\rm rt}(\chi)-\widehat
    V_{\rm rt}(\chi_\infty)]\).  The minimum \(\chi_\infty\) and the barrier
    \(\chi_{\rm bar}\) bound the metastable basin; barrier crossing
    occurs once the throat-driven endpoint of the scalar trajectory passes
    \(\chi_{\rm bar}\).}
    \label{fig:RTpotential_crit}
\end{figure}

We use a toy racetrack potential\footnote{The racetrack exponents denoted by \(q,q_1,q_2\) in this
subsection are independent shape parameters of the toy racetrack potential.
They should not be confused with the dimensionless exponential slope
\(q=\lambda/(2\alpha)\) used in
Subsection~\ref{subsec:exponential-runaway}, nor with the inverse-power
exponent \(q\) used in Subsection~\ref{subsec:inverse-power-runaway}.} built from a squared two-exponential
racetrack shape function, together with an uplift exponential,
\begin{equation}
    V_{\rm rt}(\chi;T)
    =
    T\,\widehat V_{\rm rt}(\chi) \;,
    \qquad
    \widehat V_{\rm rt}(\chi)
    =
    \mathcal W_{\rm rt}(\chi)^2
    +
    U\,e^{-q(\chi-x_0)} \;.
    \label{eq:racetrack-potential-shape}
\end{equation}
Here, \(T\) is an overall amplitude controlling the height of the racetrack
potential. The function \(\mathcal W_{\rm rt}\) 
serves as a convenient
two-exponential shape function used to generate a metastable minimum, a
barrier, and a runaway direction.  We consider
\begin{equation}
    \mathcal W_{\rm rt}(\chi)
    =
    K_0
    +
    A\,e^{-q_1(\chi-x_0)}
    -
    B\,e^{-q_2(\chi-x_0)} \;.
    \label{eq:racetrack-shape-function}
\end{equation}
The parameters \(q,q_1,q_2\) are racetrack exponents. Our
reference shape uses
\begin{equation}
    A=B=U=1.2 \;,
    \quad
    K_0=10^{-8}\;,
    \quad
    q_1=0.5\;,
    \quad
    q_2=0.04\;,
    \quad
    q=0.8\;,
    \quad
    x_0=3.375 \;,
    \label{eq:racetrack-shape-parameters}
\end{equation}
see Fig.~\ref{fig:RTpotential_crit}. This fixes the metastable minimum and
the barrier top at
\begin{equation}
    \chi_\infty
    =
    4.9977298801 \;,
    \qquad
    \chi_{\rm bar}
    =
    8.3754336144 \;.
    \label{eq:racetrack-min-bar-values}
\end{equation}
The physically relevant number is the distance from the minimum to the
barrier top,
\begin{equation}
    \Delta\chi_{\rm bar}
    \equiv
    \chi_{\rm bar}-\chi_\infty
    =
    3.3777037343 \;.
    \label{eq:racetrack-barrier-distance}
\end{equation}

Eq.~\eqref{eq:racetrack-fixed-throat-equation} is second order, so the
outer condition \(\chi(b_{\rm out})=\chi_\infty\) must be supplemented by a
condition at the inner cutoff.  Physically, this condition approximates
regular scalar flux at the regulated horizon.  The radial flux is
\begin{equation}
    \mathcal F_\chi(b)\equiv p(b)\chi_b(b) \;,
    \label{eq:racetrack-radial-flux}
\end{equation}
where subscript \(b\) denotes \(d/db\). Since \(p(b)\to0\) at the horizon, a nonzero \(\mathcal F_\chi(0)\) would
force a logarithmic singularity in \(\chi\); regularity requires
\(\mathcal F_\chi(0)=0\).  Integrating
Eq.~\eqref{eq:racetrack-fixed-throat-equation} from \(b=0\) to
\(b_{\rm in}=a\), and approximating \(\chi(b)\) by its cutoff value
\(\chi_h\) over this short interval, gives
\begin{equation}
    p(a)\chi_b(a)
    \simeq
    F_{\rm in}^{(0)}
    +
    \beta_{\rm rt}^2 V_{{\rm rt},\chi}(\chi_h)\,P_{\rm in} \;,
    \qquad
    F_{\rm in}^{(0)}
    =
    -c_\beta\,\frac{a(1-a)}{1+a} \;,
    \qquad
    P_{\rm in}
    =
    \frac{3}{2}a^2+\frac{5}{6}a^3 .
    \label{eq:racetrack-regular-flux-approx}
\end{equation}
Using \(p(a)=2a^2\) and \(\chi_b=-\chi_\tau/a\) at \(b=a\), this becomes the
Robin condition,
\begin{equation}
    \chi_\tau(\tau_{\rm max})
    =
    -\frac{1}{2a}
    \left[
        F_{\rm in}^{(0)}
        +
        \beta_{\rm rt}^2 V_{{\rm rt},\chi}(\chi_h)P_{\rm in}
    \right] \;.
    \label{eq:racetrack-regular-flux-robin}
\end{equation}
Together with \(\chi(b_{\rm out})=\chi_\infty\), this closes the
fixed-throat boundary-value problem.

As a check, switching off the racetrack potential\footnote{We make explicit that a quantity is evaluated on the massless GHS profile with the superscript \((0)\).} must reproduce the pure
GHS slope at the cutoff.  For \(\beta_{\rm rt}=1/\sqrt2\), we find
\begin{equation}
    \chi_h^{(0)}=9.2558767731,
    \qquad
    \chi_\tau(\tau_{\rm max})=0.2499500050,
    \label{eq:racetrack-ghs-check-beta-half}
\end{equation}
and for \(\beta_{\rm rt}=1\),
\begin{equation}
    \chi_h^{(0)}=11.0196589668,
    \qquad
    \chi_\tau(\tau_{\rm max})=0.3534826870 .
    \label{eq:racetrack-ghs-check-beta-one}
\end{equation}
This confirms that the boundary condition is implemented correctly.  In each
case,
\begin{equation}
    \Delta\chi_h^{(0)}
    \equiv
    \chi_h^{(0)}-\chi_\infty
\end{equation}
is the field-space distance by which the black hole alone, with no racetrack
force, pushes the scalar trajectory away from the metastable minimum at the
inner cutoff.

The barrier-crossing question can now be phrased as a comparison between two
distances in the same field coordinate.  The first is the barrier distance,
\(\Delta\chi_{\rm bar}\); the second is the excursion driven by the
black hole in the absence of the racetrack force, \(\Delta\chi_h^{(0)}\).
To scan cleanly between trapped and escaping solutions, we keep the
dimensionless racetrack shape fixed in a rescaled coordinate and stretch
only the field-space distance to the barrier by a factor \(L\).  Equivalently,
we define the stretched force by
\begin{equation}
    V_{{\rm rt},\chi}(\chi;T,L)
    =
    \frac{1}{L}\,
    V_{{\rm rt},\chi}
    \left(
        \chi_\infty+\frac{\chi-\chi_\infty}{L};\,T,1
    \right) \;,
    \qquad
    \chi_{\rm bar}(L)
    =
    \chi_\infty
    +
    L\Delta\chi_{\rm bar} .
    \label{eq:racetrack-barrier-stretch}
\end{equation}
Here \(L=1\) is the original racetrack, while larger \(L\) moves the barrier
farther away and makes escape harder.  A first, purely geometric estimate of
the critical stretching is obtained by ignoring the racetrack force and
comparing distances:
\begin{equation}
    L_{\rm crit}^{\rm geom}
    =
    \frac{\Delta\chi_h^{(0)}}{\Delta\chi_{\rm bar}} \;.
    \label{eq:racetrack-Lcrit-geom}
\end{equation}
The full fixed-throat solution below shows that this geometric estimate is
already very close to the actual threshold.

We normalize the racetrack strength by
\begin{equation}
    \kappa_{\rm rt}\equiv \frac{T}{T_{\rm crit}} \;,
    \qquad
    T_{\rm crit}\simeq 5.8563\times10^{-4} \;,
    \label{eq:racetrack-Tcrit}
\end{equation}
and solve the full nonlinear boundary-value problem,
Eqs.~\eqref{eq:racetrack-tau-equation} and
\eqref{eq:racetrack-regular-flux-robin}, at \(\kappa_{\rm rt}=1\).  Scanning
\(L\), we find the point where the modulus value at the cutoff first
coincides with the stretched barrier:
\begin{equation}
    \chi_h(L_{\rm crit})=\chi_{\rm bar}(L_{\rm crit}) \;.
    \label{eq:racetrack-threshold-condition}
\end{equation}

The results are shown in Table~\ref{tab:racetrack-fixed-throat-thresholds}. The near-equality of the last two columns verifies the criticality condition
to high precision.  The force-inclusive fixed-throat result,
\(L_{\rm crit}^{\rm fixed}\), lies below the geometric estimate by only about
\(5\times10^{-4}\) in relative terms, for both values of \(\beta_{\rm rt}\).
Thus, within the fixed-throat approximation, the crossing threshold is set
almost entirely by the field-space excursion driven by the black hole: the
racetrack force only gives a small correction to the distance criterion.

\begin{table}[t]
\centering
\begin{tabular}{ccccc}
\toprule
\(\beta_{\rm rt}\)
&
\(L_{\rm crit}^{\rm geom}\)
&
\(L_{\rm crit}^{\rm fixed}\)
&
\(\chi_h(L_{\rm crit})\)
&
\(\chi_{\rm bar}(L_{\rm crit})\)
\\
\midrule
\(1/\sqrt2\)
&
\(1.2606632280\)
&
\(1.2600247335\)
&
\(9.2537201135\)
&
\(9.2537201279\)
\\
\(1\)
&
\(1.7828470347\)
&
\(1.7819440603\)
&
\(11.0166089895\)
&
\(11.0166089870\)
\\
\bottomrule
\end{tabular}
\caption{Fixed-throat barrier-crossing thresholds for the stretched
racetrack potential at \(\kappa_{\rm rt}~=~T/T_{\rm crit}~=~1\).  The geometric
estimate \(L_{\rm crit}^{\rm geom}\) is obtained by comparing the GHS
excursion \(\Delta\chi_h^{(0)}\) with the unstretched barrier distance
\(\Delta\chi_{\rm bar}\), while \(L_{\rm crit}^{\rm fixed}\) is
obtained from the full nonlinear fixed-throat boundary-value problem.  At
threshold, the cutoff value \(\chi_h\) coincides with the stretched barrier
position \(\chi_{\rm bar}\).}
\label{tab:racetrack-fixed-throat-thresholds}
\end{table}

We now perform an independent back-reacted check.  This check is not meant
to repeat the same fixed-throat boundary-value problem with the same outer
boundary condition.  Instead, it asks whether a local back-reacted evolution
initialized near the horizon reaches a larger or smaller maximum value of
the racetrack coordinate than the fixed-throat estimate.  We start from GHS
data at the cutoff,
\[
    b_{\rm cut}=a \;,
    \qquad
    x_0=1+a \;,
\]
and integrate outward to \(x_{\rm UV}=1+b_{\rm out}=1001\), recording the
largest value reached by the racetrack coordinate, \(\chi_{\rm max}\).  In
the runs below, this maximum occurs at the starting cutoff,
\(x_{\rm max}=x_0=1+a\).  The diagnostic is therefore a direct comparison
between \(\chi_{\rm max}\) and the stretched barrier \(\chi_{\rm bar}(L)\).

For \(\beta_{\rm rt}=1/\sqrt2\),
\begin{equation}
    \chi_{\rm max}=9.2563764733,
    \qquad
    L_{\rm crit}^{\rm back}
    \equiv
    \frac{\chi_{\rm max}-\chi_\infty}{\Delta\chi_{\rm bar}}
    =
    1.2608111688 ,
    \label{eq:racetrack-back-half}
\end{equation}
and for \(\beta_{\rm rt}=1\),
\begin{equation}
    \chi_{\rm max}=11.0203656495,
    \qquad
    L_{\rm crit}^{\rm back}
    \equiv
    \frac{\chi_{\rm max}-\chi_\infty}{\Delta\chi_{\rm bar}}
    =
    1.7830562545 .
    \label{eq:racetrack-back-one}
\end{equation}

The results are compared against their fixed-throat counterparts in 
Table~\ref{tab:racetrack-fixed-back-comparison}. The sign of the shift is also physically natural: in the local back-reacted
evolution the maximum racetrack coordinate is slightly larger than in the
fixed-throat estimate, so the barrier has to be stretched slightly farther
away before crossing is avoided. The racetrack diagnostic therefore reduces to a sharp field-space
comparison.  

\begin{table}[t]
\centering
\begin{tabular}{cccc}
\toprule
\(\beta_{\rm rt}\)
&
\(L_{\rm crit}^{\rm fixed}\)
&
\(L_{\rm crit}^{\rm back}\)
&
\(\left(L_{\rm crit}^{\rm back}-L_{\rm crit}^{\rm fixed}\right)/
L_{\rm crit}^{\rm fixed}\)
\\
\midrule
\(1/\sqrt2\)
&
\(1.2600247335\)
&
\(1.2608111688\)
&
\(6.24\times10^{-4}\)
\\
\(1\)
&
\(1.7819440603\)
&
\(1.7830562545\)
&
\(6.24\times10^{-4}\)
\\
\bottomrule
\end{tabular}
\caption{Comparison between the fixed-throat and local back-reacted
barrier-crossing thresholds.  Back-reaction shifts the critical stretching
upward, but only at the per-mille level.}
\label{tab:racetrack-fixed-back-comparison}
\end{table}

Let \(\Delta\chi_h^{\rm reg}\) denote the actual distance by
which the black hole pushes the scalar trajectory away from the metastable
minimum in the racetrack coordinate, evaluated at the regulated horizon.
Depending on the approximation, \(\Delta\chi_h^{\rm reg}\) may be the pure
GHS value \(\Delta\chi_h^{(0)}\), the fixed-throat value including the
racetrack force, or the corresponding local back-reacted value.

The survival criterion is then
\begin{align}
    \Delta\chi_{\rm bar}
    \gtrsim
    \Delta\chi_h^{\rm reg}
    \quad
    &\Longrightarrow\quad
    \text{the metastable vacuum survives,}
    \label{eq:racetrack-survival-criterion}
    \\
    \Delta\chi_{\rm bar}
    \lesssim
    \Delta\chi_h^{\rm reg}
    \quad
    &\Longrightarrow\quad
    \text{the barrier is crossed and the scalar trajectory runs away.}
    \label{eq:racetrack-crossing-criterion}
\end{align}
The quantity \(\Delta\chi_h^{\rm reg}\) depends on the near-extremality
parameter, on the scalar normalization \(\beta_{\rm rt}\), and mildly on
back-reaction.  The criterion itself is robust.  In the benchmark studied
here, the fixed-throat shooting and the local back-reacted check agree that
back-reaction changes the critical stretching \(L_{\rm crit}\) only at the
\(10^{-3}\) level.  The crossing threshold is therefore controlled almost
entirely by the field-space excursion driven by the black hole.



\subsection{Axion-like periodic potentials}
\label{subsec:axion-like-periodic-potentials}

We now consider a qualitatively different class of potentials, namely
periodic axion-like potentials,
\begin{equation}
  V_a(\phi)
  =
  \Lambda_a^4
  \left[
    1-\cos\left(\frac{\phi-\phi_0}{f_a}\right)
  \right] \;.
  \label{eq:axion-potential}
\end{equation}
Here, \(\phi\) denotes the same canonically normalized EMD scalar as in the
GHS profile.  It is useful to parametrize the potential in terms of the force
amplitude,
\begin{equation}
  A_a
  \equiv
  \frac{\Lambda_a^4}{f_a}
  =
  m_a^2 f_a \;,
  \qquad
  m_a^2=\frac{\Lambda_a^4}{f_a^2} \;.
  \label{eq:axion-force-amplitude}
\end{equation}
The scalar force is then
\begin{equation}
  V_{a,\phi}(\phi)
  =
  A_a
  \sin\left(\frac{\phi-\phi_0}{f_a}\right) \;.
  \label{eq:axion-force}
\end{equation}
Unlike the monotonic potentials discussed above, this force can change sign
along the GHS throat whenever the scalar excursion samples successive
half-periods of the axion potential.  This makes the local and integrated
diagnostics inequivalent.

We use the same fixed-throat GHS background as before.  In the canonical
normalization used throughout this paper,
\begin{equation}
  2\alpha^2=1,
  \qquad
  \alpha=\frac{1}{\sqrt2} \;,
  \qquad
  \frac{M_{\rm Pl}}{2\alpha}=\frac{1}{\sqrt2}M_{\rm Pl} \;.
  \label{eq:axion-alpha-convention}
\end{equation}
The zeroth-order scalar profile is
\begin{equation}
  \phi_{\rm GHS}(b)
  =
  \Phi_\infty
  +
  \frac{M_{\rm Pl}}{2\alpha}
  \log\left(\frac{1+b}{a+b}\right) \;,
  \label{eq:axion-ghs-profile}
\end{equation}
where \(a=(r_+-r_-)/r_+\) and \(b=(r-r_+)/r_+\) as before.  In the numerical
scans below, we use \(M_{\rm Pl}=1\), \(r_+=1\), \(a=10^{-4}\), and the throat
interval \(a\leq b\leq1\).  The scalar excursion across this interval is
\begin{equation}
  \Delta\phi_{\rm throat}
  =
  \frac{1}{\sqrt2}
  \log\left(\frac{(1+a)^2}{4a}\right)
  \simeq
  5.53 \;.
  \label{eq:axion-throat-excursion}
\end{equation}
Thus, the number of axion oscillations sampled by the GHS throat is roughly
\begin{equation}
  N_{\rm osc}
  \simeq
  \frac{\Delta\phi_{\rm throat}}{2\pi f_a}
  \simeq
  \frac{0.88}{f_a/M_{\rm Pl}} \;.
  \label{eq:axion-number-oscillations}
\end{equation}
For \(f_a\sim M_{\rm Pl}\), the axion force varies slowly across the throat.
For \(f_a\ll M_{\rm Pl}\), it oscillates many times and the integrated
response can be strongly suppressed by cancellations.

Let
\begin{equation}
  \theta_\infty
  =
  \frac{\Phi_\infty-\phi_0}{f_a}
  \label{eq:axion-asymptotic-phase}
\end{equation}
be the asymptotic phase.  Evaluated on the GHS profile, the axion force is
\begin{equation}
  V_{a,\phi}\bigl(\phi_{\rm GHS}(b)\bigr)
  =
  A_a
  \sin\left[
    \theta_\infty
    +
    \frac{1}{\sqrt2 f_a}
    \log\left(\frac{1+b}{a+b}\right)
  \right] \;,
  \label{eq:axion-force-on-ghs}
\end{equation}
where we have set \(M_{\rm Pl}=1\) in the second term.  The near-throat
pointwise diagnostic is the axion force compared with the positive GHS
gauge-source magnitude,
\begin{equation}
  \eta_{\rm src}^{(a)}(b)
  =
  \frac{
    r_+^2 A_a
    \left|
    \sin\left[
      \theta_\infty
      +
      \frac{1}{\sqrt2 f_a}
      \log\left(\frac{1+b}{a+b}\right)
    \right]
    \right|
  }{
    \mathcal S_{\rm gauge}(b)
  } \;,
  \label{eq:axion-local-diagnostic}
\end{equation}
where \(S_{\rm gauge}(b)\) remains as defined in Eq. \eqref{eq:gauge-source-magnitude},
\begin{equation}
  \mathcal S_{\rm gauge}(b)
  =
  \frac{1}{\sqrt2}
  \frac{1-a}{(1+b)^3(a+b)} \;,
  \label{eq:axion-gauge-source-magnitude}
\end{equation}
under the canonical normalization \(M_{\rm Pl}=1\).  For a reference amplitude
\(A_{\rm ref}\), we define
\begin{equation}
  A_{a,\rm crit}
  =
  \frac{A_{\rm ref}}
  {
    \max_{a\leq b\leq b_{\max}^{\rm diag}}
    \eta_{\rm src}^{(a)}(b;A_{\rm ref})
  },
  \qquad
  m_{a,\rm crit}
  =
  \sqrt{\frac{A_{a,\rm crit}}{f_a}} .
  \label{eq:axion-critical-amplitude}
\end{equation}

This pointwise ratio depends strongly on the radial interval over which it is
evaluated.  Since \(\mathcal S_{\rm gauge}\sim b^{-4}\) in the asymptotic
tail, a global scan out to large \(b\) is dominated by the outer boundary and
is not a clean measure of throat destabilization.  We therefore distinguish
the near-throat criterion, \(b_{\max}^{\rm diag}=1\), from more extended
diagnostic intervals.

For the near-throat scan \(a\leq b\leq1\), scanning uniformly over the
asymptotic phase \(\theta_\infty\), we find the median critical amplitudes
shown in Table~\ref{tab:axion-throat-median-critical}.

\begin{table}[t]
\centering
\begin{tabular}{ccc}
\toprule
\(f_a/M_{\rm Pl}\)
&
\(A_{a,\rm crit}^{\rm throat,med}\)
&
\(m_{a,\rm crit}^{\rm throat,med}\)
\\
\midrule
\(1\)    & \(1.39\times10^{-1}\) & \(3.73\times10^{-1}\) \\
\(0.7\)  & \(1.37\times10^{-1}\) & \(4.43\times10^{-1}\) \\
\(0.5\)  & \(1.25\times10^{-1}\) & \(5.00\times10^{-1}\) \\
\(0.3\)  & \(1.18\times10^{-1}\) & \(6.27\times10^{-1}\) \\
\(0.1\)  & \(1.22\times10^{-1}\) & \(1.11\) \\
\(0.03\) & \(1.05\times10^{-1}\) & \(1.87\) \\
\(0.01\) & \(9.50\times10^{-2}\) & \(3.08\) \\
\bottomrule
\end{tabular}
\caption{Median critical amplitudes for axion-like periodic potentials in
the near-throat diagnostic interval \(a\leq b\leq1\), obtained by scanning
uniformly over the asymptotic phase \(\theta_\infty\).  The critical force
amplitude \(A_{a,\rm crit}\) is defined by
\(\max_{a\leq b\leq1}\eta_{\rm src}^{(a)}=1\), while
\(m_{a,\rm crit}=(A_{a,\rm crit}/f_a)^{1/2}\).}
\label{tab:axion-throat-median-critical}
\end{table}

The critical force amplitude is approximately independent of \(f_a\).  This
is expected: the pointwise diagnostic compares the local axion force
directly with the GHS gauge-source magnitude, and the force amplitude is
bounded by \(A_a\).  The corresponding critical mass therefore scales
approximately as
\begin{equation}
    m_{a,\rm crit}
    =
    \left(\frac{A_{a,\rm crit}}{f_a}\right)^{1/2}
    \propto f_a^{-1/2},
\end{equation}
up to mild phase-dependent oscillatory effects.

The dependence on the diagnostic cutoff is much stronger.  For the same
phase scan, the median critical amplitudes are of order
\begin{equation}
  A_{a,\rm crit}^{\rm med}
  \sim
  10^{-1},\quad
  10^{-4},\quad
  10^{-12},
  \qquad
  b_{\max}^{\rm diag}=1,\ 10,\ 10^3,
  \label{eq:axion-cutoff-scaling}
\end{equation}
respectively.  This hierarchy is not a new physical instability of the
throat; it reflects the rapid falloff of the gauge source in the asymptotic
region.  For this reason, the near-throat criterion is the appropriate one
for assessing whether the throat solution is locally distorted.

We also solved the fixed-throat deformation problem directly.  Writing
\begin{equation}
  \phi(b)=\phi_{\rm GHS}(b)+\delta\phi(b) \;,
\end{equation}
the nonlinear deformation satisfies, in the logarithmic coordinate
\begin{equation}
  \ell=\log b \;,
\end{equation}
the equation
\begin{equation}
  \frac{d}{d\ell}
  \left[
    Q_\ell(\ell)\frac{d\delta\phi}{d\ell}
  \right]
  =
  b(1+b)(a+b) A_a
  \sin\left[
    \frac{
      \phi_{\rm GHS}(b)+\delta\phi(b)-\phi_0
    }{f_a}
  \right],
  \qquad
  Q_\ell(\ell)=a+b \;.
  \label{eq:axion-integrated-deformation}
\end{equation}
Here, \(Q_\ell\) is only a shorthand for the differential-operator
coefficient; it is unrelated to the exponent \(q\) used in the exponential
and racetrack sections.  The initial conditions are
\begin{equation}
  \delta\phi(b=a)=0 \;,
  \qquad
  \frac{d \delta\phi}{d\log b}\bigg|_{b=a}=0 \;.
  \label{eq:axion-deformation-ic}
\end{equation}
This is a local fixed-throat initial-value test: the deformation is set to
zero at the regulated throat endpoint, and the subsequent profile measures
the accumulated response to the axion force along the GHS background.

The corresponding gauge-coupling deformation is measured by
\begin{equation}
  \Delta_g(b)
  =
  \left|
  \frac{g^2_{\rm axion}(b)}{g^2_{\rm GHS}(b)}-1
  \right|
  =
  \left|
  e^{+2\alpha \delta\phi(b)/M_{\rm Pl}}-1
  \right| \;.
  \label{eq:axion-gauge-shift}
\end{equation}
In units \(M_{\rm Pl}=1\), for small deformations,
\begin{equation}
  \Delta_g(b)
  \simeq
  2\alpha|\delta\phi(b)|
  =
  \sqrt2\,|\delta\phi(b)| \;.
  \label{eq:axion-gauge-shift-linear}
\end{equation}

At the generic phase \(\theta_\infty=\pi/2\), using
\(A_a=A_{a,\rm crit}^{\rm throat,med}\), we obtain
Table~\ref{tab:axion-integrated-response-generic-phase}.

\begin{table}[t]
\centering
\begin{tabular}{ccc}
\toprule
\(f_a/M_{\rm Pl}\)
&
\(\max_{a\leq b\leq1}|\delta\phi|\)
&
\(\max_{a\leq b\leq1}\Delta_g\)
\\
\midrule
\(1\)    & \(3.69\times10^{-2}\) & \(5.36\times10^{-2}\) \\
\(0.1\)  & \(9.76\times10^{-3}\) & \(1.39\times10^{-2}\) \\
\(0.01\) & \(1.29\times10^{-4}\) & \(1.82\times10^{-4}\) \\
\bottomrule
\end{tabular}
\caption{Fixed-throat integrated deformation induced by axion-like periodic
potentials at the generic phase \(\theta_\infty=\pi/2\).  The force
amplitude is set equal to the median near-throat local critical value
\(A_{a,\rm crit}^{\rm throat,med}\) for each \(f_a\).  The table shows the
maximum scalar deformation and the corresponding maximum fractional
gauge-coupling shift over the near-throat interval \(a\leq b\leq1\).}
\label{tab:axion-integrated-response-generic-phase}
\end{table}

Thus, for \(f_a\sim M_{\rm Pl}\), a force at the near-throat local threshold
induces only a few-percent change in the gauge coupling.  For smaller decay
constants, the pointwise force can still reach the local threshold, but the
integrated response is strongly suppressed by oscillatory cancellations along
the throat.

The contrast between slowly varying and rapidly oscillating axion forces
becomes even sharper if the amplitude is increased to
\(A_a=10A_{a,\rm crit}^{\rm throat,med}\).  Scanning the representative
phases \(\theta_\infty~=~0,\pi/2,\pi\), we find
Table~\ref{tab:axion-integrated-response-large-amplitude}.

\begin{table}[t]
\centering
\begin{tabular}{cccc}
\toprule
\(f_a/M_{\rm Pl}\)
&
\(\theta_\infty\)
&
\(\max_{a\leq b\leq1}|\delta\phi|\)
&
\(\max_{a\leq b\leq1}\Delta_g\)
\\
\midrule
\(1\) & \(0\)       & \(8.24\times10^{-1}\) & \(2.21\) \\
\(1\) & \(\pi/2\)   & \(3.62\times10^{-1}\) & \(6.68\times10^{-1}\) \\
\(1\) & \(\pi\)     & \(5.39\times10^{-1}\) & \(5.34\times10^{-1}\) \\
\(0.1\) & \(0\)     & \(2.23\times10^{-1}\) & \(3.70\times10^{-1}\) \\
\(0.1\) & \(\pi/2\) & \(1.43\times10^{-1}\) & \(2.23\times10^{-1}\) \\
\(0.1\) & \(\pi\)   & \(4.60\times10^{-2}\) & \(6.30\times10^{-2}\) \\
\(0.01\) & \(0\)     & \(2.43\times10^{-3}\) & \(3.45\times10^{-3}\) \\
\(0.01\) & \(\pi/2\) & \(1.72\times10^{-3}\) & \(2.43\times10^{-3}\) \\
\(0.01\) & \(\pi\)   & \(1.96\times10^{-3}\) & \(2.77\times10^{-3}\) \\
\bottomrule
\end{tabular}
\caption{Fixed-throat integrated deformation induced by axion-like periodic
potentials at amplitude \(A_a=10A_{a,\rm crit}^{\rm throat,med}\).  The
table compares three representative asymptotic phases and shows the maximum
scalar deformation and fractional gauge-coupling shift over the near-throat
interval \(a\leq b\leq1\).}
\label{tab:axion-integrated-response-large-amplitude}
\end{table}

For \(f_a\sim M_{\rm Pl}\), this already lies outside the small-deformation
regime: depending on the phase, the gauge-coupling shift ranges from order
one-half to larger than unity.  For \(f_a=0.1M_{\rm Pl}\), the response is
still sizable but remains below order one.  For
\(f_a=10^{-2}M_{\rm Pl}\), by contrast, oscillatory averaging keeps the
fixed-throat gauge-coupling shift at the per-mille level even at ten times
the local throat threshold.

As a back-reacted check, we also solved the radial equations for a small set
of representative axion parameters.  In this comparison, we shoot on the
horizon value of the scalar so that the axion and reference solutions share
the same exterior value at \(b=1\).  This removes the trivial offset in the
asymptotic scalar and isolates the distortion of the throat profile at fixed
exterior modulus.  For the generic phase \(\theta_\infty=\pi/2\), the matched
back-reacted diagnostic gives Table~\ref{tab:axion-backreacted-matched}.

\begin{table}[t]
\centering
\begin{tabular}{cccc}
\toprule
\(f_a/M_{\rm Pl}\)
&
\(A_a/A_{a,\rm crit}^{\rm throat,med}\)
&
\(\max_{a\leq b\leq1}|\Delta\phi_{\rm back}|\)
&
\(\max_{a\leq b\leq1}\Delta_g^{\rm back}\)
\\
\midrule
\(1\)    & \(1\)  & \(8.43\times10^{-2}\) & \(1.12\times10^{-1}\) \\
\(0.1\)  & \(1\)  & \(5.95\times10^{-2}\) & \(8.78\times10^{-2}\) \\
\(0.1\)  & \(10\) & \(2.34\times10^{-1}\) & \(3.93\times10^{-1}\) \\
\(0.01\) & \(1\)  & \(5.81\times10^{-3}\) & \(8.25\times10^{-3}\) \\
\(0.01\) & \(10\) & \(6.11\times10^{-3}\) & \(8.68\times10^{-3}\) \\
\bottomrule
\end{tabular}
\caption{Matched back-reacted diagnostic for axion-like periodic potentials
at the generic phase \(\theta_\infty~=~\pi/2\).  For each axion solution, the
horizon value of the scalar is shot so that the solution matches the
corresponding \(V=0\) reference solution at \(b=1\).  The table reports the
maximum scalar-profile distortion and the corresponding maximum fractional
gauge-coupling shift over the throat interval \(a\leq b\leq1\).}
\label{tab:axion-backreacted-matched}
\end{table}

In every run, the maximum is attained near the inner cutoff \(b=a\), so these
numbers should be read as upper-envelope diagnostics for the profile
distortion over the throat interval, not as deformations spread uniformly
across it.  Back-reaction increases the size of the deformation, most
noticeably at large \(f_a\): slowly varying axion forces can produce
order ten-percent corrections already near the local threshold, while
rapidly oscillating forces remain strongly suppressed.  In particular, for
\(f_a=10^{-2}M_{\rm Pl}\), even
\(A_a=10\,A_{a,\rm crit}^{\rm throat,med}\) keeps the matched back-reacted
gauge-coupling shift below the percent level.

The conclusion is therefore twofold.  The relevant local force scale is
\(A_a=m_a^2 f_a\), and the near-throat pointwise criterion gives
\(A_{a,\rm crit}^{\rm throat}\sim10^{-1}\) in Planck units for the present
near-extremal benchmark.  However, this pointwise criterion can be overly
pessimistic when \(f_a\ll M_{\rm Pl}\): the axion force then oscillates many
times along the throat, and the integrated deformation is strongly reduced
by cancellations.  The matched back-reacted check preserves this hierarchy.
It increases the deformation relative to the fixed-throat estimate,
especially for slowly varying forces with \(f_a~\sim~M_{\rm Pl}\), but rapidly
oscillating forces remain suppressed even at amplitudes several times above
the near-throat local threshold.  Periodic potentials can therefore be
significantly less disruptive than monotonic runaway potentials with the
same pointwise force amplitude, because their alternating sign suppresses
the accumulated scalar response.



\subsection{Supergravity-inspired inverse-power potentials}
\label{subsec:sugra-inspired-inverse-power}

We close this survey with a deliberately limited toy diagnostic.  We consider
inverse-power potentials dressed by an exponential factor, of the kind often
used as phenomenological proxies for effects in supergravity or string
effective theories.  The goal is not to derive a complete compactification
potential, nor to introduce a new universality class on the same footing as
the previous examples.  Rather, we use this simple model to test how
sensitive a regulated throat profile can be to the sign of an exponential
correction.

In this subsection, it is convenient to use a dimensionless modulus coordinate
\(s\) adapted to the gauge coupling,
\begin{equation}
    g^2(s)=\frac{1}{s} \;.
    \label{eq:sugra-gauge-coupling}
\end{equation}
Thus, large \(s\) corresponds to weak coupling.  This convention differs from
the previous subsections, where the GHS displacement was written in terms of
a canonically normalized scalar whose increase raised the local gauge
coupling.  Here, the same physical effect is described by a decrease of
\(s\) along the magnetic throat.

For the numerical benchmarks in this subsection, we use a normalized monotone
profile,
\begin{equation}
    s_{\rm norm}(b)
    =
    s_\infty
    +
    \frac{
        \log \left[
        \dfrac{b(a+b_{\rm max})}{b_{\rm max}(a+b)}
        \right]
    }{
        \log \left[
        \dfrac{b_{\rm max}(a+b_{\rm min})}{b_{\rm min}(a+b_{\rm max})}
        \right]
    } \;,
    \label{eq:sugra-finite-normalized-profile}
\end{equation}
with
\begin{equation}
    b_{\rm min}=a,
    \qquad
    a=10^{-4},
    \qquad
    b_{\rm max}=10^3 .
    \label{eq:sugra-profile-benchmark-domain}
\end{equation}
This is not the exact GHS profile used in the previous subsections.  It is a
normalized proxy with a fixed unit excursion,
\begin{equation}
    s_{\rm norm}(b_{\rm max})=s_\infty \;,
    \qquad
    s_{\rm norm}(b_{\rm min})=s_\infty-1 \;.
    \label{eq:sugra-profile-endpoints}
\end{equation}
The numerator in Eq.~\eqref{eq:sugra-finite-normalized-profile} has a
different shape from the GHS logarithm
\(\log[(1+b)/(a+b)]\), and most of its variation is concentrated near
\(b\sim O(a)\).  The results below should therefore be read as a
fixed-unit excursion sign test, not as a direct replacement for the full
unnormalized GHS throat.  In the exact GHS profile, the field range grows as
\(\log(1/a)\), and the corresponding critical normalizations would inherit an
additional dependence on that larger excursion.

For example, if \(s_\infty=5\), then
\begin{equation}
    s_{\rm norm}(b_{\rm min})=4 \;,
    \qquad
    \frac{g^2(b_{\rm min})}{g_\infty^2}
    =
    \frac{s_\infty}{s_\infty-1}
    =
    \frac54 \;.
    \label{eq:sugra-coupling-shift-s5}
\end{equation}
This corresponds to an order-one, but still perturbative, change in the local
gauge coupling.

We compare the two sign choices
\begin{equation}
    V_\pm(s)
    =
    \lambda_{\rm sg}\,s^{-\alpha}
    \exp\!\left(\pm\frac{\kappa s^2}{2}\right) ,
    \label{eq:sugra-potential-pm}
\end{equation}
and in the numerical scans set
\begin{equation}
    \alpha=1 \;,
    \qquad
    \kappa=1 \;.
    \label{eq:sugra-alpha-kappa-choice}
\end{equation}
The plus-sign potential is
\begin{equation}
    V_+(s)
    =
    \lambda_{\rm sg}\,s^{-1}
    \exp\!\left(+\frac{s^2}{2}\right).
    \label{eq:sugra-plus-potential}
\end{equation}
It is meant as a phenomenological proxy for a positive exponential dressing
of the inverse-power potential.  This is the potentially dangerous sign: if
\(s_\infty\gg1\), the exponential factor is large precisely where the
asymptotic modulus is weakly coupled.

The minus-sign potential is
\begin{equation}
    V_-(s)
    =
    \lambda_{\rm sg}\,s^{-1}
    \exp\!\left(-\frac{s^2}{2}\right).
    \label{eq:sugra-minus-potential}
\end{equation}
It has the opposite behaviour and is exponentially suppressed at weak
coupling, as expected for an instanton-like or non-perturbative correction.
The comparison between \(V_+\) and \(V_-\) isolates the sign sensitivity of
the regulated profile.

We write
\begin{equation}
    s(b)=s_{\rm norm}(b)+\Delta(b) \;.
    \label{eq:sugra-delta-def}
\end{equation}
The deformation \(\Delta\) is sourced by the scalar potential.  We use the
logarithmic coordinate
\begin{equation}
    \ell=\log b \;.
\end{equation}
After the change of variables \(b=e^\ell\), the potential term in the radial
equation,
\begin{equation}
    \frac{d}{db}\!\left[p(b)\,s'(b)\right]
    =
    w(b)\,V_{,s}(s)+\cdots ,
\end{equation}
acquires a positive weight.  In the finite-difference implementation on a
uniform \(\ell\)-grid, the kinetic operator \({\cal L}\) absorbs part of this
weight, and the net factor multiplying the potential source is
\begin{equation}
    \mathcal B_\ell
    =
    \frac{e^\ell}{1+e^\ell}
    =
    \frac{b}{1+b} \;.
    \label{eq:sugra-Bell-def}
\end{equation}
Using the deformation \(\Delta\) introduced in Eq.~\eqref{eq:sugra-delta-def}, the linearized finite-difference equation then takes the schematic form,
\begin{equation}
    {\cal L}\,\Delta
    +
    \mathcal B_\ell
    \left[
        V_{,s}\!\left(s_{\rm norm}(\ell)\right)
        -V_{,s}(s_\infty)
        +
        V_{,ss}\!\left(s_{\rm norm}(\ell)\right)\Delta(\ell)
    \right]
    =
    0 \;.
    \label{eq:sugra-linear-delta-equation}
\end{equation}
The subtraction of \(V_{,s}(s_\infty)\) keeps the asymptotic boundary value
fixed at \(s_\infty\).  Since \(\mathcal B_\ell>0\), the sign of the
deformation is controlled by the sign of the bracketed source term.

We measure the deformation by
\begin{equation}
    \epsilon_{\rm max}
    =
    \max_b
    \left|
        \frac{\Delta(b)}{s_{\rm norm}(b)}
    \right| \;.
    \label{eq:sugra-epsilon-max}
\end{equation}
The critical normalization is defined by the one-percent criterion
\begin{equation}
    \epsilon_{\rm max}=10^{-2}.
    \label{eq:sugra-one-percent-criterion}
\end{equation}
Since the linearized problem is proportional to the overall normalization
\(\lambda_{\rm sg}\), a reference run at \(\lambda_{\rm ref}\) gives
\begin{equation}
    \lambda_{\rm crit}^{\rm lin}
    =
    \lambda_{\rm ref}\,
    \frac{10^{-2}}{\epsilon_{\rm max}(\lambda_{\rm ref})} \;.
    \label{eq:sugra-linear-lambda-crit}
\end{equation}
In the scans belowm we used
\begin{equation}
    \lambda_{\rm ref}=6.08994\times10^{-2}.
    \label{eq:sugra-lambda-ref}
\end{equation}

For the plus-sign potential, the one-percent point is close enough to the
nonlinear regime that we also solved the nonlinear finite-difference problem,
\begin{equation}
    {\cal L}\,\Delta
    +
    \mathcal B_\ell
    \left[
        V_{,s}\!\left(s_{\rm norm}(\ell)+\Delta(\ell)\right)
        -
        V_{,s}(s_\infty)
    \right]
    =
    0 \;,
    \label{eq:sugra-nonlinear-delta-equation}
\end{equation}
using the linearized profile as a Newton seed.  For the minus-sign potential
in the weak-coupling regime, the deformation at \(\lambda_{\rm ref}\) is so
small that the linear estimate is already the relevant diagnostic.

Consider first the plus-sign potential.  At the benchmark value,
\begin{equation}
    s_\infty=5,
    \qquad
    g_\infty=\frac{1}{\sqrt5}\simeq0.447,
\end{equation}
the exponential factor is
\begin{equation}
    \exp\!\left(\frac{s_\infty^2}{2}\right)
    =
    \exp\!\left(\frac{25}{2}\right)
    \simeq
    2.68\times10^5 .
    \label{eq:sugra-plus-exp-factor-s5}
\end{equation}
The nonlinear continuation gives a critical normalization
\begin{equation}
    \lambda_{\rm crit}^{(+)}
    \simeq
    1.04\times10^{-1},
    \label{eq:sugra-lambda-crit-plus-s5}
\end{equation}
with
\begin{equation}
    \max_b\left|\frac{\Delta}{s_{\rm norm}}\right|
    \simeq
    9.98\times10^{-3}.
    \label{eq:sugra-plus-max-deformation}
\end{equation}
The maximum occurs at
\begin{equation}
    b_{\rm peak}^{(+)}
    \simeq
    1.7\times10^{-2},
    \label{eq:sugra-plus-bpeak}
\end{equation}
where
\begin{equation}
    s_{\rm norm}(b_{\rm peak})
    \simeq
    4.99,
    \qquad
    \Delta(b_{\rm peak})
    \simeq
    4.98\times10^{-2},
    \qquad
    s(b_{\rm peak})
    \simeq
    5.04 .
    \label{eq:sugra-plus-peak-values}
\end{equation}
The sign of the extracted deformation is consistent with the finite-difference
equation.  Because \(\mathcal B_\ell>0\), the source generated by the
plus-sign potential pushes the solution toward larger \(s\) relative to the
normalized background; the nonlinear solve gives \(\Delta>0\) at the peak.
Thus, the plus-sign potential can deform the regulated profile at the
one-percent level for a normalization of order \(10^{-1}\).  The effect is
not a pure boundary artifact: the peak lies in the exterior region at
\(b\simeq1.7\times10^{-2}\), not at the inner cutoff.

For the opposite sign, the weak-coupling region is exponentially protected.
At \(s_\infty=5\),
\begin{equation}
    \exp\!\left(-\frac{s_\infty^2}{2}\right)
    =
    \exp\!\left(-\frac{25}{2}\right)
    \simeq
    3.73\times10^{-6}.
    \label{eq:sugra-minus-exp-factor-s5}
\end{equation}
At the reference normalization \(\lambda_{\rm ref}=6.08994\times10^{-2}\),
the maximal linearized fractional deformation is only
\begin{equation}
    \epsilon_{\rm max}^{(-)}(\lambda_{\rm ref})
    \simeq
    1.22\times10^{-8}.
    \label{eq:sugra-minus-eps-ref}
\end{equation}
The corresponding one-percent critical value is
\begin{equation}
    \lambda_{\rm crit}^{(-)}(s_\infty=5)
    \simeq
    5.00\times10^4 .
    \label{eq:sugra-lambda-crit-minus-s5}
\end{equation}
A local refinement around the true maximum gives
\begin{equation}
    \epsilon_{\rm max}^{(-)}
    =
    1.21698\times10^{-8},
    \qquad
    \lambda_{\rm crit}^{(-)}
    =
    5.00414\times10^4,
    \label{eq:sugra-minus-refined}
\end{equation}
with
\begin{equation}
    b_{\rm peak}^{(-)}
    \simeq
    8.34 .
    \label{eq:sugra-minus-bpeak-s5}
\end{equation}
The convergence with the local grid was stable: increasing the refinement
grid from \(400\) to \(2400\) points changed the extracted critical value
only at relative order \(10^{-7}\) to \(10^{-8}\).

The comparison with the plus-sign potential is striking:
\begin{equation}
    \frac{\lambda_{\rm crit}^{(-)}}{\lambda_{\rm crit}^{(+)}}
    \simeq
    4.8\times10^5 .
    \label{eq:sugra-minus-plus-ratio-s5}
\end{equation}
This hierarchy is close to the expected exponential factor
\(\exp(s_\infty^2/2)\simeq2.68\times10^5\) at \(s_\infty=5\), with a
remaining factor of approximately \(1.8\).  This residual order-one factor is
not meaningful by itself; the two potentials are sampled through their
derivatives over different radial regions, rather than simply evaluated at
\(s_\infty\).

The minus-sign potential becomes less suppressed as \(s_\infty\) is lowered.
We therefore also tested stronger-coupling benchmarks, summarized in
Table~\ref{tab:minus-sugra-scan}.

\begin{table}[t]
\centering
\begin{tabular}{ccccc}
\toprule
\(s_\infty\) &
\(g_\infty=1/\sqrt{s_\infty}\) &
\(\exp(-s_\infty^2/2)\) &
\(\lambda_{\rm crit}^{(-)}\) &
\(b_{\rm peak}^{(-)}\)
\\
\midrule
\(\sqrt2\) &
\(2^{-1/4}\simeq0.841\) &
\(e^{-1}\simeq3.68\times10^{-1}\) &
\(1.77\times10^{1}\) &
\(5.11\)
\\
\(2\) &
\(0.707\) &
\(1.35\times10^{-1}\) &
\(8.39\times10^{1}\) &
\(2.04\times10^{1}\)
\\
\(5\) &
\(0.447\) &
\(3.73\times10^{-6}\) &
\(5.00\times10^{4}\) &
\(8.34\)
\\
\bottomrule
\end{tabular}
\caption{Critical normalizations for the minus-sign SUGRA-inspired potential
\(V_-(s)=\lambda_{\rm sg}s^{-1}\exp(-s^2/2)\).  The criterion is
\(\max|\Delta/s_{\rm norm}|=10^{-2}\), evaluated on the normalized unit-excursion profile \(s_{\rm norm}\).}
\label{tab:minus-sugra-scan}
\end{table}

As expected, the minus-sign potential becomes more relevant as the
asymptotic modulus is moved toward stronger coupling.  Even then, however,
the critical normalization remains much larger than in the plus-sign case at
\(s_\infty=5\).  The exponential sign therefore controls the sensitivity of
the regulated profile much more strongly than the inverse-power prefactor
alone.

The point \(s_\infty=\sqrt2\) should be interpreted cautiously.  Although it
is a useful numerical stress test, the regulated inner value is
\begin{equation}
    s_{\rm norm}(b_{\rm min})
    =
    \sqrt2-1
    \simeq
    0.414 ,
    \label{eq:sugra-sqrt2-inner-value}
\end{equation}
so \(g^2=1/s\) is already large near the cutoff.  The coupling enhancement
relative to infinity is
\begin{equation}
    \frac{g^2(b_{\rm min})}{g_\infty^2}
    =
    \frac{\sqrt2}{\sqrt2-1}
    \simeq
    3.4 .
    \label{eq:sugra-sqrt2-coupling-ratio}
\end{equation}
This is therefore not a clean perturbative weak-coupling benchmark.  The more
reliable perturbative cases are \(s_\infty=2\), and especially
\(s_\infty=5\).

The trend is nevertheless clear.  At \(s_\infty=\sqrt2\), where the
exponential suppression is only \(e^{-1}\), the one-percent deformation occurs
at
\begin{equation}
    \lambda_{\rm crit}^{(-)}
    \simeq
    17.7 .
\end{equation}
At \(s_\infty=2\), this increases to
\begin{equation}
    \lambda_{\rm crit}^{(-)}
    \simeq
    83.9 ,
\end{equation}
and by \(s_\infty=5\) it has grown to
\begin{equation}
    \lambda_{\rm crit}^{(-)}
    \simeq
    5.0\times10^4 .
\end{equation}
Thus, the minus-sign potential rapidly decouples as the asymptotic modulus is
moved into the weak-coupling regime.

The lesson of this toy diagnostic is the sign sensitivity.  The plus-sign
potential grows at weak coupling and can compete with the regulated profile
for
\begin{equation}
    \lambda_{\rm crit}^{(+)}
    \sim
    10^{-1}
    \qquad
    (s_\infty=5).
\end{equation}
It is therefore dangerous: an order-one plus-sign contribution would strongly
deform, and could remove, the altered-coupling region.  By contrast, the
minus-sign potential is exponentially suppressed at weak coupling.  For the
same benchmark,
\begin{equation}
    \lambda_{\rm crit}^{(-)}
    \sim
    5\times10^4 ,
\end{equation}
so an order-one normalization is essentially harmless.  It can become relevant
only if the prefactor is very large or if the asymptotic modulus is not far
into the weak-coupling regime.

The comparison
\begin{equation}
    \frac{\lambda_{\rm crit}^{(-)}}{\lambda_{\rm crit}^{(+)}}
    \simeq
    4.8\times10^5
\end{equation}
is the cleanest quantitative expression of this sign sensitivity.  The
conclusion is not that supergravity-inspired corrections generically destroy
the GHS exterior.  Rather, the effect is sign-dependent:
\(e^{+s^2/2}\)-type corrections are dangerous because they are enhanced at
weak coupling, while \(e^{-s^2/2}\)-type corrections are strongly suppressed
and typically do not obstruct the throat in the weak-coupling regime.  In
terms of the asymptotic gauge coupling, this hierarchy scales as
\begin{equation}
    \exp\!\left(\frac{s_\infty^2}{2}\right)
    =
    \exp\!\left(\frac{1}{2g_\infty^4}\right),
\end{equation}
and therefore becomes enormous as \(g_\infty\to0\).



\section{Phenomenological outlook: large charged black holes and moduli fields}
\label{sec:pheno-outlook}

The analysis above was deliberately local, minimal, and conservative.  It did not rely
on a specific microscopic interpretation of the gauge field or of the scalar:
the equations only require a charged dilatonic black hole, a gauge kinetic
function \(B(\phi)\), and a scalar potential \(V(\phi)\).  One possible
phenomenological interpretation is that the gauge field belongs to a hidden
\(U(1)\) sector, while the dilaton is a modulus, or another very light scalar,
controlling the corresponding effective gauge coupling.

\subsection{Hidden charge and macroscopic scalar excursions}

Astrophysical black holes are expected to be essentially neutral under
ordinary electromagnetism.  Any Standard Model electric charge is efficiently
screened or discharged by accretion of ambient plasma, and sufficiently large
electric fields can also discharge through Schwinger pair production
\cite{Zajacek:2018ycb,Gibbons:1975kk}.  Both mechanisms rely on the existence
of light charged states coupled to the black hole charge.  A hidden charge is
less constrained.  If the Standard Model is neutral under the hidden
\(U(1)\), and if the hidden sector does not contain an abundant population of
light oppositely charged particles, then the usual discharge channels can be
strongly suppressed.  In that case, a black hole which forms with hidden
charge, or accretes it, may retain it much longer than it would retain an
ordinary electric charge \cite{Cardoso:2016olt}.

This possibility is speculative, and several effects could shorten the
lifetime of the hidden charge.  Kinetic mixing with electromagnetism, light
hidden matter in the environment, or efficient hidden-sector plasma effects
would all tend to discharge the black hole.  The minimal phenomenological
assumption behind the scenario is therefore simply that the charge belongs to
a hidden gauge sector with no abundant light discharge channel.  In the body
of the paper, we used magnetically charged solutions because they lead to a
particularly simple static description.  A literal hidden magnetic charge
would require monopoles of the hidden \(U(1)\).  However, a magnetically-charged black hole
should not be viewed as
essential to the mechanism: the corresponding electric solutions obey the
same logic, with the appropriate electric-magnetic replacement, as discussed
in Section~\ref{sec:ghs-solution}.

In this interpretation, the object described by the effective theory is a
large black hole charged under a hidden gauge symmetry.  The scalar
\(\phi\) controls the hidden gauge kinetic function \(B(\phi)\), and hence the
local hidden gauge coupling.  If \(\phi\) is a modulus, the same scalar
profile may also change internal volumes, thresholds, or effective
four-dimensional couplings for any sector that couples to it.

The size of the black hole is important in a very concrete way.  The radial
scale of the altered-scalar region is set by the horizon radius \(r_+\), while
the length of the throat is controlled by the near-extremality parameter,
\[
    a=\frac{r_+-r_-}{r_+}.
\]
In the massless GHS solution, the scalar excursion over the regulated throat
grows logarithmically as \(a\to0\).  Thus, a large near-extremal hidden-charged
black hole converts the dimensionless throat profiles studied in this paper
into a macroscopic region of spacetime.  The effect is not tied to the Planck
or string length; its physical size is set by \(r_+\).

\subsection{Altered-coupling regions and possible observational handles}

In the exterior of a near-extremal charged dilatonic black hole, the scalar
can be displaced by an order-one amount over the throat region.  If ordinary
matter, or some mediator connecting ordinary matter to the hidden sector,
couples to this scalar, then the local microphysics near the black hole need
not be the same as in the asymptotic vacuum.  Gauge couplings, mass
thresholds, confinement scales, or compactification radii may all become
position dependent.

The observational consequences depend on how the probe constituents sector is embedded.
If the probe constituents are localized on a brane and the displaced modulus
affects only another brane/sector, then the altered region may be nearly invisible,
except through portals or through gravitational back-reaction.  If, on the
other hand, some of its fields, mediators, or light hidden states propagate
in the bulk or couple appreciably to the modulus, then the local shift of
compactification data can affect its physics. For an observable (Standard Model constituents) probe, scattering rates, cooling, opacity, or line
emission in the inner accretion flow will be affected.  Accretion spectra, near-horizon
emission, and possibly the dynamics of matter close to the black hole are
therefore natural places to look for indirect signatures.  A detailed
phenomenological analysis would require a specific compactification, and a discussion of 
specific couplings of the matter sector, which is beyond the scope of this paper.

Two basic theoretical questions must be answered before such observational
questions become meaningful.  First, is the altered-scalar region large in
absolute units, rather than merely in units of \(r_+\)?  For astrophysical
black holes, the answer can be yes, provided the hidden charge is retained and
the solution is sufficiently close to extremality.  Second, does the scalar
profile survive once the modulus is stabilized?  This is the more universal
question addressed in the main body of the paper.

A strictly massless modulus is not expected in a realistic compactification.
Moduli are stabilized by perturbative and non-perturbative effects, and over a
large field range their potentials need not be well approximated by a
quadratic mass term.  The question is therefore not simply whether the modulus
has a mass, but whether the force from its potential can compete with the
gauge source over the field range sampled by the black hole throat.

This is precisely what the diagnostics of
Section~\ref{sec:static-potential-classes} quantify.  A quadratic potential
leaves the throat only mildly deformed when \(m r_+\lesssim O(0.1\text{--}1)\),
but pins the scalar once the Compton wavelength becomes shorter than the
black hole scale.  Inverse-power runaways do not generate an IR-enhanced
obstruction at the bottom of the throat, although they can deform the outer
matching region at sufficiently large amplitude.  Exponential potentials are
highly sign dependent: a favourable exponential is suppressed along the
inward flow, whereas a dangerous exponential wall is amplified.  Racetrack
potentials introduce a different criterion, namely whether the black hole
excursion crosses the barrier separating the metastable minimum from the
runaway region.  Periodic axion-like potentials can have large local force
while producing a small integrated deformation, because the force changes
sign along the throat.

These are the conditions that any concrete string or supergravity
realization would have to satisfy for a macroscopic altered-modulus region to
exist.  We do not claim that a realistic compactification necessarily realizes
such a region.  The result of the paper is instead a set of sharp criteria:
given a gauge kinetic function and a scalar potential, one can decide whether
the black hole gauge source or the stabilizing potential controls the throat.

\subsection{The special case of quintessence-like scalars}

A particularly light scalar can arise if the field is associated with dark
energy, as in quintessence models (see reviews \cite{Andriot:2026lac,Cicoli:2023opf}).  In such a case, the potential curvature is
naturally of order
\[
    m_{\rm eff}^2 \sim H_0^2 \;,
\]
and the slope is of order \(H_0^2 M_{\rm Pl}\), up to model-dependent
slow-roll factors.  For any black hole whose size is much smaller than the
Hubble radius,
\[
    r_+\ll H_0^{-1},
\]
one has
\begin{equation}
    m_{\rm eff}^2 r_+^2
    \sim
    H_0^2 r_+^2
    \ll 1 \;.
\end{equation}
The scalar is then effectively massless on the horizon scale.  The potential
does not pin the field over the near-horizon region, and the local throat
profile is expected to be very close to the massless GHS solution, with
corrections suppressed by powers of \(H_0 r_+\).

The cosmological motion of the asymptotic scalar value is also negligible on
the dynamical timescale of an astrophysical black hole.  The black hole can
therefore be treated quasi-statically, with the asymptotic value of the
quintessence field fixed during the local evolution.  In this sense,
quintessence-like potentials are among the least obstructive examples: the
potential is simply too shallow to compete with the gauge source on black hole
scales.

This does not by itself make the scenario observable.  A quintessence field
may couple extremely weakly to visible matter, and environmental effects near
an astrophysical black hole may obscure any scalar-induced change of local
microphysics.  It also does not solve the separate question of whether the
black hole can retain a sufficiently large hidden charge over cosmological
time.  The point is narrower: if such a charge is present, a quintessence-like
potential is not the limiting factor for the existence of a gauge-sourced
scalar throat.

\subsection{Outlook}

Whether a macroscopic altered-coupling region survives is never decided by
the mere presence of a potential.  It is decided by how the potential's
force, sign, or barrier structure compares to the gauge-sourced excursion
over the field range that the throat actually samples.  A Compton wavelength
shorter than, or comparable to, the horizon scale can pin the scalar and
erase the throat; a dangerous-orientation exponential wall amplifies toward
the bottom; a racetrack barrier fails once the excursion is large enough.
Inverse powers, minus-sign non-perturbative corrections, and quintessence
are, by contrast, much less disruptive.

For hidden-sector black holes, this is a genuine phenomenological window: if
the scalar also touches visible-sector couplings, or visible fields live in
the bulk, accreting matter tests the local variation of the underlying
effective theory directly.  The constraints that follow are model-dependent,
but the physics is concrete enough to be worth testing, which is reason
enough to treat macroscopic charged throats as more than a formal solution.




\section{Conclusions}
\label{sec:conclusions}

Charged dilatonic black holes provide a simple and robust mechanism for generating large scalar excursions outside a horizon.  In the massless GHS solution, the magnetic gauge source can drive in the appropriate limit the scalar logarithmically along the throat, and the local gauge coupling changes by an order-one, or even parametrically large, amount as one approaches the regulated near-horizon
region.  If the scalar is interpreted as a modulus, this gives a concrete realization of an excursion in moduli space induced by a black hole, and
potentially of an altered-coupling or decompactified region surrounding a charged black hole \cite{Sen:2025ljz}.

The purpose of this paper was to ask how much of this picture survives once
the scalar is not exactly massless.  This is the physically relevant question
for moduli: in any realistic compactification, the scalar should feel a
potential.  We therefore treated the GHS throat as a gauge-sourced background
and studied how a variety of scalar potentials compete with the gauge source.
The central issue is not simply whether a potential is present, but how its force behaves along the particular scalar trajectory traced by the black hole throat.

\subsection*{Fixed-throat diagnostics and back-reacted checks}

We first formulated a set of fixed-throat diagnostics.  The most local one is the pointwise ratio between the potential force and the GHS gauge source,
\begin{equation}
    \eta_{\rm src}(b)
    =
    \frac{
        r_+^2 |V_{,\phi}(\phi_{\rm GHS}(b))|
    }{
        \mathcal S_{\rm gauge}(b)
    } \;,
\end{equation}
where \(\mathcal S_{\rm gauge}\) is the positive magnitude of the gauge source on the GHS background.  This diagnostic identifies where the potential force first becomes locally comparable to the force supporting the massless
throat.  We supplemented it with local and cumulative flux diagnostics, because a small local force can still accumulate over a long radial interval, whereas an oscillatory force can be large pointwise but integrate to a small net deformation.  Finally, we solved directly for the induced scalar
deformation and the corresponding gauge-coupling shift whenever this was the most transparent diagnostic.

For several representative potentials we compared these fixed-throat estimates with back-reacted exterior evolutions of the coupled Einstein--Maxwell--scalar system.  These back-reacted checks serve two purposes.  First, they identify which fixed-throat criteria are genuinely local to the throat and which are artifacts of a long regulated tail or of UV matching.  Second, they give a direct estimate of the actual deformation
of the scalar profile and gauge coupling once the geometry is allowed to respond.

\subsection*{Summary of the potential classes}

Table~\ref{tab:conclusions-summary} summarizes the main outcome for each class of potential studied in this paper.

\begin{table}[t]
\centering
\begin{tabular}{p{3.1cm}p{3.5cm}p{6.8cm}}
\toprule
Potential & Control parameter & Main outcome \\
\midrule
Quadratic &
\(\mu=mr_+\) &
Throat mildly deformed for \(\mu\lesssim O(0.1\text{--}1)\); pinned for
\(\mu\gtrsim O(1)\). \\[2pt]
\\[2pt]
Shifted exponential &
\(\mu=mr_+\) &
Same order-one crossover, with \(\mu\simeq0.3\) giving a ten-percent
back-reacted gauge-coupling deformation in the benchmark scan; the restoring
force saturates rather than grows without bound. \\[2pt]
\\[2pt]
Exponential \hspace{1cm} runaway &
\(q=\lambda/(2\alpha)\) &
Favourable roll: force suppressed toward the bottom and the throat is
protected.\hspace{1cm}Dangerous wall: local source enhanced for \(q>1\), cumulative
bottom impulse enhanced only for \(q>2\); global UV matching can fail even
when the local source ratio is still small. \\[2pt]
\\[2pt]
Inverse-power \hspace{1cm}runaway &
\(q\), amplitude \(\Lambda\) &
Force weakens down the throat; the crossing amplitude tends to an \(O(1)\)
constant as \(a\to0\) rather than vanishing.  The deep throat is not
IR-destabilized, although the outer matching region can deform for large
amplitudes. \\[2pt]
\\[2pt]
Racetrack &
barrier stretching \(L\) &
Survival is controlled almost entirely by the geometric ratio
\(L_{\rm crit}^{\rm geom}=\Delta\chi_h^{(0)}/\Delta\chi_{\rm bar}\);
the racetrack force and back-reaction shift this only at the per-mille
level. \\[2pt]
\\[2pt]
Axion-like periodic &
\(A_a=m_a^2f_a\) &
Near-throat critical force amplitude
\(A_{a,\rm crit}^{\rm throat}\sim10^{-1}\) in Planck units, with mild
dependence on \(f_a\); for \(f_a\ll M_{\rm Pl}\), oscillatory cancellation
strongly suppresses the net deformation. \\[2pt]
\\[2pt]
SUGRA-inspired sign test &
\(\lambda_{\rm crit}^{(\pm)}\) &
At \(s_\infty=5\),
\(\lambda_{\rm crit}^{(+)}\sim10^{-1}\) while
\(\lambda_{\rm crit}^{(-)}\sim5\times10^4\): an \(e^{+s^2/2}\) correction
is dangerous, while an \(e^{-s^2/2}\) correction is essentially harmless. \\
\bottomrule
\end{tabular}
\caption{Summary of the fixed-throat and back-reacted results for the scalar
potentials studied in this paper.}
\label{tab:conclusions-summary}
\end{table}

The quadratic stabilizing potential is the simplest benchmark: a fixed
Compton wavelength \(m^{-1}\) competing with the gauge source.  Long regulated fixed-throat tails are sensitive to small masses through the asymptotic falloff of the profile, but the back-reacted exterior test gives the physically more local criterion.  The throat/exterior region is only mildly deformed as long as
\begin{equation}
    m r_+ \lesssim O(0.1\text{--}1) \;,
\end{equation}
and the GHS-like excursion is erased once \(m r_+\gtrsim O(1)\).  Thus, a massive modulus can follow the black hole gauge source only if it is light on the scale of the black hole itself.

The shifted exponential stabilizing potential gives a closely related but slightly softer test.  We used
\begin{equation}
    V_{\rm sh.exp.}(\varphi)
    =
    \frac{\mu^2 M_{\rm Pl}^2}{\lambda^2 r_+^2}
    \left[
        e^{-\lambda\varphi/M_{\rm Pl}}
        -1
        +\lambda\frac{\varphi}{M_{\rm Pl}}
    \right] \;,
    \qquad
    \mu\equiv m r_+ \;,
\end{equation}
where \(m^2=V_{,\varphi\varphi}(0)\) is the scalar mass at the minimum.  Near its minimum, this reduces to the quadratic potential, but its restoring force saturates at large displacement rather than growing without bound.  We found that the local throat crossover again occurs at an order-one value of
\(\mu=m r_+\), and that in our numerical studies the back-reacted exterior scan gives a ten-percent gauge-coupling deformation at
\begin{equation}
    \mu\simeq 0.3 \;,
\end{equation}
in the same range as the quadratic benchmark.  The important lesson is that the \(\sqrt a\)-type scalings visible in long-tail fixed-throat scans measure the massive asymptotic tail, not the local breakdown of the GHS throat itself.

Exponential runaway potentials behave differently because there is no single
Compton wavelength controlling the problem.  We parametrized them as
\begin{equation}
    V(\varphi)
    =
    V_\infty
    \exp\!\left[
        -\sigma\lambda\frac{\varphi}{M_{\rm Pl}}
    \right] \;,
    \qquad
    \lambda>0 \;,
    \qquad
    \sigma=\pm1 \;,
\end{equation}
where \(\varphi=\phi-\Phi_\infty\) is the GHS scalar displacement.  The
dimensionless slope measured in units of the GHS logarithmic flow is
\begin{equation}
    q=\frac{\lambda}{2\alpha} \;.
\end{equation}
The sign \(\sigma=+1\) is the favourable orientation: as the scalar moves
inward along the GHS throat, it rolls down the exponential and the potential
force is suppressed toward the bottom.  The sign \(\sigma=-1\) is the
dangerous orientation: the black hole flow pushes the scalar up an
exponential wall, and the force is amplified.  For this dangerous
orientation, the bottom source is locally enhanced for
\begin{equation}
    q>1 \;,
\end{equation}
whereas the cumulative impulse from the bottom of the throat is power-law
enhanced only for
\begin{equation}
    q>2 \;.
\end{equation}
Between these thresholds, the pointwise source can become large near the
cutoff even while the integrated bottom impulse remains finite.  The
UV-matched fixed-throat scans also revealed a global effect: preserving the
same asymptotic modulus can force the boundary-value problem onto a strongly
deformed branch, or cause the weak branch connected to GHS to disappear.  The
back-reacted initial-value tests confirm the qualitative hierarchy: \(q=1\)
is mild, while \(q=2\) is much more restrictive.

Inverse-power runaways provide the opposite limiting behaviour.  They were written in terms of a positive dimensionless
variable \(u\), related affinely to the canonical EMD displacement by
\begin{equation}
    u(\phi)
    =
    u_\infty+\frac{\phi-\Phi_\infty}{M_{\rm Pl}} \;,
    \qquad
    u_\infty\equiv u(\Phi_\infty)>0 \;,
\end{equation}
as
\begin{equation}
    V(u)
    =
    \Lambda M_{\rm Pl}^4 u^{-q},
    \qquad
    q>0 \;.
\end{equation}
Along the magnetic GHS flow, \(\varphi\) increases as one moves inward, and
therefore \(u\) also increases.  The potential and its force then decrease
toward the bottom of the throat:
since \(u\) is an affine function of the canonical displacement,
\[
    \frac{\partial u}{\partial\varphi}=\frac{1}{M_{\rm Pl}} \;,
\]
the force along the canonical EMD direction is
\begin{equation}
    V_{,\varphi}
    =
    V_{,u}\frac{\partial u}{\partial\varphi}
    =
    -q\Lambda M_{\rm Pl}^3 u^{-q-1},
    \qquad
    |V_{,\varphi}|
    =
    q\Lambda M_{\rm Pl}^3 u^{-q-1} \;.
\end{equation}
With our sign convention, this force points in the same direction as the
magnetic gauge source, pushing the scalar further along the runaway, but its
magnitude is progressively weakened in the deep throat.  The local
obstruction is therefore smallest near the bottom and largest near the outer
edge of the regulated interval. The crossing amplitude on \(a\le b\le1\) tends to an \(O(1)\) constant as
\(a\to0\), rather than becoming parametrically small.  On a deeper subregion
\[
    a\le b\le b_{\rm cut} \;,
    \qquad
    a\ll b_{\rm cut}\ll1 \;, 
\]
where \(b_{\rm cut}\) marks the outer edge of the deep-throat region, the
amplitude required to compete with the gauge source grows as
\begin{equation}
    b_{\rm cut}^{-1}
    \left(\log\frac1{b_{\rm cut}}\right)^{q+1} \;.
\end{equation}
Thus, the deeper one probes into the throat, the larger the inverse-power
normalization must be in order to compete locally with the magnetic gauge source. The fixed-throat solutions confirm this local picture: even at the crossing
amplitude, the integrated deformation at the throat exit remains only a few
percent.  The back-reacted initial-value test does show visible deformations
in the outer matching region for sufficiently large amplitudes, especially
for shallow powers, but this does not signal an IR-enhanced instability of
the throat.  Inverse-power potentials are therefore mild in the specific
sense that they do not attack the bottom of a long throat.

Racetrack potentials change the question again.  The issue is not whether a
small deformation remains perturbative, but whether the black hole-driven
scalar excursion reaches the top of the finite barrier separating the
metastable minimum from the runaway direction.  By expressing the same EMD
scalar trajectory in a racetrack coordinate \(\chi\), we reduced the
fixed-throat problem to a field-space comparison.  In our convention, the
magnetic GHS flow pushes \(\chi\) toward larger values as one moves inward
along the throat.  Thus, the relevant question is whether the regulated
horizon value \(\chi_h\) remains below the barrier location
\(\chi_{\rm bar}\), or instead reaches and crosses it:
\[
    \chi_h < \chi_{\rm bar}
    \quad \Longrightarrow \quad
    \text{the modulus remains trapped in the metastable basin,}
\]
\[ 
    \chi_h \geq \chi_{\rm bar}
    \quad \Longrightarrow \quad
    \text{the barrier is crossed.}
\]

In the benchmark studied here, the
critical stretching of the barrier is extremely well approximated by the
geometric estimate
\begin{equation}
    L_{\rm crit}^{\rm geom}
    =
    \frac{\Delta\chi_h^{(0)}}{\Delta\chi_{\rm bar}} \;.
\end{equation}
Including the racetrack force shifts this threshold only at the per-mille
level, and the back-reacted check changes it by a comparable amount.  Thus,
for this class of examples, the survival of the metastable vacuum is governed
almost entirely by the black hole-induced field-space excursion relative to the distance from the minimum to the barrier top.

Periodic axion-like potentials illustrate another possibility: the local
force can be large while the integrated effect is small.  We considered
\begin{equation}
    V_a(\phi)
    =
    \Lambda_a^4
    \left[
        1-\cos\left(\frac{\phi-\phi_0}{f_a}\right)
    \right] \;,
\end{equation}
so that the force along the GHS trajectory is
\begin{equation}
    V_{a,\phi}(\phi_{\rm GHS})
    =
    A_a
    \sin\left(\frac{\phi_{\rm GHS}-\phi_0}{f_a}\right) \;,
    \qquad
    A_a=\frac{\Lambda_a^4}{f_a}=m_a^2f_a \;.
\end{equation}
The local diagnostic is therefore controlled by the force amplitude \(A_a\).
For the near-throat interval in our benchmark, the critical value is roughly
\begin{equation}
    A_{a,\rm crit}^{\rm throat}\sim10^{-1}
\end{equation}
in Planck units, with only mild dependence on \(f_a\).

The new feature is that the sign of the force changes along the throat.  If
\(f_a\sim M_{\rm Pl}\), the phase varies slowly over the GHS excursion and
the axion force behaves, over the throat interval, almost like a monotonic
force.  It can then produce order-ten-percent or larger gauge-coupling shifts
near the local threshold.  If instead \(f_a\ll M_{\rm Pl}\), the GHS
excursion covers many axion periods.  The pointwise force may still reach the
local threshold, but successive oscillations contribute with alternating
signs, so the accumulated scalar deformation is strongly reduced.  This
suppression is visible both in the fixed-throat solutions and in the matched
back-reacted check, and it persists even at amplitudes several times above
the near-throat local threshold.

Periodic potentials can therefore be significantly less disruptive than
monotonic potentials with the same pointwise force amplitude.  The relevant
distinction is not the maximum force alone, but whether that force maintains
a coherent sign over the field range sampled by the black hole throat.

The final example was a deliberately limited supergravity-inspired sign
test: inverse powers dressed by \(e^{\pm s^2/2}\), evaluated on a normalized
fixed-unit excursion profile.  This was not meant as a complete
compactification potential, but as a diagnostic of weak-coupling sign
sensitivity.  The result is very sharp.  A plus-sign correction, enhanced at
weak coupling, can deform the regulated profile at the one-percent level for
a normalization
\begin{equation}
    \lambda_{\rm crit}^{(+)}
    \sim
    10^{-1}
    \qquad
    (s_\infty=5) \;,
\end{equation}
whereas the minus-sign correction is exponentially suppressed and requires
\begin{equation}
    \lambda_{\rm crit}^{(-)}
    \sim
    5\times10^4
\end{equation}
for the same benchmark.  The ratio
\begin{equation}
    \frac{\lambda_{\rm crit}^{(-)}}{\lambda_{\rm crit}^{(+)}}
    \simeq
    4.8\times10^5
\end{equation}
is the cleanest quantitative expression of this sign sensitivity.  The
lesson is not that supergravity-inspired corrections generically destroy the
GHS exterior; rather, their sign and weak-coupling behaviour are decisive. The two signs lead to parametrically separated critical normalizations, and leave no ambiguous intermediate regime for this diagnostic.
In the particular benchmark studied here, we see that changing the sign produces results that differ by almost six orders of magnitude.


Taken together, these examples give a qualitative hierarchy of threats to the
GHS throat.  Quadratic stabilizing potentials erase the throat when the
Compton wavelength becomes shorter than the horizon scale.  Dangerous
exponential walls can be amplified toward the bottom of a long throat.
Racetrack potentials can be destabilized if the black hole-induced excursion exceeds
the barrier distance.  Slowly varying axion potentials can produce sizeable
deformations if their force is large enough.  By contrast, inverse-power
runaways, rapidly oscillating axions, and minus-sign non-perturbative
corrections are much less disruptive in the regimes identified above.

The decisive criterion is always the same: one must compare the
gauge-sourced trajectory of the black hole throat with the force,
periodicity, or barrier structure of the scalar potential over the actual
field range sampled by the solution.  A potential that is large in some
abstract region of field space may be harmless if it is weak along the
black hole trajectory.  Conversely, a potential that looks innocuous near the
asymptotic vacuum can become dangerous if the throat drives the scalar into a
region where the force is exponentially enhanced, where cancellations are
absent, or where a finite barrier is crossed.

\subsection*{Phenomenological outlook and open directions}

The phenomenological implications are necessarily model-dependent.  If the
charged black hole carries only hidden-sector charge and the scalar couples
only to hidden fields, the altered-coupling region may be difficult to probe
except through gravitational effects or portal interactions.  If instead the
scalar also controls visible-sector couplings, or if visible fields propagate
in the extra-dimensional bulk, matter accreting through the altered-modulus
region could experience modified microphysics.  This may include shifted
couplings, masses, thresholds, cooling rates, or interaction strengths
diluted by a larger internal volume.  In such scenarios, accretion-disk
spectra and near-horizon emission provide possible observational handles on
the size and persistence of a macroscopic decompactified region that can be strongly constrained in known black holes.

Several directions remain open.  The first is to replace the representative
potentials studied here by potentials derived from explicit compactifications,
with all moduli and gauge sectors included consistently.  The second is to
solve the full UV-matched back-reacted boundary-value problem for the
exponential and racetrack cases, rather than relying on local exterior tests.
The third is to couple the scalar throat to realistic accretion environments,
in order to determine whether the altered-modulus region can leave observable
signatures.

More broadly, the analysis suggests that charged black holes can act as local
probes of moduli potentials.  By forcing the scalar across a controlled region
of field space, they test which potentials allow a macroscopic
altered-coupling throat to survive, and which ones eliminate it.

\section*{Acknowledgments}
AC acknowledges the support of the Initiative Physique des Infinis (IPI), a research training programmme of Idex SUPER at Sorbonne Universit\'e.

\bibliographystyle{apsrev4-2}
\bibliography{GHSphi}

@article{Cicoli:2023opf,
    author = "Cicoli, Michele and Conlon, Joseph P. and Maharana, Anshuman and Parameswaran, Susha and Quevedo, Fernando and Zavala, Ivonne",
    title = "{String cosmology: From the early universe to today}",
    eprint = "2303.04819",
    archivePrefix = "arXiv",
    primaryClass = "hep-th",
    doi = "10.1016/j.physrep.2024.01.002",
    journal = "Phys. Rept.",
    volume = "1059",
    pages = "1--155",
    year = "2024"
}

@article{Garfinkle:1990qj,
    author = "Garfinkle, David and Horowitz, Gary T. and Strominger, Andrew",
    title = "{Charged black holes in string theory}",
    reportNumber = "UCSB-TH-90-66",
    doi = "10.1103/PhysRevD.43.3140",
    journal = "Phys. Rev. D",
    volume = "43",
    pages = "3140",
    year = "1991",
    note = "[Erratum: Phys.Rev.D 45, 3888 (1992)]"
}

@article{Horowitz:1991cd,
    author = "Horowitz, Gary T. and Strominger, Andrew",
    title = "{Black strings and P-branes}",
    reportNumber = "UCSBTH-91-06",
    doi = "10.1016/0550-3213(91)90440-9",
    journal = "Nucl. Phys. B",
    volume = "360",
    pages = "197--209",
    year = "1991"
}

@article{Gregory:1992kr,
    author = "Gregory, Ruth and Harvey, Jeffrey A.",
    title = "{Black holes with a massive dilaton}",
    eprint = "hep-th/9209070",
    archivePrefix = "arXiv",
    reportNumber = "EFI-92-49",
    doi = "10.1103/PhysRevD.47.2411",
    journal = "Phys. Rev. D",
    volume = "47",
    pages = "2411--2422",
    year = "1993"
}

@article{Ferrara:1995ih,
    author = "Ferrara, Sergio and Kallosh, Renata and Strominger, Andrew",
    title = "{N=2 extremal black holes}",
    eprint = "hep-th/9508072",
    archivePrefix = "arXiv",
    reportNumber = "CERN-TH-95-211, SU-ITP-95-16",
    doi = "10.1103/PhysRevD.52.R5412",
    journal = "Phys. Rev. D",
    volume = "52",
    pages = "R5412--R5416",
    year = "1995"
}

@article{Ferrara:1996dd,
    author = "Ferrara, Sergio and Kallosh, Renata",
    title = "{Supersymmetry and attractors}",
    eprint = "hep-th/9602136",
    archivePrefix = "arXiv",
    reportNumber = "UCLA-96-TEP-8, CERN-TH-96-53, SU-ITP-96-8",
    doi = "10.1103/PhysRevD.54.1514",
    journal = "Phys. Rev. D",
    volume = "54",
    pages = "1514--1524",
    year = "1996"
}

@article{Ferrara:1997tw,
    author = "Ferrara, Sergio and Gibbons, Gary W. and Kallosh, Renata",
    title = "{Black holes and critical points in moduli space}",
    eprint = "hep-th/9702103",
    archivePrefix = "arXiv",
    reportNumber = "CERN-TH-97-017, CERN-TH-97-17, DAMTP-R-97-09, SU-ITP-97-05",
    doi = "10.1016/S0550-3213(97)00324-6",
    journal = "Nucl. Phys. B",
    volume = "500",
    pages = "75--93",
    year = "1997"
}

@article{Moore:1998pn,
    author = "Moore, Gregory W.",
    title = "{Arithmetic and attractors}",
    eprint = "hep-th/9807087",
    archivePrefix = "arXiv",
    reportNumber = "YCTP-P17-98",
    journal = " ",
      month = "7",
    year = "1998"
}

@article{Denef:2000nb,
    author = "Denef, Frederik",
    title = "{Supergravity flows and D-brane stability}",
    eprint = "hep-th/0005049",
    archivePrefix = "arXiv",
    doi = "10.1088/1126-6708/2000/08/050",
    journal = "JHEP",
    volume = "08",
    pages = "050",
    year = "2000"
}

@article{Delgado:2022dkz,
    author = "Delgado, Matilda and Montero, Miguel and Vafa, Cumrun",
    title = "{Black holes as probes of moduli space geometry}",
    eprint = "2212.08676",
    archivePrefix = "arXiv",
    primaryClass = "hep-th",
    reportNumber = "IFT-UAM/CSIC-22-151",
    doi = "10.1007/JHEP04(2023)045",
    journal = "JHEP",
    volume = "04",
    pages = "045",
    year = "2023"
}

@misc{Delgado:2025crl,
    author = "Delgado, Matilda and Reymond, S{\'e}bastien and Van Riet, Thomas",
    title = "{Black Holes, Moduli Stabilisation and the Swampland}",
    eprint = "2504.02645",
    archivePrefix = "arXiv",
    primaryClass = "hep-th",
    reportNumber = "MPP-2025-62",
    month = "4",
    year = "2025"
}

@misc{Sen:2025ljz,
    author = "Sen, Ashoke",
    title = "{How to Create a Flat Ten or Eleven Dimensional Space-time in the Interior of an Asymptotically Flat Four Dimensional String Theory}",
    eprint = "2503.00601",
    archivePrefix = "arXiv",
    primaryClass = "hep-th",
    month = "3",
    year = "2025"
}

@article{Danielsson:2006xw,
    author = "Danielsson, Ulf H. and Johansson, Niklas and Larfors, Magdalena",
    title = "{The World next door: Results in landscape topography}",
    eprint = "hep-th/0612222",
    archivePrefix = "arXiv",
    reportNumber = "UUITP-23-06",
    doi = "10.1088/1126-6708/2007/03/080",
    journal = "JHEP",
    volume = "03",
    pages = "080",
    year = "2007"
}

@article{Green:2006nv,
    author = "Green, Daniel R. and Silverstein, Eva and Starr, David",
    title = "{Attractor explosions and catalyzed vacuum decay}",
    eprint = "hep-th/0605047",
    archivePrefix = "arXiv",
    reportNumber = "SLAC-PUB-11846, SU-ITP-06-13",
    doi = "10.1103/PhysRevD.74.024004",
    journal = "Phys. Rev. D",
    volume = "74",
    pages = "024004",
    year = "2006"
}

@misc{Sen:2025bmj,
    author = "Sen, Ashoke",
    title = "{Are Moduli Vacuum Expectation Values or Parameters?}",
    eprint = "2502.07883",
    archivePrefix = "arXiv",
    primaryClass = "hep-th",
    month = "2",
    year = "2025"
}

@misc{Andriot:2026lac,
    author = "Andriot, David",
    title = "{Dark energy from string theory: an introductory review}",
    eprint = "2603.25797",
    archivePrefix = "arXiv",
    primaryClass = "hep-th",
    month = "3",
    year = "2026"
}

@article{Zajacek:2018ycb,
    author = "Zaja{\v{c}}ek, Michal and Tursunov, Arman and Eckart, Andreas and Britzen, Silke",
    title = "{On the charge of the Galactic centre black hole}",
    eprint = "1808.07327",
    archivePrefix = "arXiv",
    primaryClass = "astro-ph.GA",
    doi = "10.1093/mnras/sty2182",
    journal = "Mon. Not. Roy. Astron. Soc.",
    volume = "480",
    number = "4",
    pages = "4408--4423",
    year = "2018"
}

@article{Gibbons:1975kk,
    author = "Gibbons, G. W.",
    title = "{Vacuum Polarization and the Spontaneous Loss of Charge by Black Holes}",
    doi = "10.1007/BF01609829",
    journal = "Commun. Math. Phys.",
    volume = "44",
    pages = "245--264",
    year = "1975"
}

@article{Cardoso:2016olt,
    author = "Cardoso, Vitor and Macedo, Caio F. B. and Pani, Paolo and Ferrari, Valeria",
    title = "{Black holes and gravitational waves in models of minicharged dark matter}",
    eprint = "1604.07845",
    archivePrefix = "arXiv",
    primaryClass = "hep-ph",
    doi = "10.1088/1475-7516/2016/05/054",
    journal = "JCAP",
    volume = "05",
    pages = "054",
    year = "2016",
    note = "[Erratum: JCAP 04, E01 (2020)]"
}

@article{Horne:1992bi,
    author = "Horne, James H. and Horowitz, Gary T.",
    title = "{Black holes coupled to a massive dilaton}",
    eprint = "hep-th/9210012",
    archivePrefix = "arXiv",
    reportNumber = "UCSBTH-92-17",
    doi = "10.1016/0550-3213(93)90621-U",
    journal = "Nucl. Phys. B",
    volume = "399",
    pages = "169--196",
    year = "1993"
}

@article{Gibbons:1987ps,
    author = "Gibbons, G. W. and Maeda, Kei-ichi",
    title = "{Black Holes and Membranes in Higher Dimensional Theories with Dilaton Fields}",
    reportNumber = "UTAP-48-87, LPTENS-87-10",
    doi = "10.1016/0550-3213(88)90006-5",
    journal = "Nucl. Phys. B",
    volume = "298",
    pages = "741--775",
    year = "1988"
}

@article{Garfinkle:1991qj,
    title = {Charged black holes in string theory},
    volume = {43},
    doi = {10.1103/PhysRevD.43.3140},
    journal = {Phys. Rev. D},
    author = {Garfinkle, David and Horowitz, Gary T. and Strominger, Andrew},
    year = {1991},
    note = {Number: UCSB-TH-90-66},
    pages = {3140},
}
\end{document}